\documentclass[12pt]{article}
 \usepackage{ifpdf}
 \usepackage{epsfig}
 \usepackage[latin1]{inputenc}
 \usepackage{amsfonts}
 \usepackage{amsthm}
 \usepackage{amssymb, latexsym, amsmath}
 \usepackage{amsbsy}
 \usepackage{amstext}
 \usepackage{amsopn}
 \usepackage{amsgen}
 \usepackage{graphicx}
 \usepackage[normalem]{ulem}
 \usepackage[usenames, dvipsnames]{color}
 \usepackage{verbatim}
 \usepackage{color}
 \usepackage[T1]{fontenc}
 \usepackage[scale=0.8]{geometry}
\usepackage{fancyhdr}
\usepackage{txfonts}
\usepackage{bm}
\usepackage[bottom]{footmisc}

 \usepackage{authblk} 

\definecolor{mypink}{rgb}{0.858, 0.188, 0.478}
\definecolor{mygrey}{rgb}{0.55, 0.68, 0.55}

 \newcommand{\s}{\nobreak\hspace{.11em}\nobreak}

 \newcommand{\inc}{{\it i}}

 \newcommand{\be}{\begin{equation}}
 \newcommand{\ee}{\end{equation}}
 \newcommand{\ba}{\begin{eqnarray}}
 \newcommand{\ea}{\end{eqnarray}}
 \newcommand{\bs}{\begin{subequations}}
 \newcommand{\es}{\end{subequations}}
 \newcommand{\efbold}{\mbox{{\boldmath $
 f$}}}
 \newcommand{\robold}{\mbox{{\boldmath $
 \rho$}}}
 \newcommand{\erbold}{\mbox{{\boldmath $
 r$}}}
  
 \newcommand{\rbold}{\mbox{{\boldmath $
 r$}}}
 
 \newcommand{\mubold}{\mbox{{\boldmath $\vec \mu$}}}

 \newcommand{\pbold}{\mbox{{\boldmath $
 p$}}}

 \newcommand{\ddotrobold}{\ddot{\bf {\mbox{\boldmath $
 {\boldmath \rho}$}} }}

 \newcommand{\doterbold}{\dot{\textbf {\mbox{\boldmath $
 {\boldmath r}$}}}}
 \newcommand{\Fbold}{\mbox{\boldmath $
 {\boldmath{F}}$}}

\renewcommand{\vec}[1]{\bm{#1}}

\newcommand{\hvec}[1]{\hat{\vec{#1}}}
\newcommand{\mat}[1]{\varmathbb{#1}}
\newcommand{\trans}[1]{{}^\mathrm{T}{#1}}
\newcommand{\In}{\mat I_\mathrm{n}}
\newcommand{\Id}{\mat I\mathrm{d}}
\newcommand{\Frac}[2]{\displaystyle\frac{#1}{#2}}
\newcommand{\mulLagrange}{\Lambda}

\begin{document}
  \title{
         ${{\,}^{^{^{
         {\rm{Published~in:}}
                   ~~Celestial~Mechanics~and~Dynamical~Astronomy
        \,~131\,:\,30\,~(2019)
                  }}}}$\\
 {\Large{\textbf{{{Tidal evolution of the Keplerian elements}
 ~\\}
            }}}}


\author{
                         {\Large{Gwena\"el Bou\'e}}\\
                                          {\small{ASD/IMCCE, Observatoire de Paris, Universit\'e PSL, CNRS, Sorbonne Universit\'e, 77 avenue Denfert-Rochereau, Paris 75014 France}}\\
                                          {\small{e-mail:
                                             ~gwenael.boue$\s$@$\s$obspm.fr~$\,$
                                          }}\\
                                   ~\\
                                     {\Large{and}}\\
                                   ~\\
                                        {\Large{Michael Efroimsky}}\\
                                          {\small{US Naval Observatory, Washington DC 20392 USA}}\\
                                          {\small{e-mail:
                                            ~michael.efroimsky$\s$@$\s$navy.mil~$\,$
                                             }}\\
}

     \date{}

 \maketitle

    \begin{abstract}

 We address the expressions for the rates of the Keplerian orbital elements within a two-body problem perturbed by the tides in both partners. Formulae for these rates have appeared in the literature in various forms, at times with errors. We reconsider, from scratch, the derivation of these rates and arrive at the Lagrange-type equations which, in some details, differ from the corresponding equations obtained previously by Kaula (1964).

 We also write down detailed expressions for $\,da/dt\,$, $\,de/dt\,$ and $\,di/dt\,$, to order $\,e^4\,$. They differ from Kaula's expressions which contain a redundant factor of $\,M/(M+M^{\,\prime})\,$, with $\,M\,$ and $\,M^{\,\prime}\,$ being the masses of the primary and the secondary. As Kaula was interested in the Earth-Moon system, this redundant factor was close to unity and was unimportant in his developments. This factor, however, must be removed when Kaula's theory is applied to a binary composed of partners of comparable masses.

 We have found that, while it is legitimate to simply sum the primary's and secondary's inputs in $\,da/dt\,$ or $\,de/dt\,$, this is not the case for $\,di/dt\,$. So our expression for $\,di/dt\,$ differs from that of Kaula in two regards. First, the contribution due to the dissipation in the secondary averages out when the apsidal precession is uniform. Second, we have obtained an additional term which emerges owing to the conservation of the angular momentum: a change in the inclination of the orbit causes a change of the primary's plane of equator.

    \end{abstract}

 \section{Motivation}

 The Darwin-Kaula theory of bodily tides is a fundamental development with many ramifications. It provides the means for calculating spin-orbit evolution of planets and moons, including their entrapment in spin-orbit resonances (e.g, Correia, Laskar \& de Surgy 2003; Correia \& Laskar 2003;
 Makarov, Berghea \& Efroimsky 2012; Noyelles et al. 2014) and the final obliquities (Cunha et al. 2015; Ferraz-Mello et al. 2008). In the situations where tidal heating is intensive, an approach based on this theory gives the key to the thermal histories of celestial bodies (e.g., Peale \& Cassen 1978; Efroimsky \& Makarov 2014; Efroimsky 2018; Makarov et al. 2018). This theory also enables one to calculate the influence of the lunisolar tides on the orbital motion of an artificial spacecraft (Pucacco \& Lucchesi 2018).

 In the current paper, we address the orbit evolution  within a two-body problem perturbed by the tides in both partners. Specifically, we are interested in the tidal rates $\,da/dt\,$, $\,de/dt\,$, and $\,di/dt\,$ to order $\,e^4\,$ (the symbols being used as defined in Table 1). Formulae for these rates appeared in the literature in various forms, usually to a lower precision and sometimes with mistakes. So we compare our expressions with those suggested in some other publications, including the cornerstone work by Kaula (1964).

 For the additional tidal potential of a disturbed body, Kaula (1964) developed an expansion valid for an arbitrary rheology (i.e., for an arbitrary frequency-dependence of the quality function $\,k_l/Q_l\,$). Kaula's derivation was terse and omitted several steps as self-evident. We accurately fill in these gaps, and point out a step at which Kaula made a tacit approximation $\,M\gg M^{\,\prime}\,$, where $\,M\,$ and $\,M^{\,\prime}\,$ are the masses of the primary and the secondary. Owing to this assumption, Kaula's expressions for the orbital elements' rates contain a redundant factor of $\,M/(M+M^{\,\prime})\,$. Since Kaula was concerned with the Earth-Moon system, this approximation made little difference
 as the factor was close to unity. However, in the case of a binary composed of bodies of comparable masses, this redundant factor must be removed from these expressions.

 We also reexamine from scratch Kaula's derivation of the rates $\,da/dt\,$, $\,de/dt\,$, and $\,di/dt\,$.  We find that, while it is legitimate to simply sum the primary's and secondary's inputs in $\,da/dt\,$ or $\,de/dt\,$, this is not the case for $\,di/dt\,$.
 It turns out that in the expression for the primary's $\,di/dt\,$ the contribution due to the dissipation in the secondary averages out when the apsidal precession is uniform. Also, in that expression we obtain an additional term emerging from the conservation of the angular momentum: a change in the inclination of the orbit causes a change of the primary's plane of equator. For these two reasons, our formula for $\,di/dt\,$ differs considerably from that of Kaula.


 \begin{table*}
 \centering
 \begin{minipage}{190mm}
  \centering
  \caption{Symbol key.}
  \label{description}
  \begin{tabular}{@{}lll@{}}
  \hline
   Variable & Explanation & Reference \\
 \hline
 $\robold$ & inertial position of the primary  &   \\[2pt]
 $\robold^{\,\prime}$ & inertial position of the secondary  &   \\[2pt]
 $\erbold$ & position of the secondary in the primary's equatorial frame & eqn (\ref{relative})  \\[2pt]
  $\erbold^{\,\prime}$ & position of the primary in the secondary's equatorial frame & eqn (\ref{relative})  \\[2pt]


 $U$ & tidal potential of the deformed primary &   \\[2pt]
 $\Fbold$  &  tidal force acting on the secondary &   \\[2pt]
            &  due to the primary's deformation & eqn (\ref{ga})  \\[2pt]

 $U^{\,\prime}$ & tidal potential of the deformed secondary &   \\[2pt]
 $\Fbold^{\,\prime}$  &  tidal force acting on the primary &   \\[2pt]
            &  due to the secondary's deformation & eqn (\ref{do})  \\[2pt]

 $M$ & mass of the primary &   \\[2pt]
 $M^{\,\prime}$ & mass of the secondary &   \\[2pt]

 $R$ & radius of the primary &   \\[2pt]
 $R^{\,\prime}$ & radius of the secondary &   \\[2pt]

   $\rho$  &   mean density of the primary                    &                               \\[2pt]
   $\rho^{\,\prime}$  &   mean density of the secondary                    &                               \\[2pt]

 $\theta$          &  rotation angle of the primary    &             \\[2pt]
 $\theta^{\,\prime}$  &  rotation angle of the secondary  &             \\[2pt]

 $a$ & semimajor axis of the mutual orbit &    \\[2pt]
 $e$ & eccentricity of the mutual orbit &     \\[2pt]
 $i$ & orbit inclination on the primary's equator &        \\[2pt]
 $i^{\,\prime}$ & orbit inclination on the secondary's equator &     \\[2pt]
 $\Omega$ & longitude of the ascending node on the primary's equator &   \\[2pt]
 $\Omega^{\,\prime}$ & longitude of the ascending node on the secondary's equator &   \\[2pt]
 $\omega$ & argument of the pericentre on the primary's equator &   \\[2pt]
 $\omega^{\,\prime}$ & argument of the pericentre on the secondary's equator &   \\[2pt]

 $\mathcal{M}$ & mean anomaly  &   \\[2pt]
 $n\,\equiv\,\dot{\mathcal{M}\,}$ & anomalistic mean motion  &   \\[2pt]

 $F_{lmp}(i)$ & inclination functions &  eqn (\ref{A21})  \\[2pt]
 $G_{lpq}(e)$ & eccentricity function &  eqn (\ref{A21})  \\[2pt]

 $\omega_{lmpq}$ & Fourier modes of the tides in the primary & eqn (\ref{omega}) \\[2pt]
 $\chi_{lmpq}\,\equiv\,|\,\omega_{lmpq}\,|\,$ & forcing frequencies excited in the primary &  \\[2pt]
 $\epsilon_l\,=\,\epsilon_l(\omega_{lmpq})$ & tidal phase lags in the primary &  \\[2pt]
 $k_l\,=\,k_l(\omega_{lmpq})$ & dynamical Love numbers of the primary &   \\[2pt]
 $K_l(\omega_{lmpq})\,\equiv\,k_l(\omega_{lmpq})\,\sin\epsilon_l(\omega_{lmpq})$ & quality functions of the primary &   \\[2pt]

 $\omega^{\,\prime}_{lmpq}$ & Fourier modes of the tides in the secondary & eqn (\ref{omegaprime}) \\[2pt]
 $\chi^{\,\prime}_{lmpq}\,\equiv\,|\,\omega^{\,\prime}_{lmpq}\,|$ & forcing frequencies excited in the secondary &  \\[2pt]
 $\epsilon^{\,\prime}_l\,=\,\epsilon^{\,\prime}_l(\omega^{\,\prime}_{lmpq})$ & tidal phase lags in the secondary &  \\[2pt]
 $k^{\,\prime}_l\,=\,k_l(\omega_{lmpq})$ & dynamical Love numbers of the secondary  &   \\[2pt]
 $K_l^{\,\prime}(\omega_{lmpq})\,\equiv\,k_l^{\,\prime}(\omega_{lmpq})\,\sin\epsilon_l^{\,\prime}(\omega_{lmpq})$ & quality functions of the primary &   \\[2pt]

 ${\cal{G}}$ & gravitational constant & \\[2pt]

\end{tabular}
\end{minipage}
\end{table*}

 \section{Basics\label{section1}}

 \subsection{The two-body problem perturbed by tides\label{section1.1}}

 Consider two near-spherical bodies. One, called ``planet'' or ``primary'', has a mass $\,M\,$ and an inertial position $\,\robold\,$. Another, named
 ``secondary'', has a mass $\,M^{\,\prime}\,$ and is residing in $\,\robold^{\,\prime}\,$. We are interested in the orbital evolution of this system, with tides in both partners taken into account. Within this setting, Kaula (1964) expressed the perturbing gravitational potential through the Keplerian elements of the mutual orbit,
 thus allowing him to describe the evolution of the system by Lagrange's planetary equations. Three caveats are in order regarding Kaula's development.

 First of all, by definition of a mutual orbit, the position of the secondary is measured with respect to the primary's centre of mass. Let $\,\Fbold\,$ be the force exerted by the primary on the secondary. By virtue of Newton's third law of motion, the secondary simultaneously exerts a force $\,-\,\Fbold\,$ on the primary. Hence, the mutual acceleration reads as
 \begin{equation}
 {\vec{a}}\;=\;\ddotrobold^{\,\prime}\;-\;\ddotrobold\;=\;\frac{\Fbold}{M^{\,\prime}}\;-\;\frac{(-\,\Fbold)}{M}\;=\;\frac{M\,+\,M^{\,\prime}}{M\;M^{\,\prime}}\;\Fbold\,\;.
 \label{eq.accForce}
 \end{equation}
 In the limit of the secondary being a test particle ($\,M\,\gg\,M^{\,\prime}\,$), the above relation becomes simply $\,\vec{a}\,=\,\Fbold / M^{\,\prime}\,$, which was the tacit approximation accepted by Kaula. We in our study shall rely on the general expression (\ref{eq.accForce}) and
 therefore shall employ the {\em reduced mass} $\,\beta\,$ defined as
 \begin{equation}
 \beta\;=\;\frac{M\;M^{\,\prime}}{M\,+\,M^{\,\prime}}\,\;.
 \end{equation}

 Secondly, it should be noted that within Kaula's formalism the elements $\,(a,\,e,\,i,\,{\cal M},\,\omega,\,\Omega)\,$ of the mutual orbit are defined in an inertially fixed frame coinciding with the primary's equator at the instant when the equations of motion are computed. To simplify the interpretation of the Keplerian elements, we here assume that they are defined in a frame coprecessing with the primary's equator.\,\footnote{~In this frame, the role of the origin of longitude is played by the descending node of the primary's equator on an inertial plane.}
 The perturbing forces showing up in this setting include the inertial forces associated with the non-Galilean nature of the coprecessing frame.
 An important feature of these forces is that they depend not only on the positions but also on velocities.
 As was pointed out in Efroimsky (2005a,b), for such kind of disturbances the Lagrange- and Delaunay-type planetary equations in their standard form render orbital elements which are not osculating. To be more precise, if we wish the orbital element to osculate the orbit defined in a noninertial frame, not only must we amend the disturbing function, but we should also insert in the Lagrange- and Delaunay-type equations additional terms that are not part of the disturbing function. If, however, we choose only to amend the disturbing function, the orbital equations will give us Keplerian orbital elements which will not osculate with the orbit defined in the noninertial frame, but will instead osculate with the orbit as seen in the inertial frame.\,\footnote{~Such orbital elements are sometimes called \,{\it{contact elements}}, in order to distinguish them from their osculating counterparts.
 For example, the semimajor axis $\,a_{osc}\,$ and eccentricity $\,e_{osc}\,$ osculating in the noninertial frame wherein the orbit is defined
 are linked via $\,\beta\,(\erbold\times\doterbold)\,=\,\beta\,\sqrt{{\cal G}(M+M^{\,\prime})\,a_{osc}\,(1-e_{osc}^2)}\,\hat{\bf{w}}_{osc}\,$ to the relative position $\,\erbold\,$ and the velocity $\,\doterbold\,$ in that same noninertial frame.  At the same time, the contact elements $\,a\,$ and $\,e\,$ (those rendered by the Lagrange- or Delaunay-type equations with only the disturbing function amended) are connected through
 $\,\erbold\times\pbold\,=\,\beta\,\sqrt{{\cal G}(M+M^{\,\prime})\,a\,(1-e^2)}\,\hat{\bf{w}}\,$ with the relative position $\,\erbold\,$ and the momentum $\,\pbold\,$ which is equal to the reduced mass multiplied by the velocity in the inertial frame: $\,\pbold\,=\,\beta\,(\doterbold\,+\,\mubold\times\erbold)\;$.

 In these formulae, $\,\mubold\,$ is the precession rate of the noninertial frame relative the inertial one, while $\,\hat{\bf{w}}_{osc}\,$ and $\,\hat{\bf{w}}\,$ denote the unit vector in the direction of the orbital angular momentum, as seen in the noninertial and inertial frames, correspondingly.

 For comprehensive treatment, see Efroimsky (2005a,b) and references therein. \label{footnote}}

 While the treatment in the coprecessing frame has its advantages, there exists alternatives to it: both the orbital motion and the primary's spin can be described in the Laplace plane (e.g., Bou{\'{e}}, Correia \& Laskar 2016; Rubincam 2016), or the primary's spin can be reckoned from the orbital plane whose Keplerian elements are given in the Laplace plane (e.g., N{\'{e}}ron de Surgy \& Laskar 1997; Correia, Laskar \& N{\'{e}}ron de Surgy 2003; Correia \& Laskar 2003; Correia \& Laskar 2010).

 Lastly, a complete description of motion requires not only the aforementioned set of the Keplerian elements relative to the primary's equator, but also a set of the  elements $\,(a^{\,\prime},\,e^{\,\prime},\,i^{\,\prime},\,{\cal M}^{\,\prime},\,\omega^{\,\prime},\,\Omega^{\,\prime})\,$ relative to the secondary's equator.\,\footnote{~As we shall see below, in the paragraph after equation (\ref{equality}), the instantaneous ellipses, as seen from the primary's and secondary's equator, will have the same shape, with $\,a=a^{\,\prime}\,$ and $\,e=e^{\,\prime}\,$ (and also with $\,{\cal{M}}={\cal{M^{\,\prime}}}\,$). Generally, though, these equalities are not obligatory~---~see an example in Footnote 2 above.
 } Therefore, the equations of motion must satisfy the relations existing between these two sets.

 With all these details taken into account, we now derive the Lagrange-type planetary equations compatible with Kaula's (1964) formalism.

 \subsection{Lagrangian formalism \label{sec.lagrange}}

 \subsubsection{Lagrangian function}

\begin{figure}
 \vspace{0.05mm}
 \centering
 \begin{center}
\includegraphics[angle=0,width=1.1\linewidth
 ,natwidth=290,natheight=242
]{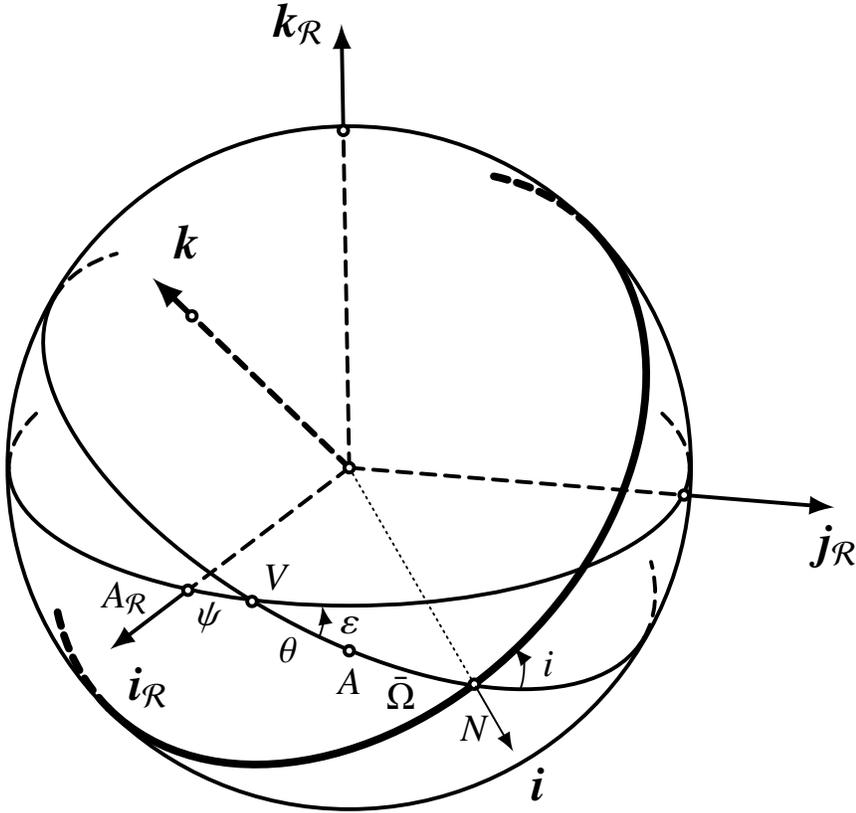}
 \vspace{5mm}
 \caption{\label{fig.euler} Definition of the Euler (3,-1,3) angles $(\psi, \varepsilon, \theta)$.
The inertial frame is denoted with $(\vec i_\mathcal{R}, \vec j_\mathcal{R}, \vec k_\mathcal{R})$, and
the reference plane $(\vec i_\mathcal{R}, \vec j_\mathcal{R})$ is represented by the horizontal
great circle. The origin of longitude, labelled $A_\mathcal{R}$, is the intersection of $\vec
i_\mathcal{R}$ with the unit sphere. The projection of the primary's equatorial plane on the unit
sphere is the great circle whose descending node on the reference plane is $V$.  The point $N$ is the orbit's ascending node on the equator, while
$A$ defines a reference meridian. The intersection of the orbital plane with the unit
 sphere is represented by the thick great circle passing through $N$. The orientation of the primary on the reference plane is parameterised by three angles: the precession angle $\psi$ measured between $A_\mathcal{R}$ and $V$, the tilt $\varepsilon$  of the figure axis, and the rotation angle $\theta$ defining the angular separation of $A$ from $V$.  Analogous quantities $(\psi', \varepsilon', \theta')$ are defined for the secondary (not shown in the figure).  When the orbit is described in the corotating frame of the primary, its longitude of ascending node, $\bar\Omega$, is the angular separation of $N$ from $A$, while its inclination on the primary's equator is $i$.
}
 \end{center}
\end{figure}

 Let $\In$ be the primary's inertia matrix and $\,{\mat{R}}\,$ the rotation matrix, function of the Euler (3,-1,3) 
 angles $\, \Theta = (\psi, \varepsilon, \theta)\,$, describing orientation in an inertial frame. More explicitly,
\begin{equation}
\mat R\,\equiv\,\mat R_3(\psi)\;\mat R_1(-\varepsilon)\;\mat R_3(\theta) ,
\label{eq.rotation}
\end{equation}
where $\mat R_1$ and $\mat R_3$ represent the rotation matrices around the first and third axis,
respectively:
\begin{equation}
\mat R_1(\varphi) = \begin{bmatrix}
1 & 0 & 0 \\
0 & \cos\varphi & -\sin\varphi \\
0 & \sin\varphi &  \cos\varphi
\end{bmatrix} , \qquad
\mat R_3(\varphi) = \begin{bmatrix}
\cos\varphi & -\sin\varphi & 0 \\
\sin\varphi &  \cos\varphi & 0 \\
0 & 0 & 1
\end{bmatrix} .
\end{equation}
 The angle $\psi$ is thus the longitude of the descending node of the equator with respect
to the inertial frame, $\varepsilon$ the inclination of the spin axis in the same inertial frame, and
$\theta$ the rotation angle around the body's figure axis (see Figure~\ref{fig.euler}).
 This convention is chosen for $\theta$ to be reckoned from the descending node of the equator, as in Kaula (1964).
 We similarly denote by $\,\In^{\,\prime}\,$ and $\,{\mat{R}}^{\,\prime}(\,\vec \Theta^{\,\prime}=(\psi^{\,\prime},\varepsilon^{\,\prime},\theta^{\,\prime})\,)\,$ the
inertia and rotation matrices of the secondary, respectively. The planetocentric position of the
secondary will be denoted with $\vec r$, when expressed in the body-fixed frame of the
primary, or with $\vec r^{\,\prime}$, when expressed in the body-fixed frame of the secondary :
\begin{equation}
\vec r = \trans{\mat R}(\vec \rho^{\,\prime}-\vec \rho)\;\;, \qquad
\vec r^{\,\prime} = \trans{\mat R^{\,\prime}}(\vec \rho^{\,\prime}-\vec \rho)\;\,,
\label{relative}
\end{equation}
 where $\trans{(\cdot)}$ denotes the transposition operator.
 Notice that we do not strictly follow Kaula's convention because, to represent the orbit, we are employing the corotating frames instead of the coprecessing ones. At the end of the derivation, we explain how to switch between these two classes of frames.

 For the state of the system to be entirely defined, we also introduce the primary's and secondary's angular velocities $\,\vec \Omega\,$ and
 $\,\vec \Omega^{\,\prime}\,$, respectively. These vectors are expressed in their respective body-fixed frame.

 The Lagrangian $\,{\cal L} = T-V\,$ of the system is a function of $\,(\vec\Omega,\,\vec\Omega^{\,\prime},\,\dot{\vec r\,},\,\vec r,\,\vec r^{\,\prime})\,$. Specifically, the kinetic energy $\,T(\vec\Omega,\,\vec\Omega^{\,\prime},\,\dot{\vec r},\,\vec r)\,$ is given by
 \begin{equation}
 T\;=\;\frac{1}{2}\,\vec \Omega \cdot \In \vec \Omega\,+\,\frac{1}{2}\,\vec \Omega^{\,\prime} \cdot \In^{\,\prime} \vec \Omega^{\,\prime}
 \,+\,\frac{1}{2}\;\beta\;\|\dot{\vec r\,}\,+\,\vec\Omega\times\vec r\,\|^2\,\;,
\label{eq.T}
 \end{equation}
 where $\,\beta = MM^{\,\prime}/(M+M^{\,\prime})\,$ is the reduced mass of the system. As for the potential energy $\,V(\vec r, \vec r^{\,\prime})\,$, we decompose it into the point mass potential energy
$\,V_0(\vec r) = - {\cal G}MM^{\,\prime}/r\,$
and a perturbation $V_1(\vec r, \vec r^{\,\prime})$, whose expression will be specified later on:
\begin{equation}
V(\vec r, \vec r^{\,\prime}) = V_0(\vec r) + V_1(\vec r, \vec r^{\,\prime}) \ .
\label{eq.V}
\end{equation}
 To this Lagrangian function, we have to add the constraint $\,{\mat{R}} \vec r = {\mat{R}}^{\,\prime} \vec r^{\,\prime}\,$ which links two expressions of the same quantity $\,\vec \rho^{\,\prime}-\vec \rho\,$. This constraint will enter the Lagrangian, accompanied with Lagrange multipliers	 $\,\vec\mulLagrange\in\mathbb{R}^3\,$,  leading to a new Lagrangian
 \begin{equation}
 {\cal F} = {\cal L} + \vec \mulLagrange \cdot ({\mat{R}}\vec r - {\mat{R}}^{\,\prime}\vec r^{\,\prime})\,\;.
 \end{equation}
 Note that this Lagrangian also depends on the additional variables $\,(\vec\Theta,\,\vec\Theta^{\,\prime})\,$.

 \subsubsection{Spin operator}

 To derive the Euler-Poincar\'e-Lagrange equations of motion for this Lagrangian, let us first introduce the {\em spin
 operator} $\hvec J\,\equiv\,\mat J\,\partial / \partial \vec \Theta\,$, where $\,\mat J\,$ is a matrix yet to be defined (e.g., Bou\'e 2017; Bou\'e et al. 2017). On the one hand, by definition of the spin operator,\,\footnote{~For a detailed introduction in the theory of the spin operator, see Varshalovich et al. (1988).} the time derivative of an arbitrary function $\,f(\vec \Theta)\,$ can be written as
 \begin{equation}
 \frac{d}{dt}\,f(\vec\Theta)\,=\,\vec\Omega\cdot\hvec J(f) = \vec\Omega \cdot \mat J \frac{\partial
f}{\partial \vec\Theta}\;\,,
 \label{eq.spin1}
 \end{equation}
 with $\vec\Omega$ the angular velocity expressed in the same frame as $\,\hvec J\,$, which here is the
body-fixed frame. On the other, applying the chain rule for the time derivative of $\,f(\vec\Theta)\,$,
we get
 \begin{equation}
 \frac{d}{dt}\,f(\vec\Theta)\,=\,\dot{\vec \Theta} \cdot \frac{\partial f}{\partial \vec\Theta}\,\;.
 \label{eq.spin2}
 \end{equation}
 By identification of (\ref{eq.spin1}) and (\ref{eq.spin2}), we deduce that $\mat J$ is the matrix
such that $\dot{\vec\Theta} = \trans{\mat J}\,\vec\Omega$, or equivalently, $\vec\Omega = \trans{\mat
J}^{-1}\dot{\vec \Theta}$.

 Knowing that the components of $\vec \Omega$ in the body-fixed frame (rotated with respect to an inertial frame according to (\ref{eq.rotation})) are
\begin{eqnarray}
\Omega_X &=& -\dot\psi \sin\varepsilon\sin\theta - \dot\varepsilon \cos\theta , \\
\Omega_Y &=& -\dot\psi \sin\varepsilon\cos\theta + \dot\varepsilon \sin\theta , \\
\Omega_Z &=& \dot\psi\cos\varepsilon + \dot\theta ,
\end{eqnarray}
we get
\begin{equation}
\mat J = \begin{bmatrix}
\displaystyle -\frac{\sin\theta}{\sin\varepsilon} & -\cos\theta & \sin\theta \cot\varepsilon \\[0.8em]
\displaystyle -\frac{\cos\theta}{\sin\varepsilon} &  \sin\theta & \cos\theta \cot\varepsilon \\[0.8em]
0 & 0 & 1
\end{bmatrix} .
\end{equation}
The matrix $\mat J^{\,\prime}$ is equivalently defined for the secondary.

In the following, we also have to determine the image of the function $\mat R \mapsto \vec \mulLagrange \cdot \mat R \vec
r$ by the spin operator $\hvec J$ which, by definition, evaluates the variation of a function under infinitesimal rotation of the primary (therefore the rotation matrix $\mat R$ alone is affected by the operator $\hvec J$ and $\vec r$ shall be taken constant in this calculation). Using $\trans{\mat R}\dot{\mat R}\,\vec v = \vec\Omega \times\vec v$ for any vector $\vec
v\in\mathbb{R}^3$, we get
\begin{equation}
\frac{d}{dt}(\vec \mulLagrange \cdot \mat R \vec r) = \vec \mulLagrange \cdot \dot{\mat R}\vec r
              = \vec \mulLagrange \cdot \mat R\trans{\mat R}\dot{\mat R}\vec r
              = (\trans{\mat R}\vec \mulLagrange) \cdot \left(\vec \Omega \times \vec r\right)
              = \vec \Omega  \cdot\left(\vec r\times (\trans{\mat R}\vec \mulLagrange) \right) .
\label{eq.spin3}
\end{equation}
Identifying (\ref{eq.spin1}) with (\ref{eq.spin3}), we obtain
\begin{equation}
\hvec J (\vec \mulLagrange \cdot \mat R \vec r) = \vec r\times (\trans{\mat R}\vec \mulLagrange) .
\label{eq.spin4}
\end{equation}

 \subsubsection{Equations of motion}

 Using the matrices $\mat J$ and $\mat J^{\,\prime}$ defined hereinabove, the Euler-Poincar\'e-Lagrange equations of motion read (e.g., Bou\'e 2017; Bou\'e et al. 2017)
 \begin{eqnarray}
 \label{eq.dFdOm}
 \frac{d}{dt}\frac{\partial \cal F}{\partial \vec\Omega} &=& \frac{\partial \cal F}{\partial\vec \Omega} \times\vec\Omega + \mat J\frac{\partial \cal F}{\partial \vec \Theta} , \\
 \frac{d}{dt}\frac{\partial \cal F}{\partial \vec\Omega^{\,\prime}} &=& \frac{\partial \cal F}{\partial\vec \Omega^{\,\prime}} \times\vec\Omega^{\,\prime}
 + \mat J^{\,\prime}\frac{\partial \cal F}{\partial \vec \Theta^{\,\prime}} , \\
 \frac{d}{dt}\frac{\partial \cal F}{\partial \dot{\vec r\,}} &=& \frac{\partial \cal F}{\partial\vec r} , \\
 \frac{d}{dt}\frac{\partial \cal F}{\partial \dot{\vec r\,}^{\,\prime}} &=& \frac{\partial \cal F}{\partial\vec r^{\,\prime}} = \vec 0\,\;.
 \label{eq.dFdr'}
 \end{eqnarray}
Notice that the
Euler-Lagrange equation (\ref{eq.dFdr'}) is equal to zero. This is due to the fact that the
Lagrangian does not depend on $\dot{\vec r}^{\,\prime}$. Let us now rewrite the equations of motion
(\ref{eq.dFdOm}-\ref{eq.dFdr'}) in terms of the original Lagrangian $\cal L$ :
\begin{eqnarray}
\label{eq.dLdOmmu}
\frac{d}{dt}\frac{\partial \cal L}{\partial \vec\Omega} &=& \frac{\partial \cal L}{\partial\vec \Omega} \times\vec\Omega
+ \mat J\frac{\partial \cal L}{\partial \vec \Theta}
+ \vec r \times \left(\trans{\mat R}\vec \mulLagrange\right), \\
\label{eq.dLdOmmu'}
\frac{d}{dt}\frac{\partial \cal L}{\partial \vec\Omega^{\,\prime}} &=& \frac{\partial \cal L}{\partial\vec \Omega^{\,\prime}} \times\vec\Omega^{\,\prime}
+ \mat J^{\,\prime}\frac{\partial \cal L}{\partial \vec \Theta^{\,\prime}}
- \vec r^{\,\prime} \times\left(\trans{\mat R^{\,\prime}}\vec\mulLagrange\right), \\
\label{eq.dLdrmu}
\frac{d}{dt}\frac{\partial \cal L}{\partial \dot{\vec r}} &=& \frac{\partial \cal L}{\partial\vec r} + \trans{\mat R}\vec \mulLagrange, \\
\frac{d}{dt}\frac{\partial \cal L}{\partial \dot{\vec r}^{\,\prime}} &=&
\frac{\partial \cal L}{\partial\vec r^{\,\prime}} - \trans{\mat R^{\,\prime}}\vec\mulLagrange = \vec 0 ,
\label{eq.dLdr'}
\end{eqnarray}
with, for the problem studied in this paper, $\partial {\cal L}/\partial\vec\Theta = \partial {\cal
L}/\partial\vec\Theta^{\,\prime} = \vec 0$.
To derive the first two equations, we made use of relation (\ref{eq.spin4}).

From the equation of motion (\ref{eq.dLdr'}), we determine the expression of the Lagrange multiplier,
namely,
\begin{equation}
\vec\mulLagrange = \mat R^{\,\prime} \frac{\partial \cal L}{\partial \vec r^{\,\prime}} .
\end{equation}
Substituting this expression in the other equations of motion (\ref{eq.dLdOmmu}-\ref{eq.dLdrmu}), we get
\begin{eqnarray}
\label{eq.dLdOm}
\frac{d}{dt}\frac{\partial \cal L}{\partial \vec\Omega} &=& \frac{\partial \cal L}{\partial\vec \Omega} \times\vec\Omega
+ \mat J\frac{\partial \cal L}{\partial \vec \Theta}
+ \trans{\mat R}\mat R^{\,\prime} \left(\vec r^{\,\prime} \times \frac{\partial\cal L}{\partial \vec r^{\,\prime}}\right), \\
\label{eq.dLdOm'}
\frac{d}{dt}\frac{\partial \cal L}{\partial \vec\Omega^{\,\prime}} &=& \frac{\partial \cal L}{\partial\vec \Omega^{\,\prime}} \times\vec\Omega^{\,\prime}
+ \mat J^{\,\prime}\frac{\partial \cal L}{\partial \vec \Theta^{\,\prime}}
- \vec r^{\,\prime} \times \frac{\partial \cal L}{\partial \vec r^{\,\prime}}, \\
\frac{d}{dt}\frac{\partial \cal L}{\partial \dot{\vec r\,}} &=& \frac{\partial \cal L}{\partial\vec r} + \trans{\mat R}\mat R^{\,\prime} \frac{\partial \cal L}{\partial \vec r^{\,\prime}} .
\label{eq.dLdr2}
\end{eqnarray}
In the first equation of motion, we made use of the relations $\vec r = \trans{\mat R}\mat R^{\,\prime} \vec
r^{\,\prime}$ and $(\mat R\vec u \times \mat R\vec v) = \mat R(\vec u\times\vec v)$. In the last equation of
motion, we recognise the force $\vec F^{\,\prime} \equiv \partial {\cal L} / \partial \vec r^{\,\prime}$ written in the
body-fixed frame of the primary thanks to the rotation matrix $\trans{\mat R}\mat R^{\,\prime}$. Moreover, in
the first two equations of motion, we observe the presence of the torque $\vec r^{\,\prime}\times\vec F^{\,\prime}$
expressed in the body-fixed frame of the primary in equation (\ref{eq.dLdOm}) and in the body-fixed frame
of the secondary in equation (\ref{eq.dLdOm'}).

 Inserting the expressions for the kinetic energy (\ref{eq.T}) and the potential energy (\ref{eq.V}) in equation (\ref{eq.dLdr2}), we get
\begin{equation}
\ddot{\vec r} + \frac{{\cal G}(M+M^{\,\prime})}{r^3}\vec r =
-  \dot{\vec \Omega}\times\vec r
- 2\vec\Omega\times\dot{\vec r}
-  \vec\Omega\times(\vec\Omega\times\vec r) -
\frac{1}{\beta}\left(
\frac{\partial V_1}{\partial \vec r} + \trans{\mat R}{\mat R^{\,\prime}}\frac{\partial V_1}{\partial \vec r^{\,\prime}}
\right) .
\label{eq.ddotr1}
\end{equation}
 Were the right-hand side equal to zero, we would have obtained the equation of motion of the classical two-body problem. But here this is not the case. The first three terms of the right-hand side account for the inertial forces of the non-Galilean frame in which $\vec r$ is expressed, while the last two terms represent the perturbation induced by tides. It is common to name the quantity $\,-\,V_1/\beta\,$ as the {\em perturbing function} and to denote it with $\,\cal R\,$. In this notation, equation (\ref{eq.ddotr1}) reads:
\begin{equation}
\ddot{\vec r} + \frac{{\cal G}(M+M^{\,\prime})}{r^3}\vec r =
-  \dot{\vec \Omega}\times\vec r
- 2\vec\Omega\times\dot{\vec r}
-  \vec\Omega\times(\vec\Omega\times\vec r) +
\left(
\frac{\partial \cal R}{\partial \vec r} + \trans{\mat R}{\mat R^{\,\prime}}\frac{\partial \cal R}{\partial \vec r^{\,\prime}}
\right) .
\label{eq.ddotr}
\end{equation}

\subsection{Hamiltonian formalism}

To get the Hamiltonian form of the equations of motion, we apply a Legendre transformation on the
Lagrangian. Let $\vec \Sigma$, $\vec \Pi^{\,\prime}$ and $\vec p$ be the generalised momenta given by
\begin{eqnarray}
\vec\Sigma &\equiv& \frac{\partial \cal L}{\partial \vec \Omega} = \In \vec\Omega + \beta \vec r \times
(\dot{\vec r} + \vec\Omega \times\vec r) , \\
\vec\Pi^{\,\prime} &\equiv& \frac{\partial \cal L}{\partial \vec \Omega^{\,\prime}} = \In^{\,\prime} \vec \Omega^{\,\prime} , \\
\vec p &\equiv& \frac{\partial \cal L}{\partial \dot{\vec r}} = \beta ( \dot{\vec r} + \vec\Omega \times\vec r ) .
\end{eqnarray}
$\vec \Sigma=\vec\Pi + \vec \Gamma$ is the sum of the angular momenta of the primary ($\vec\Pi$)
and of the orbit ($\vec\Gamma = \vec r\times\vec p$), $\vec \Pi^{\,\prime}$ is the angular momentum of the secondary and
$\vec p$ is the linear orbital momentum with respect to the inertial frame but expressed in the
body-fixed frame of the primary.

The Hamiltonian ${\cal H} \equiv \vec \Sigma \cdot\vec \Omega + \vec\Pi^{\,\prime}\cdot\vec\Omega^{\,\prime} + \vec p
\cdot\dot{\vec r} - {\cal L}$ can be written as ${\cal H} = T + V$ with
\begin{equation}
T = \frac{1}{2}(\vec \Sigma - \vec r\times\vec p)\cdot\In^{-1}(\vec \Sigma - \vec r\times\vec p)
  + \frac{1}{2}\vec \Pi^{\,\prime} \cdot \In^{\prime-1}\vec\Pi^{\,\prime} + \frac{\|\vec p\|^2}{2\beta} .
\end{equation}
The equations of motion deduced from the Legendre transformation are
\begin{eqnarray}
\label{eq.dPi}
\frac{d\vec\Sigma}{dt} &=& \vec \Sigma \times \frac{\partial \cal H}{\partial \vec \Sigma}
 - \mat J \;\frac{\partial \cal H}{\partial \vec \Theta}
- \trans{\mat R}\mat R^{\,\prime}
\left(\vec r^{\,\prime}\times\frac{\partial \cal H}{\partial \vec r^{\,\prime}}\right) , \\
\label{eq.dPi'}
\frac{d\vec\Pi^{\,\prime}}{dt} &=& \vec \Pi^{\,\prime} \times \frac{\partial \cal H}{\partial \vec \Pi^{\,\prime}}
 - \mat J^{\,\prime}\;\frac{\partial \cal H}{\partial \vec \Theta^{\,\prime}}
+ \vec r^{\,\prime}\times\frac{\partial \cal H}{\partial \vec r^{\,\prime}} , \\
\frac{d\vec p}{dt} &=& -\frac{\partial \cal H}{\partial \vec r} - \trans{\mat R}\mat R^{\,\prime}
\frac{\partial \cal H}{\partial \vec r^{\,\prime}} .
\label{eq.dtr}
\end{eqnarray}
 The kinematic equations of motion are
\begin{eqnarray}
\label{eq.dTheta}
\frac{d\vec\Theta}{dt} &=& \trans{\mat J}\frac{\partial \cal H}{\partial \vec \Pi} , \\
\label{eq.dTheta'}
\frac{d\vec\Theta^{\,\prime}}{dt} &=& \trans{\mat J^{\,\prime}}\frac{\partial \cal H}{\partial \vec \Pi^{\,\prime}} , \\
\label{eq.drt}
\frac{d\vec r}{dt} &=& \frac{\partial \cal H}{\partial \vec p} , \\
\frac{d\vec r^{\,\prime}}{dt} &=& \trans{\mat R^{\,\prime}}\mat R \left(\frac{\partial \cal H}{\partial \vec p} - \vec r \times \frac{\partial \cal H}{\partial \vec \Sigma}\right) + \vec r^{\,\prime} \times \frac{\partial \cal H}{\partial \vec \Pi^{\,\prime}} .
\end{eqnarray}
 Recall that these expressions are general. In our case, the two-body problem
perturbed by tides, $\,{\cal H}\,$ is independent of $\,\vec\Theta\,$ and $\,\vec\Theta^{\,\prime}\,$. So the partial derivatives of the Hamiltonian over these two quantities in equations (\ref{eq.dPi}) and (\ref{eq.dPi'}) become zero.

\subsection{Elliptical elements}

 \begin{figure}
 \vspace{0.05mm}
  \centering
  \begin{center}
\includegraphics[angle=0,width=1.1\linewidth
 ,natwidth=290,natheight=242
]{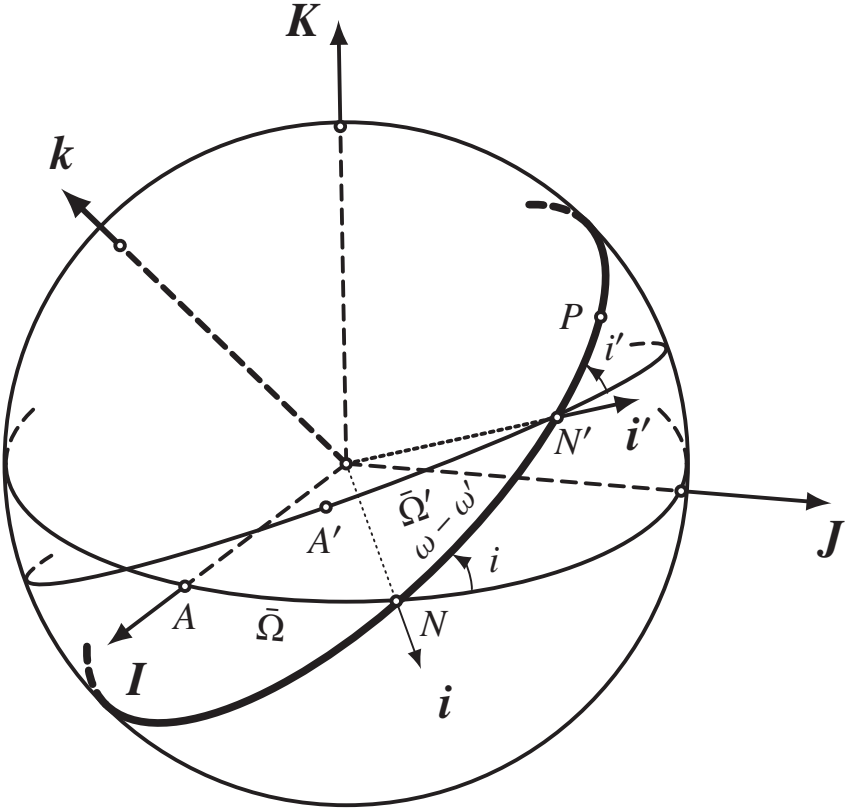}
 \vspace{5mm}
 \caption{\label{fig.angles} Orientation of the orbit as seen from both partners. The projection of the
primary's equatorial plane on the unit sphere is the great circle passing through $A$ -- the
reference meridian -- and $N$ -- the orbit ascending node. Equivalent points are defined on the projection of
the secondary's equator on the unit sphere and are denoted $A^{\,\prime}$ and $N^{\,\prime}$, respectively. The
intersection of the orbital plane with the unit sphere is represented by the thick great circle
passing through $N$, $N^{\,\prime}$ and $P$ -- the direction of the pericentre. When the orbit is described in
the corotating frame of the primary, its longitude of ascending node, denoted $\bar\Omega$, is
the angular separation of $N$ from $A$. Otherwise, in the corotating frame of the secondary, the longitude
of ascending node, denoted $\bar\Omega^{\,\prime}$, is measured between $A^{\,\prime}$ and $N^{\,\prime}$. The angle between $N$ and
$N^{\,\prime}$ is equal to the difference $\omega-\omega^{\,\prime}$ between the two arguments of the orbit pericentre
reckoned from $N$ and $N^{\,\prime}$, respectively. $i$ and $i^{\,\prime}$ are the orbital inclinations with respect to the
primary's and secondary's equator, respectively.}
 \end{center}
\end{figure}

 In Kaula's work, the equations of motion are written in terms of elliptical elements ($a$, $e$, $i$, ${\cal{M}}$, $\omega$, $\Omega$) reckoned from an inertial frame instantaneously comoving with the primary's precessing equator.\,\footnote{~The frame employed by Kaula (1964) should be termed ``instantaneously comoving'', not coprecessing, because in his equations of motion the inertial forces were omitted.}
This set of variables
becomes singular at zero inclination or zero eccentricity. We will nevertheless provide the
equations of motion in this set of variables for an easier comparison with previous works. Here, we
define elliptical elements $(a,e,i,{\cal M},\omega,\bar\Omega)$ (represented in
Figure~\ref{fig.angles}) as a change of variable from the
conjugated variables $(\vec p, \vec r)$. Therefore, they describe the instantaneous ellipse $\cal E$
constructed from the position vector $\vec r$ and its inertial velocity $\dot{\vec r} + \vec\Omega\times\vec r$,
both defined in the primary body-fixed frame. Our set of Keplerian elements differs from Kaula's
by the frame in which it is defined. This choice only impacts the longitude of the ascending node, hence
the introduction of a new symbol $\bar\Omega$. The two longitudes of ascending node are related to each
other by
 \begin{equation}
 \bar\Omega = \Omega - \theta\,\;.
 \label{equality}
 \end{equation}
As for the couple $(\vec p, \vec r)$, to the position vector $\vec r^{\,\prime}$ we associate the
elliptical elements $(a^{\,\prime}, e^{\,\prime}, i^{\,\prime}, {\cal M}^{\,\prime},
\omega^{\,\prime},$ $\bar\Omega^{\,\prime})$ of the ellipse $\cal E^{\,\prime}$ defined by $\vec r^{\,\prime}$
and its inertial velocity, both expressed in the secondary body-fixed frame. Thus, both ellipses $\cal E$ and $\cal E^{\,\prime}$ are
the same up to a rotation, i.e., $a=a^{\,\prime}$, $e=e^{\,\prime}$ and ${\cal M}={\cal
M}^{\,\prime}$. To get the relation between $(\bar\Omega^{\,\prime}, i^{\,\prime}, \omega^{\,\prime})$
and $(\bar\Omega, i, \omega)$, we consider the following function with values in $\mathrm{SO}(3)$ :
\begin{equation}
\mat F = \mat R_3(\bar\Omega)\mat R_1(i)\mat R_3(\omega-\omega^{\,\prime}) \mat R_1(-i^{\,\prime}) \mat R_3(-\bar\Omega^{\,\prime}) \;.
\end{equation}
We have $\mat F = \trans{\mat R}\mat R^{\,\prime}$. Hence, for fixed orientation of the bodies (as in
equations (\ref{eq.dtr}) and (\ref{eq.drt})), we get $d\mat F = \vec 0$ and therefore $d\mat F\,\trans{\mat F} = \vec 0$. Let us denote by $\vec K$, $\vec i$, $\vec k$,
$\vec i^{\,\prime}$, $\vec K^{\,\prime}$ the unit vectors of the rotations of angle $\bar\Omega$, $i$, $\omega-\omega^{\,\prime}$, $-i^{\,\prime}$,
and $-\bar\Omega^{\,\prime}$, respectively. $\vec K$ and $\vec K^{\,\prime}$ are the primary's and secondary's figure axes,
respectively, $\vec i$ and $\vec i^{\,\prime}$ are the directions of the orbit ascending node relative to the
primary's and the secondary's equatorial plane, respectively, and $\vec k$ is the orbit normal (see
Figure~\ref{fig.angles}). In the orbital reference frame, we have in particular
\begin{equation}
\vec k = \begin{pmatrix}
0 \\ 0 \\ 1
\end{pmatrix} , \quad
\vec i = \begin{pmatrix}
1 \\ 0 \\ 0
\end{pmatrix} , \quad
\vec i^{\,\prime} = \begin{pmatrix}
\cos(\omega-\omega^{\,\prime}) \\
\sin(\omega-\omega^{\,\prime}) \\
0
\end{pmatrix} , \quad
\vec K = \begin{pmatrix}
0 \\
\sin i \\
\cos i
\end{pmatrix} , \quad
\vec K^{\,\prime} = \begin{pmatrix}
-\sin i^{\,\prime} \sin(\omega-\omega^{\,\prime}) \\
 \sin i^{\,\prime} \cos(\omega-\omega^{\,\prime}) \\
 \cos i^{\,\prime}
\end{pmatrix} .
\end{equation}
The product $d\mat F\,\trans{\mat F}$ belongs to the Lie algebra $\mathrm{so}(3)$. Let $d\mat X
\equiv \mat R_1(-i)\mat R_3(-\bar\Omega)\,d\mat F \,\trans{\mat F}\,\mat R_3(\bar \Omega) \mat R_1(i)$ be its
expression in the orbit frame. We have (see Appendix~\ref{sec.dX})
\begin{equation}
d\mat X = \hvec K d\bar\Omega + \hvec i d i + \hvec k (d\omega -d\omega^{\,\prime}) - \hvec i^{\,\prime} di^{\,\prime} - \hvec K^{\,\prime}
d\bar\Omega^{\,\prime}\,\;,
\end{equation}
where the hat over any vector $\vec v = (v_x, v_y, v_z)$ denotes the skew-symmetric
matrix\footnote{The skew-symmetric matrix is so defined that for any two vectors $\vec a, \vec
b\in\mathbb{R}^3$, their vector product $\vec a\times \vec b$ is equal to $\hvec a \vec b$.}
\begin{equation}
\hvec v = \begin{bmatrix}
0 & -v_z & v_y \\
v_z & 0 & -v_x \\
-v_y & v_x & 0
\end{bmatrix} .
\end{equation}
Applying to $\,d\mat X = \vec 0\,$ the canonical bijection from $\mathrm{so}(3)$ to $\mathbb{R}^3$ (i.e., the inverse of the
hat application), we get
\begin{equation}
\vec K d\bar\Omega + \vec i di + \vec k d\omega = \vec K^{\,\prime} d\bar\Omega^{\,\prime} + \vec i^{\,\prime} di^{\,\prime} + \vec k d\omega^{\,\prime} .
\end{equation}
We now replace the vectors $\vec K$, $\vec i$, $\vec k$, $\vec i^{\,\prime}$, $\vec K^{\,\prime}$ by their coordinates
and get
\begin{equation}
\begin{bmatrix}
0      & 1 & 0  \\
\sin i & 0 & 0  \\
\cos i & 0 & 1
\end{bmatrix} \begin{pmatrix}
d\bar\Omega \\ d i \\ d\omega
\end{pmatrix} = \begin{bmatrix}
-\sin i^{\,\prime}\sin(\omega-\omega^{\,\prime}) & \cos(\omega-\omega^{\,\prime}) & 0 \\
 \sin i^{\,\prime}\cos(\omega-\omega^{\,\prime}) & \sin(\omega-\omega^{\,\prime}) & 0 \\
 \cos i^{\,\prime} & 0 & 1
\end{bmatrix} \begin{pmatrix}
d\bar\Omega^{\,\prime} \\ di^{\,\prime} \\ d\omega^{\,\prime}
\end{pmatrix} .
\end{equation}
We finally deduce the Jacobian $\mat A_1$ of the transformation $(\bar\Omega^{\,\prime}, i^{\,\prime}, \omega^{\,\prime}) \rightarrow (\bar\Omega, i, \omega)$, which reads
\begin{equation}
\mat A_1 \equiv \frac{\partial(\bar\Omega, i, \omega)}{\partial(\bar\Omega^{\,\prime}, i^{\,\prime}, \omega^{\,\prime})} = \begin{bmatrix}
\displaystyle\frac{\sin i^{\,\prime} \cos(\omega-\omega^{\,\prime})}{\sin i} & \displaystyle\frac{\sin(\omega-\omega^{\,\prime})}{\sin i} & 0 \\[1.0em]
-\sin i^{\,\prime}\sin(\omega-\omega^{\,\prime}) & \cos(\omega-\omega^{\,\prime}) & 0 \\[0.8em]
\cos i^{\,\prime} - \sin i^{\,\prime}\cot i\cos(\omega-\omega^{\,\prime}) & -\cot i \sin(\omega-\omega^{\,\prime}) & 1
\end{bmatrix} ,
\end{equation}
and thus,
\begin{equation}
\mat A \equiv \frac{\partial(a, e, i, {\cal M}, \omega, \bar\Omega)}{\partial(a^{\,\prime}, e^{\,\prime}, i^{\,\prime}, {\cal M}^{\,\prime}, \omega^{\,\prime}, \bar\Omega^{\,\prime})} =
\begin{bmatrix}
1 & 0 & 0 & 0 & 0 & 0 \\
0 & 1 & 0 & 0 & 0 & 0 \\
0 & 0 & \cos(\omega-\omega^{\,\prime}) & 0 & 0 & -\sin i^{\,\prime}\sin(\omega-\omega^{\,\prime}) \\
0 & 0 & 0 & 1 & 0 & 0 \\
0 & 0 & -\cot i\sin(\omega-\omega^{\,\prime}) & 0 & 1 & \cos i^{\,\prime} - \sin i^{\,\prime}\cot i\cos(\omega-\omega^{\,\prime}) \\
0 & 0 & \Frac{\sin(\omega-\omega^{\,\prime})}{\sin i} & 0 & 0 & \Frac{\sin i^{\,\prime}\cos(\omega-\omega^{\,\prime})}{\sin i}
\end{bmatrix} .
\end{equation}

\subsection{Equations of motion of the Keplerian elements}

Planetary equations of motion for $(a, e, i, {\cal M}, \omega, \bar\Omega)$ are deduced from
the canonical equations of motion satisfied by Delaunay variables (rescaled by the reduced mass)
\begin{equation}
\left\{
\begin{array}{ll}
L = \beta \sqrt{{\cal G}(M+M^{\,\prime})a} , \qquad & l = {\cal M} ,\\[0.3em]
G = L \sqrt{1-e^2} , \qquad & g = \omega , \\[0.3em]
H = G \cos i , \qquad & h = \bar\Omega .
\end{array}
\right.
\end{equation}
Let $\vec X = \trans{(L, G, H)}$, $\vec x = \trans{(l, g, h)}$, $\vec Y = \trans{(a, e, i)}$ and $\vec y = \trans{({\cal M},
\omega, \bar\Omega)}$. We have
\begin{equation}
\frac{d}{dt}\begin{pmatrix}
\vec X \\[1em] \vec x
\end{pmatrix} = \begin{bmatrix}
\vec 0 & -\Id \\[1em]
\Id & \vec 0
\end{bmatrix}
\begin{pmatrix}
\displaystyle\frac{\partial \cal H}{\partial \vec X} \\[1em]
\displaystyle\frac{\partial \cal H}{\partial \vec x}
\end{pmatrix} ,
\end{equation}
thus
\begin{equation}
\frac{d}{dt}\begin{pmatrix}
\vec Y \\[1em] \vec y
\end{pmatrix} = \mat M^{-1}\begin{bmatrix}
\vec 0 & -\Id \\[1em]
\Id & \vec 0
\end{bmatrix} \trans{\mat M}^{-1}
\begin{pmatrix}
\displaystyle\frac{\partial \cal H}{\partial \vec Y} \\[1em]
\displaystyle\frac{\partial \cal H}{\partial \vec y}
\end{pmatrix} ,
\end{equation}
where $\mat M$ is the Jacobian defined as
\begin{equation}
\mat M \equiv \frac{\partial (\vec X, \vec x)}{\partial (\vec Y, \vec y)} = \begin{bmatrix}
\mat M_1 & \vec 0 \\
\vec 0 & \Id
\end{bmatrix} , \qquad
\mat M_1 = \begin{bmatrix}
\displaystyle\frac{L}{2a} & 0 & 0 \\[0.8em]
\displaystyle\frac{G}{2a} & -\displaystyle\frac{Ge}{1-e^2} & 0 \\[0.8em]
\displaystyle\frac{H}{2a} & -\displaystyle\frac{He}{1-e^2} & -G\sin i
\end{bmatrix} .
\end{equation}
The result, written in a matrix form, reads
\begin{equation}
\frac{d}{dt}\begin{pmatrix}
a        \\[0.5em]
e        \\[0.5em]
i        \\[0.5em]
{\cal M} \\[0.5em]
\omega   \\[0.5em]
\bar\Omega
\end{pmatrix} =
\begin{bmatrix}
0 & 0 & 0 & -\Frac{2a}{L} & 0 & 0                           \\[0.6em]
0 & 0 & 0 & -\Frac{1-e^2}{Le} & \Frac{1-e^2}{Ge} & 0        \\[0.6em]
0 & 0 & 0 & 0 & -\Frac{\cos i}{G\sin i} & \Frac{1}{G\sin i} \\[0.6em]
\Frac{2a}{L} & \Frac{1-e^2}{Le} & 0 & 0 & 0 & 0             \\[0.6em]
0 & -\Frac{1-e^2}{Ge} & \Frac{\cos i}{G\sin i} & 0 & 0 & 0  \\[0.6em]
0 & 0 & -\Frac{1}{G\sin i} & 0 & 0 & 0
\end{bmatrix} \begin{pmatrix}
\partial{\cal H}/\partial a        \\[0.5em]
\partial{\cal H}/\partial e        \\[0.5em]
\partial{\cal H}/\partial i        \\[0.5em]
\partial{\cal H}/\partial {\cal M} \\[0.5em]
\partial{\cal H}/\partial \omega   \\[0.5em]
\partial{\cal H}/\partial \bar\Omega
\end{pmatrix} .
\label{eq.Poisson}
\end{equation}
 These are the classical planetary equations in the form of Lagrange.  Let us denote by $\,\mat B\,$ the Poisson matrix, i.e., the matrix standing before the gradient of the Hamiltonian in equation (\ref{eq.Poisson}). In the problem under consideration, an adjustment to this equation has to be made. Owing to the constraint between $\vec r$ and $\vec r^{\,\prime}$, we have to add the contribution from $\,(\bar\Omega^{\,\prime}, i^{\,\prime}, \omega^{\,\prime})\,$ to the time derivative of the state vector, see equation (\ref{eq.dtr}). This is done through the medium of the Jacobian $\,\mat A\,$ as follows (see Appendix~\ref{sec.demoEq56})
\begin{equation}
\frac{d}{dt}\begin{pmatrix}
a \\ e \\ i \\ {\cal M} \\ \omega \\ \bar\Omega
\end{pmatrix} = \mat B
\begin{pmatrix}
\partial {\cal H} / \partial a \\
\partial {\cal H} / \partial e \\
\partial {\cal H} / \partial i \\
\partial {\cal H} / \partial {\cal M} \\
\partial {\cal H} / \partial \omega \\
\partial {\cal H} / \partial \bar\Omega
\end{pmatrix}
+ \mat A \mat B^{\,\prime} \begin{pmatrix}
0 \\
0 \\
\partial {\cal H} / \partial i^{\,\prime} \\
0 \\
\partial {\cal H} / \partial \omega^{\,\prime} \\
\partial {\cal H} / \partial \bar\Omega^{\,\prime}
\end{pmatrix}\;\; ,
\end{equation}
 where $\,\mat B^{\,\prime}\,$ is the equivalent of the matrix $\,\mat B\,$ but written as a function of $\,(a,\,e,\,i^{\,\prime})\,$
 instead of $\,(a,\,e,\,i)\,$. After some algebra, we arrive at
 \begin{eqnarray}
 \frac{da}{dt} &=&
 -\frac{2a}{L}\frac{\partial \cal H}{\partial \cal M} , \\
 \frac{de}{dt} &=&
 -\frac{1-e^2}{Le}\frac{\partial \cal H}{\partial \cal M}
 +\frac{1-e^2}{Ge}\left(\frac{\partial \cal H}{\partial \omega}
                      + \frac{\partial \cal H}{\partial \omega^{\,\prime}}\right) , \\
 \frac{di}{dt} &=&
  \frac{1}{G\sin i}\left(\frac{\partial \cal H}{\partial \bar\Omega}
            -\cos i \frac{\partial \cal H}{\partial \omega}\right)
 +\frac{\sin(\omega-\omega^{\,\prime})}{G}\frac{\partial \cal H}{\partial i^{\,\prime}}
 +\frac{\cos(\omega-\omega^{\,\prime})}{G\sin i^{\,\prime}}\left(\frac{\partial \cal H}{\partial \bar\Omega^{\,\prime}}
 -      \cos i^{\,\prime}                              \frac{\partial \cal H}{\partial \omega^{\,\prime}}\right) , \\
 \frac{d\cal M}{dt} &=&
  \frac{2a}{L}\frac{\partial \cal H}{\partial a}
 +\frac{1-e^2}{Le}\frac{\partial \cal H}{\partial e} , \\
 \frac{d\omega}{dt} &=&
 -\frac{1-e^2}{Ge}\frac{\partial \cal H}{\partial e}
 +\frac{\cos i}{G\sin i}\frac{\partial \cal H}{\partial i}
 +\frac{\cos i \cos(\omega-\omega^{\,\prime})}{G\sin i}\frac{\partial \cal H}{\partial i^{\,\prime}}
 +\frac{\cos i\sin(\omega-\omega^{\,\prime})}{G\sin i\sin i^{\,\prime}}\left(\cos i^{\,\prime}\frac{\partial \cal H}{\partial \omega^{\,\prime}}
 -                                           \frac{\partial \cal H}{\partial \bar\Omega^{\,\prime}}\right) , \qquad \\
 \frac{d\bar\Omega}{dt} &=&
 -\frac{1}{G\sin i}\frac{\partial \cal H}{\partial i}
 -\frac{\cos(\omega^{\,\prime}-\omega)}{G\sin i}\frac{\partial \cal H}{\partial i^{\,\prime}}
 -\frac{\sin(\omega-\omega^{\,\prime})}{G\sin i\sin i^{\,\prime}}\left(\cos i^{\,\prime}\frac{\partial \cal H}{\partial \omega^{\,\prime}}
 -                                           \frac{\partial \cal H}{\partial \bar\Omega^{\,\prime}}\right) \;\;.
\end{eqnarray}

\subsection{Perturbed two-body problem\label{label}}

Here, we split the Hamiltonian as ${\cal H} = {\cal H}_0 + V_1$ with ${\cal H}_0 = T+V_0$, i.e.,
%
\begin{equation}
{\cal H}_0 =
  \frac{1}{2}(\vec \Sigma-\vec r\times\vec p) \cdot \In^{-1} (\vec \Sigma-\vec r\times\vec p)
+ \frac{1}{2}\vec \Pi^{\,\prime}\cdot\In^{\prime-1}\vec\Pi^{\,\prime}
+ \frac{\|\vec p\|^2}{2\beta}
- \frac{{\cal G}MM^{\,\prime}}{r}
\end{equation}
We shall now write this Hamiltonian in terms of the elliptical elements $(a,e,i,{\cal
M},\omega,\bar\Omega)$. First, we recognise the Keplerian energy of the two-body problem
\begin{equation}
\frac{\|\vec p\|^2}{2\beta}
- \frac{{\cal G}MM^{\,\prime}}{r} = -\frac{{\cal G}MM^{\,\prime}}{2a} .
\end{equation}
Then, $\vec r\times\vec p$ is the orbital angular momentum, thus
\begin{equation}
\vec r \times\vec p \equiv \vec \Gamma = G\begin{pmatrix}
\sin i\sin\bar\Omega \\
-\sin i\cos\bar\Omega \\
\cos i
\end{pmatrix} .
\label{eq.Gamma}
\end{equation}
Moreover, we recall that $\In^{-1}(\vec \Sigma - \vec r\times\vec p)$ is equal to the primary's
angular velocity $\vec\Omega = \trans{(\Omega_X, \Omega_Y, \Omega_Z)}$.
The gradient of the Hamiltonian ${\cal H}_0$ is then
\begin{eqnarray}
&&
\frac{\partial {\cal H}_0}{\partial a}
 = \frac{{\cal G}MM^{\,\prime}}{2a^2}
  -\frac{G}{2a}\left((\Omega_X\sin\bar\Omega - \Omega_Y\cos\bar\Omega)\sin i + \Omega_Z\cos i\right) ,
\\ &&
\frac{\partial {\cal H}_0}{\partial e}
 = \frac{G e}{1-e^2}\left((\Omega_X\sin\bar\Omega - \Omega_Y\cos\bar\Omega)\sin i + \Omega_Z\cos i\right) ,
\\ &&
\frac{\partial {\cal H}_0}{\partial i}
 = -G\left((\Omega_X\sin\bar\Omega-\Omega_Y\cos\bar\Omega)\cos i-\Omega_Z\sin i\right) ,
\\ &&
\frac{\partial {\cal H}_0}{\partial \cal M}
 = 0 ,
\\ &&
\frac{\partial {\cal H}_0}{\partial \omega}
 = 0 ,
\\ &&
\frac{\partial {\cal H}_0}{\partial \bar\Omega}
 = -G(\Omega_X\cos\bar\Omega+\Omega_Y\sin\bar\Omega)\sin i .
\end{eqnarray}
Let $n = ({\cal G}(M+M^{\,\prime})/a^3)^{1/2}$ be the Keplerian mean motion. The equations of motion become
\begin{eqnarray}
\label{eq.daV1}
\frac{da}{dt} &=&
 -\frac{2a}{L}\frac{\partial V_1}{\partial \cal M} , \\
\frac{de}{dt} &=&
 -\frac{1-e^2}{Le}\frac{\partial V_1}{\partial \cal M}
 +\frac{1-e^2}{Ge}\left(\frac{\partial V_1}{\partial \omega}
                      + \frac{\partial V_1}{\partial \omega^{\,\prime}}\right) , \\
\frac{di}{dt} &=&
 -(\Omega_X\cos\bar\Omega+\Omega_Y\sin\bar\Omega)
 +\frac{1}{G\sin i}\left(\frac{\partial V_1}{\partial \bar\Omega}
            -\cos i \frac{\partial V_1}{\partial \omega}\right)
 +\frac{\sin(\omega-\omega^{\,\prime})}{G}\frac{\partial V_1}{\partial i^{\,\prime}}
\nonumber
\\ &&
 +\frac{\cos(\omega-\omega^{\,\prime})}{G\sin i^{\,\prime}}\left(\frac{\partial V_1}{\partial \bar\Omega^{\,\prime}}
 -      \cos i^{\,\prime}                              \frac{\partial V_1}{\partial \omega^{\,\prime}}\right) , \\
\frac{d\cal M}{dt} &=&
  n
 +\frac{2a}{L}\frac{\partial V_1}{\partial a}
 +\frac{1-e^2}{Le}\frac{\partial V_1}{\partial e} , \\
\frac{d\omega}{dt} &=&
 -\frac{\Omega_X\sin\bar\Omega-\Omega_Y\cos\bar\Omega}{\sin i}
 -\frac{1-e^2}{Ge}\frac{\partial V_1}{\partial e}
 +\frac{\cos i}{G\sin i}\frac{\partial V_1}{\partial i}
 +\frac{\cos i \cos(\omega-\omega^{\,\prime})}{G\sin i}\frac{\partial V_1}{\partial i^{\,\prime}}
\nonumber
\\ &&
 +\frac{\cos i\sin(\omega-\omega^{\,\prime})}{G\sin i\sin i^{\,\prime}}\left(\cos i^{\,\prime}\frac{\partial V_1}{\partial \omega^{\,\prime}}
 -                                           \frac{\partial V_1}{\partial \bar\Omega^{\,\prime}}\right) , \\
\frac{d\bar\Omega}{dt} &=&
 (\Omega_X\sin\bar\Omega-\Omega_Y\cos\bar\Omega)\cot i -\Omega_Z
 -\frac{1}{G\sin i}\frac{\partial V_1}{\partial i}
 -\frac{\cos(\omega^{\,\prime}-\omega)}{G\sin i}\frac{\partial V_1}{\partial i^{\,\prime}}
\nonumber
\\ &&
 -\frac{\sin(\omega-\omega^{\,\prime})}{G\sin i\sin i^{\,\prime}}\left(\cos i^{\,\prime}\frac{\partial V_1}{\partial \omega^{\,\prime}}
 -                                           \frac{\partial V_1}{\partial \bar\Omega^{\,\prime}}\right) .
\label{eq.dOmV1}
\end{eqnarray}
Using the expressions $L=\beta n a^2$ and $G = L(1-e^2)^{1/2}$ and substituting $V_1$ by $-\beta \cal R$, these equations of
motion read
\begin{eqnarray}
\label{eq.da}
\frac{da}{dt} &=&
  \frac{2}{n a}\frac{\partial \cal R}{\partial \cal M} , \\
\frac{de}{dt} &=&
  \frac{1-e^2}{n a^2 e}\frac{\partial \cal R}{\partial \cal M}
 -\frac{(1-e^2)^{1/2}}{n a^2 e}\left(\frac{\partial \cal R}{\partial \omega}
                      + \frac{\partial \cal R}{\partial \omega^{\,\prime}}\right) , \\
\frac{di}{dt} &=&
 -(\Omega_X\cos\bar\Omega+\Omega_Y\sin\bar\Omega)
 -\frac{1}{n a^2 (1-e^2)^{1/2}\sin i}\left(\frac{\partial \cal R}{\partial \bar\Omega}
            -\cos i \frac{\partial \cal R}{\partial \omega}\right)
\nonumber
\\ &&
 -\frac{\sin(\omega-\omega^{\,\prime})}{n a^2 (1-e^2)^{1/2}}\frac{\partial \cal R}{\partial i^{\,\prime}}
 -\frac{\cos(\omega-\omega^{\,\prime})}{n a^2 (1-e^2)^{1/2} \sin i^{\,\prime}}\left(\frac{\partial \cal R}{\partial \bar\Omega^{\,\prime}}
 -      \cos i^{\,\prime}                              \frac{\partial \cal R}{\partial \omega^{\,\prime}}\right) , \\
\frac{d\cal M}{dt} &=&
  n
 -\frac{2}{n a}\frac{\partial \cal R}{\partial a}
 -\frac{1-e^2}{n a^2 e}\frac{\partial \cal R}{\partial e} , \\
\frac{d\omega}{dt} &=&
 -\frac{\Omega_X\sin\bar\Omega-\Omega_Y\cos\bar\Omega}{\sin i}
 +\frac{(1-e^2)^{1/2}}{n a^2 e}\frac{\partial \cal R}{\partial e}
 -\frac{\cos i}{n a^2 (1-e^2)^{1/2}\sin i}\frac{\partial \cal R}{\partial i}
\nonumber
\\ &&
 -\frac{\cos i \cos(\omega-\omega^{\,\prime})}{n a^2 (1-e^2)^{1/2}\sin i}\frac{\partial \cal R}{\partial i^{\,\prime}}
 -\frac{\cos i\sin(\omega-\omega^{\,\prime})}{n a^2 (1-e^2)^{1/2}\sin i\sin i^{\,\prime}}\left(\cos i^{\,\prime}\frac{\partial \cal R}{\partial \omega^{\,\prime}}
 -                                           \frac{\partial \cal R}{\partial \bar\Omega^{\,\prime}}\right) , \\
\frac{d\bar\Omega}{dt} &=&
 (\Omega_X\sin\bar\Omega-\Omega_Y\cos\bar\Omega)\cot i -\Omega_Z
 +\frac{1}{n a^2 (1-e^2)^{1/2} \sin i}\frac{\partial \cal R}{\partial i}
\nonumber
\\ &&
 +\frac{\cos(\omega^{\,\prime}-\omega)}{n a^2 (1-e^2)^{1/2}\sin i}\frac{\partial \cal R}{\partial i^{\,\prime}}
 +\frac{\sin(\omega-\omega^{\,\prime})}{n a^2 (1-e^2)^{1/2} \sin i\sin i^{\,\prime}}\left(\cos i^{\,\prime}\frac{\partial \cal R}{\partial \omega^{\,\prime}}
 -                                           \frac{\partial \cal R}{\partial \bar\Omega^{\,\prime}}\right) .
\label{eq.dOm}
\end{eqnarray}
 Had we defined the Keplerian elements in the coprecessing frame of the primary rather than in
its corotating frame, the longitude of the ascending node would have been $\Omega = \bar\Omega +
\theta$. Then, in all the equations of motion, $\bar\Omega$ would have been replaced with $\Omega -
\theta$, and in particular the last equation of motion would have become
\begin{equation}
\begin{split}
\frac{d\Omega}{dt} =& (\Omega_X\sin(\Omega-\theta)-\Omega_Y\cos(\Omega-\theta))\cot i - (\Omega_Z-\dot\theta)
 +\frac{1}{n a^2 (1-e^2)^{1/2}\sin i}\frac{\partial \cal R}{\partial i}
\\ &
 +\frac{\cos(\omega^{\,\prime}-\omega)}{n a^2 (1-e^2)^{1/2} \sin i}\frac{\partial \cal R}{\partial i^{\,\prime}}
 +\frac{\sin(\omega-\omega^{\,\prime})}{n a^2 (1-e^2)^{1/2} \sin i\sin i^{\,\prime}}\left(\cos i^{\,\prime}\frac{\partial \cal R}{\partial \omega^{\,\prime}}
 -                                           \frac{\partial \cal R}{\partial \Omega^{\,\prime}}\right) .
\end{split}
\end{equation}
Let us now compare equations (\ref{eq.da}-\ref{eq.dOm}) with the well-known Lagrange planetary equations.
As the Keplerian elements are here defined with respect to a moving frame, the
resulting equations of motion for $i$, $\omega$ and $\bar\Omega$ contain driving terms which are
functions of the components $(\Omega_X, \Omega_Y, \Omega_Z)$ of the angular velocity $\vec\Omega$ of
this frame. Furthermore, we observe that the equations of motion are not symmetric in
$(i,\omega,\bar\Omega)$ and in $(i^{\,\prime},\omega^{\,\prime},\bar\Omega^{\,\prime})$. This is due
to the rotation matrix $\trans{\mat R}\mat R^{\,\prime}$ between the frames in which the two sets of
Keplerian elements are defined. We nevertheless recover the lost symmetry in the equations of motion
when this matrix $\trans{\mat R}\mat R^{\,\prime}$ becomes a single rotation around the third axis
(i.e., when $i=i^{\,\prime}$ and $\omega=\omega^{\,\prime}$).

\subsection{Rotation equations of motion}

 Equations (\ref{eq.da}-\ref{eq.dOm}) are the equivalent of the Lagrange planetary equations. For completeness, we now provide the explicit equations of motion for the rotation of the two bodies, given by equations (\ref{eq.dPi}-\ref{eq.dPi'}) and (\ref{eq.dTheta}-\ref{eq.dTheta'}). While these equations should involve the spin operators represented by the matrices $\,\mat J\,$ and $\,\mat J^{\,\prime}\,$, we discard those terms as the perturbing potential energy is independent of $\,\vec\Theta\,$ and $\,\vec\Theta^{\,\prime}\,$. The equations also involve the orbital angular momentum operator $\,\hvec L^{\,\prime} = \vec r^{\,\prime} \times \partial_{\vec r^{\,\prime}}\,$ expressed by a matrix $\,\mat L^{\,\prime}\,$ such that
 \begin{equation}
     \hvec L^{\,\prime}(f) = \mat L^{\,\prime}
 \begin{pmatrix}
 \partial f / \partial \bar\Omega^{\,\prime} \vspace{4mm}
 ~\\
 \partial f / \partial i^{\,\prime} \vspace{4mm}
 ~\\
 \partial f / \partial \omega^{\,\prime}
 \end{pmatrix}
 \end{equation}
 for all functions $\,f(a^{\,\prime},\,e^{\,\prime},\,i^{\,\prime},\,{\cal M}^{\,\prime},\,\omega^{\,\prime},\,\bar\Omega^{\,\prime})\,$. Applying the same approach as in Section~\ref{sec.lagrange}, we arrive at
\begin{equation}
\mat L^{\,\prime} = \begin{bmatrix}
-\sin\bar\Omega^{\,\prime}\cot i^{\,\prime} & \cos\bar\Omega^{\,\prime} & \Frac{\sin\bar\Omega^{\,\prime}}{\sin i^{\,\prime}} \\[0.8em]
\cos\bar\Omega^{\,\prime}\cot i^{\,\prime} & \sin\bar\Omega^{\,\prime} & -\Frac{\cos\bar\Omega^{\,\prime}}{\sin i^{\,\prime}} \\[0.8em]
1 & 0 & 0
\end{bmatrix} .
\end{equation}
Developing the matrix products, we obtain the following dynamical equations of motion:
 \begin{eqnarray}
 \label{eq.dSigmax}
 \frac{d\Sigma_X}{dt} &=&
 \Sigma_Y\,\Omega_Z\,-\,\Sigma_Z\,\Omega_Y
 \,+\,\frac{\cos\bar\Omega\,\sin(\omega-\omega^{\,\prime})\,+\,\sin\bar\Omega\,\cos i\,\cos(\omega-\omega^{\,\prime})}{\sin i^{\,\prime}}\;\frac{\partial V_1}{\partial \bar\Omega^{\,\prime}}
 \nonumber \\ &&
 - \left(\cos\bar\Omega\,\cos(\omega-\omega^{\,\prime})-\sin\bar\Omega\,\cos i\,\sin(\omega-\omega^{\,\prime})\right)\frac{\partial V_1}{\partial i^{\,\prime}}
\nonumber \\ &&
- \left((\cos\bar\Omega\sin(\omega-\omega^{\,\prime})+\sin\bar\Omega\cos i\cos(\omega-\omega^{\,\prime}))\cot i^{\,\prime} + \sin\bar\Omega\sin
  i\right)\,\frac{\partial V_1}{\partial \omega^{\,\prime}}\;\,,
\\
\label{eq.dSigmay}
\frac{d\Sigma_Y}{dt} &=&
  \Sigma_Z\,\Omega_X\,-\,\Sigma_X\,\Omega_Z\,
+\,\frac{\sin\bar\Omega\,\sin(\omega-\omega^{\,\prime})\,-\,\cos\bar\Omega\,\cos i\,\cos(\omega-\omega^{\,\prime})}{\sin i^{\,\prime}}\;\frac{\partial
V_1}{\partial \bar\Omega^{\,\prime}}
\nonumber \\ &&
-\,\left(\sin\bar\Omega\cos(\omega-\omega^{\,\prime})+\cos\bar\Omega\cos i\sin(\omega-\omega^{\,\prime})\right)\;\frac{\partial V_1}{\partial
  i^{\,\prime}}
\nonumber \\ &&
\;-\;\left((\sin\bar\Omega\sin(\omega-\omega^{\,\prime})-\cos\bar\Omega\cos i\cos(\omega-\omega^{\,\prime}))\cot i^{\,\prime}-\cos\bar\Omega\sin
i\right)\;\frac{\partial V_1}{\partial \omega^{\,\prime}} ,
 \\
\label{eq.dSigmaz}
\frac{d\Sigma_Z}{dt} &=&
  \Sigma_X\,\Omega_Y\,-\,\Sigma_Y\,\Omega_X
 \;-\;\frac{\sin i\cos(\omega-\omega^{\,\prime})}{\sin i^{\,\prime}}\frac{\partial V_1}{\partial \Omega^{\,\prime}}
 \;-\;\sin i\sin(\omega-\omega^{\,\prime})\;\frac{\partial V_1}{\partial i^{\,\prime}}
\nonumber \\ &&
+\;\left(\sin i\cos(\omega-\omega^{\,\prime})\cot i^{\,\prime} - \cos i\right)\;\frac{\partial V_1}{\partial \omega^{\,\prime}} ,
\\
\label{eq.dPix'}
\frac{d\Pi^{\,\prime}_X}{dt} &=&
   \Pi^{\,\prime}_Y\;\Omega^{\,\prime}_Z\;-\;\Pi^{\,\prime}_Z\Omega^{\,\prime}_X
\;-\;\sin\bar\Omega^{\,\prime}\cot i^{\,\prime}\;\frac{\partial V_1}{\partial \bar\Omega^{\,\prime}}
\;+\;\cos\bar\Omega^{\,\prime}\;\frac{\partial V_1}{\partial i^{\,\prime}}
\;+\;\frac{\sin\bar\Omega^{\,\prime}}{\sin i^{\,\prime}}\;\frac{\partial V_1}{\partial \omega^{\,\prime}} , \\
\label{eq.dPiy'}
\frac{d\Pi^{\,\prime}_Y}{dt} &=&
\Pi^{\,\prime}_Z\,\Omega^{\,\prime}_X\,-\;\Pi^{\,\prime}_X\;\Omega^{\,\prime}_Z
 \,+\;\cos\bar\Omega^{\,\prime}\,\cot i^{\,\prime}\;\frac{\partial V_1}{\partial \bar\Omega^{\,\prime}}
 \;+\;\sin\bar\Omega^{\,\prime}\;\frac{\partial V_1}{\partial i^{\,\prime}}
 \;-\;\frac{\cos\bar\Omega^{\,\prime}}{\sin i^{\,\prime}}\;\frac{\partial V_1}{\partial \omega^{\,\prime}} ,\\
 \label{eq.dPiz'}
\frac{d\Pi^{\,\prime}_Z}{dt} &=&
  \Pi^{\,\prime}_X\,\Omega^{\,\prime}_Y\,-\;\Pi^{\,\prime}_Y\,\Omega^{\,\prime}_X
 \,+\;\frac{\partial V_1}{\partial \bar\Omega^{\,\prime}} ,
\end{eqnarray}
 and the associated kinematic equations of motion:
 \begin{eqnarray}
 \frac{d\psi}{dt} &=& -\frac{\sin\theta}{\sin\varepsilon}\,\Omega_X -
 \frac{\cos\theta}{\sin\varepsilon}\,\Omega_Y\,\;, \\
 \frac{d\varepsilon}{dt} &=& -\cos\theta\;\Omega_X + \sin\theta\;\Omega_Y\,\;,  \\
 \frac{d\theta}{dt} &=& \sin\theta\cot\varepsilon\;\Omega_X + \cos\theta\cot\varepsilon\;\Omega_Y + \Omega_Z\,\;,  \\
 \frac{d\psi^{\,\prime}}{dt} &=& -\frac{\sin\theta^{\,\prime}}{\sin\varepsilon^{\,\prime}}\,\Omega^{\,\prime}_X -
 \frac{\cos\theta^{\,\prime}}{\sin\varepsilon^{\,\prime}}\,\Omega^{\,\prime}_Y\,\;,  \\
 \frac{d\varepsilon^{\,\prime}}{dt} &=& -\cos\theta^{\,\prime}\;\Omega^{\,\prime}_X + \sin\theta^{\,\prime}\;\Omega^{\,\prime}_Y\,\;, \\
 \frac{d\theta^{\,\prime}}{dt} &=& \sin\theta^{\,\prime}\cot\varepsilon^{\,\prime}\;\Omega^{\,\prime}_X + \cos\theta^{\,\prime}\cot\varepsilon^{\,\prime}\;\Omega^{\,\prime}_Y + \Omega^{\,\prime}_Z\;\,.
 \end{eqnarray}
 Recall that in these equations, $\,\vec\Omega = \In^{-1}(\vec\Sigma- \vec r \times\vec p)\,$ and $\,\vec \Omega^{\,\prime} = {\In^{\,\prime}}^{\,-1} \vec \Pi^{\,\prime}\,$.

 \subsection{The gyroscopic approximation and the approximation of constant inertia matrices}

 In the equations of motion derived above, the inertial forces are expressed in terms of the components of the rotation vectors $\,\vec \Omega\,$ and $\,\vec \Omega^{\,\prime}\,$. To simplify these dependencies, we carry out two steps.
  \begin{itemize}

  \item[(a)~] We employ the gyroscopic approximation, i.e., assume that the rotation about the axis of maximal inertia is much faster than any
              change in this axis' orientation. Mathematically, this implies:
  \ba
  |\,\dot\theta\,|\,\gg\,|\dot\psi|\,,\;\,|\,\dot \varepsilon|\qquad\mbox{and}\qquad
  |\,\dot\theta^{\,\prime}\,|\,\gg\,|\,\dot\psi^{\,\prime}\,|\,,\;\,|\,\dot\varepsilon^{\,\prime}\,|\,\;.
   \label{gyro}
  \ea

  \item[(b)~] We neglect the variation of the inertia matrices in the expressions for the angular momenta $\,\vec\Pi\,$ and $\,\vec \Pi^{\,\prime}\,$. This approximation is nontrivial because, after all, the tidal theory is about deformation. So one always needs
      to justify accurately why on one occasion the deformations must be taken into account and neglected on other. This justification is provided in Frouard \& Efroimsky (2017b, Section 3.1 and Footnote 6).
  \end{itemize}

 Let us denote the partners' principal moments of inertia with $\,(A,\,B,\,C)\,$ and $\,(A^{\,\prime},
 \,B^{\,\prime},\,C^{\,\prime})\,$. Then the angular momentum of the primary can be written as
 \begin{eqnarray}
 \label{eq.PiX}
 &&\Pi_X = A \Omega_X = A\left(-\dot\psi\sin\varepsilon\sin\theta-\dot\varepsilon\cos\theta\right)\,\;, \\
 &&\Pi_Y = B \Omega_Y = B\left(-\dot\psi\sin\varepsilon\cos\theta+\dot\varepsilon\sin\theta\right)\,\;, \\
 &&\Pi_Z = C \Omega_Z = C\left(\dot\psi\cos\varepsilon + \dot\theta\right)\,\;.
 \label{eq.PiZ}
 \end{eqnarray}
 In the afore-explained approximation, its rate becomes
 \begin{eqnarray}
 \label{eq.dotPiX}
 &&\dot\Pi_X \approx A \Omega_Z \Omega_Y\,\;, \\
 &&\dot\Pi_Y \approx-B \Omega_Z \Omega_X\,\;, \\
 &&\dot\Pi_Z = C \dot\Omega_Z\,\;.
 \label{eq.dotPiZ}
 \end{eqnarray}
 Similar formulae will be valid for the secondary.

 The rate of the angular momentum,
 \begin{equation}
 \dot{\vec\Pi} = \dot{\vec \Sigma} - \dot{\vec \Gamma}
              = \dot{\vec \Sigma} - \left(
                 \frac{\partial\vec\Gamma}{\partial a} \frac{da}{dt}
               + \frac{\partial\vec\Gamma}{\partial e} \frac{de}{dt}
               + \frac{\partial\vec\Gamma}{\partial i} \frac{di}{dt}
               + \frac{\partial\vec\Gamma}{\partial \Omega} \frac{d\Omega}{dt} \right)
 \end{equation}
 will, owing to equations (\ref{eq.Gamma}), (\ref{eq.daV1}-\ref{eq.dOmV1}), and (\ref{eq.dSigmax}-\ref{eq.dSigmay}), be equal to
\begin{eqnarray}
\label{eq.ratePiX}
&&\dot\Pi_X = \Pi_Y\Omega_Z - \Pi_Z\Omega_Y
 - \sin\bar\Omega\cot i \frac{\partial V_1}{\partial \bar\Omega}
 + \cos\bar\Omega\frac{\partial V_1}{\partial i}
 + \frac{\sin\bar\Omega}{\sin i}\frac{\partial V_1}{\partial \omega} , \qquad \\
&&\dot\Pi_Y = \Pi_Z\Omega_X - \Pi_X\Omega_Z
 + \cos\bar\Omega \cot i\frac{\partial V_1}{\partial\bar\Omega}
 + \sin\bar\Omega\frac{\partial V_1}{\partial i}
 - \frac{\cos\bar\Omega}{\sin i}\frac{\partial V_1}{\partial\omega}
, \qquad \\
&&\dot\Pi_Z = \Pi_X\Omega_Y - \Pi_Y\Omega_X + \frac{\partial V_1}{\partial \bar\Omega} .
\label{eq.ratePiZ}
\end{eqnarray}
 Naturally, the evolution rate of $\,\vec\Pi\,$ has a form analogous to that of
 $\,\vec\Pi^{\,\prime}\,$, equations (\ref{eq.dPix'}-\ref{eq.dPiz'}). Inserting the exact expressions (\ref{eq.PiX} - \ref{eq.PiZ})
 and the approximate expressions (\ref{eq.dotPiX} - \ref{eq.dotPiZ}) in equations (\ref{eq.ratePiX} - \ref{eq.ratePiZ}), we exclude the
 components of the angular momentum, to be left with the components of the angular velocity only:
\begin{eqnarray}
\label{eq.OmegaX}
&&\Omega_X \approx
\frac{1}{(C-A+B)\Omega_Z} \left(
 - \cos\bar\Omega \cot i\frac{\partial V_1}{\partial\bar\Omega}
 - \sin\bar\Omega\frac{\partial V_1}{\partial i}
 + \frac{\cos\bar\Omega}{\sin i}\frac{\partial V_1}{\partial\omega} \right), \qquad \\
\label{eq.OmegaY}
&&\Omega_Y \approx
\frac{1}{(C+A-B)\Omega_Z} \left(
 - \sin\bar\Omega\cot i \frac{\partial V_1}{\partial \bar\Omega}
 + \cos\bar\Omega\frac{\partial V_1}{\partial i}
 + \frac{\sin\bar\Omega}{\sin i}\frac{\partial V_1}{\partial \omega} \right), \qquad \\
&&  \dot\Omega_Z \approx \frac{1}{C}\frac{\partial V_1}{\partial \bar\Omega} ,
\label{eq.OmegaZ}
\end{eqnarray}
with $\Omega_Z \approx \dot\theta$. Similarly, for the secondary we have
\begin{eqnarray}
&&\Omega^{\,\prime}_X \approx
\frac{1}{(C^{\,\prime}-A^{\,\prime}+B^{\,\prime})\Omega^{\,\prime}_Z} \left(
 - \cos\bar\Omega^{\,\prime} \cot i^{\,\prime}\frac{\partial V_1}{\partial\bar\Omega^{\,\prime}}
 - \sin\bar\Omega^{\,\prime}\frac{\partial V_1}{\partial i^{\,\prime}}
 + \frac{\cos\bar\Omega^{\,\prime}}{\sin i^{\,\prime}}\frac{\partial V_1}{\partial\omega^{\,\prime}} \right), \qquad \\
&&\Omega^{\,\prime}_Y \approx
\frac{1}{(C^{\,\prime}+A^{\,\prime}-B^{\,\prime})\Omega^{\,\prime}_Z} \left(
 - \sin\bar\Omega^{\,\prime}\cot i^{\,\prime} \frac{\partial V_1}{\partial \bar\Omega^{\,\prime}}
 + \cos\bar\Omega^{\,\prime}\frac{\partial V_1}{\partial i^{\,\prime}}
 + \frac{\sin\bar\Omega^{\,\prime}}{\sin i^{\,\prime}}\frac{\partial V_1}{\partial \omega^{\,\prime}} \right), \qquad \\
&&  \dot\Omega^{\,\prime}_Z \approx \frac{1}{C^{\,\prime}}\frac{\partial V_1}{\partial \bar\Omega^{\,\prime}} .
\end{eqnarray}
We can now rewrite the equations of motion (\ref{eq.da}-\ref{eq.dOm}) where $\Omega_X$ and
$\Omega_Z$ are substituted by their expressions (\ref{eq.OmegaX},\ref{eq.OmegaY}). Furthermore, we
make the approximations $C \gg (B-A)$ and $\Omega_Z \approx \dot\theta$ and we replace
$\bar\Omega$ with $(\Omega - \theta)$. We get
\begin{eqnarray}
\label{eq.dadt}
\frac{da}{dt} &=&
  \frac{2}{n a}\frac{\partial \cal R}{\partial \cal M} , \\
\label{eq.dedt}
\frac{de}{dt} &=&
  \frac{1-e^2}{n a^2 e}\frac{\partial \cal R}{\partial \cal M}
 -\frac{(1-e^2)^{1/2}}{n a^2 e}\left(\frac{\partial \cal R}{\partial \omega}
                      + \frac{\partial \cal R}{\partial \omega^{\,\prime}}\right) , \\
\label{eq.didt}
\frac{di}{dt} &=&
 \frac{\beta}{C\dot\theta\sin i}\left(\frac{\partial \cal R}{\partial \omega}
             - \cos i\frac{\partial \cal R}{\partial \Omega}\right)
 -\frac{1}{n a^2 (1-e^2)^{1/2}\sin i}\left(\frac{\partial \cal R}{\partial \Omega}
            -\cos i \frac{\partial \cal R}{\partial \omega}\right)
\nonumber
\\ &&
 -\frac{\sin(\omega-\omega^{\,\prime})}{n a^2 (1-e^2)^{1/2}}\frac{\partial \cal R}{\partial i^{\,\prime}}
 -\frac{\cos(\omega-\omega^{\,\prime})}{n a^2 (1-e^2)^{1/2} \sin i^{\,\prime}}\left(\frac{\partial \cal R}{\partial \Omega^{\,\prime}}
 -      \cos i^{\,\prime}                              \frac{\partial \cal R}{\partial \omega^{\,\prime}}\right) , \\
\label{eq.dMdt}
\frac{d\cal M}{dt} &=&
  n
 -\frac{2}{n a}\frac{\partial \cal R}{\partial a}
 -\frac{1-e^2}{n a^2 e}\frac{\partial \cal R}{\partial e} , \\
\label{eq.domegadt}
\frac{d\omega}{dt} &=&
 -\frac{\beta}{C\dot\theta\sin i}\frac{\partial \cal R}{\partial i}
 +\frac{(1-e^2)^{1/2}}{n a^2 e}\frac{\partial \cal R}{\partial e}
 -\frac{\cos i}{n a^2 (1-e^2)^{1/2}\sin i}\frac{\partial \cal R}{\partial i}
\nonumber
\\ &&
 -\frac{\cos i \cos(\omega-\omega^{\,\prime})}{n a^2 (1-e^2)^{1/2}\sin i}\frac{\partial \cal R}{\partial i^{\,\prime}}
 -\frac{\cos i\sin(\omega-\omega^{\,\prime})}{n a^2 (1-e^2)^{1/2}\sin i\sin i^{\,\prime}}\left(\cos i^{\,\prime}\frac{\partial \cal R}{\partial \omega^{\,\prime}}
 -                                           \frac{\partial \cal R}{\partial \Omega^{\,\prime}}\right) , \\
\label{eq.dOmegadt}
\frac{d\Omega}{dt} &=&
 \frac{\beta \cos i}{C\dot\theta\sin i}\frac{\partial \cal R}{\partial i}
+\frac{\beta \cos\varepsilon}{C\dot\theta\sin\varepsilon} \left(
 \sin\Omega\cot i\frac{\partial\cal R}{\partial\Omega}-\cos\Omega\frac{\partial\cal R}{\partial i}
-\frac{\sin\Omega}{\sin i}\frac{\partial\cal R}{\partial\omega}
\right)
 +\frac{1}{n a^2 (1-e^2)^{1/2} \sin i}\frac{\partial \cal R}{\partial i}
\nonumber
\\ &&
 +\frac{\cos(\omega^{\,\prime}-\omega)}{n a^2 (1-e^2)^{1/2}\sin i}\frac{\partial \cal R}{\partial i^{\,\prime}}
 +\frac{\sin(\omega-\omega^{\,\prime})}{n a^2 (1-e^2)^{1/2} \sin i\sin i^{\,\prime}}\left(\cos i^{\,\prime}\frac{\partial \cal R}{\partial \omega^{\,\prime}}
 -                                           \frac{\partial \cal R}{\partial \Omega^{\,\prime}}\right) .
\label{eq.dOm1}
\end{eqnarray}
To close the system, we have to add the equations
\begin{eqnarray}
&&
\label{eq.depsilon}
\frac{d\varepsilon}{dt}\;=\;-\;\frac{\beta}{C\;\dot\theta}\left(\cos\Omega\;\cot i\;\frac{\partial \cal R}{\partial\Omega}
 \;+\;\sin\Omega\;\frac{\partial \cal R}{\partial i}
 \;-\;\frac{\cos\Omega}{\sin i}\;\frac{\partial\cal R}{\partial\omega}\right)\;\,,
\\ &&
\label{eq.dtheta}
\frac{d^2\theta}{dt^2}\;=\;-\;\frac{\beta}{C}\;\frac{\partial \cal R}{\partial\Omega}\;\,.
\end{eqnarray}
 Be mindful that in equation (\ref{eq.dOm1}) we employed the relation $\,\Omega_Z - \dot\theta =
 \dot\psi\cos\varepsilon = -\,(\Omega_X\sin\theta + \Omega_Y\cos\theta) \cot\varepsilon\,$, while in
 equation ~(\ref{eq.depsilon}) the relation $\dot\varepsilon = -\,\Omega_X\cos\theta+\Omega_y\sin\theta$ was used.
 Besides, we would emphasise that equations (\ref{eq.dadt}-\ref{eq.dtheta}) have been obtained under the gyroscopic
 approximation, for which reason it would be illegitimate to consider the limit of $\,\dot\theta\,\longrightarrow\,0\,$.
 Therefore, the presence of $\,\dot\theta\,$ in the denominator in equations (\ref{eq.didt}) and (\ref{eq.domegadt} - \ref{eq.dOmegadt}) produce no singularity. In the cases where the gyroscopic approximation is invalid, one has to rely on the equations of
 motion (\ref{eq.da}-\ref{eq.dOm}) instead.

\subsection{Comparison with Kaula (1964)
\label{sec.Kaula1}}

 In his 1964 paper, Kaula was mainly interested in the evolution of the semimajor axis, the
eccentricity and the inclination. The derivation of the equations (38) in Kaula (1964) should thus only be
compared with our formulae (\ref{eq.dadt}), (\ref{eq.dedt}) and (\ref{eq.didt}). Although the
evolution rates of $a$ and $e$ are identical in both approaches up to the definition of the
disturbing function (see Section \ref{sec.Kaula2}), the two expressions of the time derivative of the inclination
$i$ display important dissimilarities. This difference of behaviour between $(a,e)$ on the one hand
and $i$ on the other is due to the fact that only $i$ is affected by a rotation of the reference
frame. Let us recall that in Kaula (1964) elliptical elements are defined with respect to a fixed reference frame
coinciding with the primary's equator at the time when the equations of motion are evaluated,
whereas our set of equations (\ref{eq.dadt}-\ref{eq.dOmegadt}) is written in the coprecessing frame
of the primary. Therefore, equation (\ref{eq.didt}) contains an inertial force leading to a term in
$1/(C\dot\theta)$ which is absent in Kaula (1964, eqn 38). But these two equations have
 also distinct dependencies with respect to the primed Keplerian elements.
 In fact, Kaula erroneously assumed that the planetary and satellite contributions in $di/dt$ could simply be summed up.
 Within our formalism, this naive (and wrong) assumption is equivalent to omission of the rotation $\trans{\mat R}\mat R^{\,\prime}$ in
 equations (\ref{eq.ddotr1}) and (\ref{eq.ddotr}). Under this omission, the components of both forces ($-\partial V_1/\partial \vec r$ and $-\partial V_1/\partial \vec r^{\,\prime}$) would have been illegitimately summed up, ignoring the fact that they were written in different coordinate systems: by construction, $\partial V_1/\partial \vec r$ is expressed in the primary's frame, while $\partial V_1/\partial \vec r^{\,\prime}$ is expressed in the secondary's.

 \section{Tidal potential energy}
\label{sec.TidalPotentialEnergy}

\subsection{General expression}
 Within the Darwin-Kaula theory (Kaula 1964, Efroimsky 2012, Efroimsky \& Makarov 2013), it is taken into account that in the general case the secondary body ``feels'' the tides which may be generated in the primary not only by the secondary itself but also by some other perturber located in $\,\erbold^{\,*}\,$. Then in an arbitrary exterior point $\,\erbold\,$ (which is implied to be the position of the secondary), the tidally deformed planet generates an additional tidal potential $\,U(\erbold,\,\erbold^{\,*})\,$, both vectors $\,\erbold\,$ and $\,\erbold^{\,*}\,$ being planetocentric and parametrised by their Keplerian elements $(a, e, i, {\cal M}, \omega, \Omega)$ and $(a^*, e^*, i^*, {\cal M}^*, \omega^*, \Omega^*)$, respectively.
 In a situation where the secondary coincides with the perturber (and, thereby, is ``feeling'' the tides it itself is causing in the primary), the potential of the secondary in this field is equal to the value of $\,U(\erbold,\,\erbold^{\,*})\,$ taken for $\,\erbold^{\,*}=\,\erbold\,$. We, however, shall also need the gradient of the potential. To calculate it, we start out from a general expression with $\,\erbold^{\,*}\neq\erbold\,$, then differentiate with respect to
 $\,\erbold\,$, and only thereafter set $\,\erbold^{\,*}\,$ and $\,\erbold\,$ equal. Hence the tidal force (expressed in the inertial frame) acting on the perturber due to the distortion of the primary is
 \ba
 \Fbold_\mathrm{T}\;=\;-\;M^{\,\prime}\;\mat R\,\frac{\partial U(\erbold,\,\erbold^{\;*})}{\partial \erbold}\Big|_{\erbold^{\;*}=\erbold}\;.
 \label{ga}
 \ea

 Likewise, the tidally deformed secondary generates the additional tidal potential
$\,U^{\,\prime}(\,-\,\erbold^{\,\prime},\,\erbold^{\,\star})\,$, where $\;-\,\erbold^{\,\prime}\,$
is pointing from the centre of mass of the secondary to that of the primary, while
$\,\erbold^{\,\star}\,$ is pointing from the centre of mass of the secondary to that of the
fictitious perturber to be identified with the primary body, both vectors being expressed in the
body-fixed frame of the secondary. Accordingly, the tidal force (expressed in the inertial frame)
acting on the primary due to the tidal distortion of the secondary is \footnote{~Deriving the
right-hand side expression of equation (\ref{do}), we substituted $\,\erbold^{\,\star}\,$ with
$\,-\erbold^{\,\star}\,$.  This change is acceptable because $\,\erbold^{\,\star}\,$ does not show
up in the final answer anyway~---~after the differentiation, $\,\erbold^{\,\star}\,$ must be set
equal to $\,\erbold^{\,\prime}\,$.
}
 \ba
 \Fbold_\mathrm{T}^{\;\prime}\,=\;-\;M\;\mat R^{\,\prime}\,\frac{\partial
          U^{\,\prime}(\,-\,\erbold^{\;\prime},\,\erbold^{\;\star})
 }{\partial
 (-\erbold^{\;\prime})
 }\Big|_{\erbold^{\;\star}=-\erbold^{\;\prime}}
 \,=\;
 M\;\mat R^{\,\prime}\,\frac{\partial
           U^{\;\prime}(\,-\,\erbold^{\;\prime},\,-\,\erbold^{\;\star})
 }{\partial\erbold^{\;\prime}}\Big|_{\erbold^{\;\star}=\erbold^{\;\prime}}
  \;\;\;.
 \label{do}
 \ea
 We endow this force with a prime, because it emerges owing to the distortion of the secondary.

 $\Fbold_0\,$ being the Newtonian force (written in the inertial frame), the equations of the orbital motion in the inertial frame are:~\footnote{~Up to notation, our equations (\ref{5} - \ref{6}) agree with equations (97) in Ferraz-Mello et al. (2008). The negative
signs of the arguments in the second term in our equation (\ref{do}) correspond to the $\pi$-rotation in equation (99) in
Ferraz-Mello et al. (2008).}
 \ba
 M^{\,\prime}\,\ddot{\robold\,}^{\,\prime}\,=\;
 -\;\Fbold_0
 \;+\;\Fbold_\mathrm{T}\;-\;\Fbold_\mathrm{T}^{\;\prime}\;\;,
 \label{5}
 \ea
 \ba
 M\,\ddot{\robold\,}\,=\;
 \Fbold_0
 \;+\;\Fbold_\mathrm{T}^{\;\prime}\;-\;\Fbold_\mathrm{T}\;\;.
 \label{6}
 \ea
 Together, they render:
  \ba
 \nonumber
 \frac{M\;M^{\,\prime}}{M\,+\,M^{\,\prime}}\;\left(\ddot{\robold\,}^{\,\prime}\,-\;\ddot{\robold\,} \right)&=&-\;\Fbold_0\;+\;\Fbold_\mathrm{T}\;-\;\Fbold_\mathrm{T}^{\;^{\,\prime}}\\
 \label{}\\
 \nonumber
 &=&-\;{\cal G}\,\frac{M^{\,\prime}\,M}{\,|\,\robold\,'\,-\;\robold\,|^3}\,\left({\robold}^{\,\prime}\,-\;{\robold} \right)\,+\;\Fbold_\mathrm{T}\;-\;\Fbold_\mathrm{T}^{\;'}
 \label{}
 \ea
 or, equivalently:
 \ba
 \nonumber
 \left(\ddot{\robold\,}^{\,\prime}-\ddot{\robold}\right)+{\cal G}\,\frac{M+M^{\,\prime}}{|\robold^{\,\prime}-{\robold} |^3}\,\left({\robold}^{\,\prime}-{\robold}\right)&=&\frac{M+M^{\,\prime}}{M\;M^{\,\prime}}\,\left(\Fbold_\mathrm{T}
 -\Fbold_\mathrm{T}^{\;\prime}\right)\\
 \label{equation}\\
 \nonumber
 &=&-\;\frac{M+M^{\,\prime}}{M\;M^{\,\prime}}\,\left[M^{\,\prime}\,\mat R\,\frac{\partial U(\erbold,\,\erbold^{\;*})}{\partial \erbold}\Big|_{\erbold^{\;*}=\erbold}
 +\,M\;\mat R^{\,\prime}\,\frac{\partial U^{\,\prime}(-\,\erbold^{\;\prime},\,-\,\erbold^{\;\star})}{\partial \erbold^{\;\prime}}\Big|_{\erbold^{\;\star}=\erbold^{\;\prime}}\right]
 \;\;,\;\qquad
 \ea
 where we employed expressions (\ref{ga}) and (\ref{do}). Recall that the vectors $\,\erbold\,$ and $\,\erbold^{\;\prime}\,$ are defined in corotating frames and are related to the inertial vector $\,\robold^{\,\prime}-{\robold}\,$ via formulae (\ref{relative}).

 The contributions from the two partners enter our expression (\ref{equation}) in a symmetric manner, despite the negative signs of the arguments in the second term on the right-hand side.  The easiest way to understand the origin of these negative signs is to imagine a situation where both partners are non-rotating (i.e., maintain a constant orientation with respect to an inertial frame). In this case, both rotation matrices in the definition (\ref{relative}) of $\,\erbold\,$ and $\,\erbold^{\;\prime}\,$ can be chosen equal to the unity matrix, and we simply have $\,\erbold\,=\,\erbold^{\;\prime}\,=\,\robold^{\;\prime}-\,{\robold}\,$. The fact that both these vectors point from the primary to the secondary
 explains the difference between the arguments' signs in the two gradients in (\ref{equation}).

Let us now write equation ~(\ref{equation}) in the corotating frame of the primary, i.e., with $\,\vec \rho-\vec \rho^{\,\prime}\,$ replaced by $\,\mat R\,\vec r\,$. Successive differentiations of the rotation matrix $\,\mat R\,$ with respect to time produce the classical inertial forces:
 \ba
 \nonumber
 \ddot{\rbold\,}\,+\,\frac{{\cal G}(M+M^{\,\prime})}{r^3}\,{\rbold}\;=&-&\frac{M+M^{\,\prime}}{M\;M^{\,\prime}}
  \;\Bigg[M^{\,\prime}\,\frac{\partial U(\erbold,\,\erbold^{\;*})}{\partial \erbold}\Big|_{\erbold^{\;*}=\erbold}
 +\,M\;\trans{{\mat R}}{\mat R}^{\,\prime}\,\frac{\partial U\,'(-\,\erbold^{\;\prime},\,-\,\erbold^{\,\star})}{\partial \erbold^{\;\prime}}\Big|_{\erbold^{\,\star}=\erbold^{\;\prime}}\Bigg]\qquad\qquad\\
 &\;& \label{equa} \\
 &-& {\vec\Omega}\times({\vec\Omega}\times\rbold) - 2{\vec\Omega}\times\dot\rbold -
    \dot{{\vec\Omega}}\times \vec r
 \;\;.\,\qquad
 \nonumber
\ea
A direct comparison with equation~(\ref{eq.ddotr1}) shows that the total tidal potential energy of the
system is
\begin{equation}
 V_1\,=\;M^{\,\prime}\;U(\erbold,\,\erbold^{\;*})\,\Big|_{\erbold^{\;*}=\erbold} +\;M\;U^{\;\prime}(\,-\,\erbold^{\,\prime},\,-\,\erbold^{\,\star})\,\Big|_{\erbold^{\;\star}=\erbold^{\;\prime}}\,\;,
\end{equation}
and that the disturbing function, which should be inserted in the Lagrange- or Delaunay-type
planetary equations, is related to the physical potential energy via $\,{\cal{R}}\,=\,-\,{V_1}/{\beta}\,$. This gives us:
\begin{equation}
 {\cal{R}}\;=\;-\;\frac{M+M^{\,\prime}}{M\;M^{\;\prime}}\,\left[\,M^{\;\prime}\,{U(\erbold,\,\erbold^{\;*})}\,\Big|_{\erbold^{\;*}=\erbold}
 +\;M\;{U^{\;\prime}(-\,\erbold^{\,\prime},\,-\,\erbold^{\,\star})}\,\Big|_{\erbold^{\,\star}=\erbold^{\,\prime}}\,\right]\,\;,
 \label{eq.R}
\end{equation}
 where it is implied that in the planetary equations the differentiation of $\,{\cal{R}}\,$ should be carried out before $\,\erbold^{\,*}\,$ (resp. $\,\erbold^{\,\star}\,$) is set equal to $\,\erbold\,$ (resp. $\,\erbold^{\,\prime}\,$).

 \subsection{Comparison with Kaula (1964) \label{sec.Kaula2}}

 In his developments, however, Kaula (1964) used the approximation
 \ba
 \mbox{Kaula}~(1964)\,:\qquad {\cal{R}}\;\approx\;-\;\left[\,
 U(\erbold,\,\erbold^{\,*})\Big|_{\erbold^{\,*}=\erbold}\;+\;\frac{M}{M^{\,\prime}} \;U^{\,\prime}(\,-\,\erbold^{\,\prime},\,-\,\erbold^{\,\star})\Big|_{\erbold^{\,\star}=\erbold^{\,\prime}}
 \,\right]\,\;.
 \label{Kaula}
 \ea
  Thence, Kaula's expressions for the orbital elements' tidal rates acquired a redundant factor of $\,M/(M+M^{\,\prime})\,$.
  Tolerable for the Earth-Moon system (which Kaula was having in mind), this approximation is unacceptable for a binary comprising partners of comparable masses. So, Kaula's expressions for the rates must be multiplied by $\,(M+M^{\,\prime})/M\,$, to compensate for that oversight.\,\footnote{~Aside from that, in the general case it is necessary to take into account the tidally-generated change in the orientation of the equator. As we shall see below, this will yield an additional term in the expression for $\,di/dt\,$, see equation (\ref{eq.explicitdidt}).}

 This redundant factor of $\,M/(M+M^{\,\prime})\,$ has become a source of inaccuracy in many publications. At the same time, the overall factor is given correctly in some works, such as Ferraz-Mello et al. (2003) or Ferraz-Mello et al. (2008).

 \subsection{Expansion of the additional tidal potential}

 Let the perturber reside in the point $\,\erbold^{\,*}\,$ relative to the centre of a deformable near-spherical primary. In an exterior point $\,\erbold\,$, the tidally deformed body generates the additional tidal potential calculated by Kaula (1964) \footnote{~A partial sum of this series, with $\,l,\,|q|,\,|j|\,\leq\,2\,$ and $\,p=h=0\,$, was developed by Darwin (1879). In modern notation, his derivation is discussed by Ferraz-Mello et al. (2008).  Mind that in {\it{Ibid}}. the convention on the meaning of the notations $\,\erbold\,$ and $\,\erbold^{\,*}\,$ is opposite to ours.}
 \ba
 \nonumber
 U(\erbold,\,\erbold^{\,*})\;=\;
 -\;\sum_{l=2}^{\infty}\,\left(\frac{R}{a}\right)^{\textstyle{^{l+1}}}\frac{{\cal G}\,M^{\,\prime}}{a^*}\,\left(\frac{R}{a^*}\right)^{\textstyle{^l}}\sum_{m=0}^{l}\,\frac{(l - m)!}{(l + m)!}\;\left(2- \delta_{0m}\right)\sum_{p=0}^{l}F_{lmp}(\inc^*)~\qquad~\qquad~\qquad\qquad\quad\\
                                   \label{21}
                                   \label{A21}\\
                                    \nonumber
 \sum_{q=-\infty}^{\infty}G_{lpq}(e^*)
 \sum_{h=0}^{\it l}F_{lmh}(\inc)\sum_{j=-\infty}^{\infty}
 G_{lhj}(e)\;k_{l}(\omega_{lmpq})\,\cos\left[
 \left(v_{lmpq}^*-m\theta^*\right)-
 \left(v_{lmhj}-m\theta\right)-
 \epsilon_l(\omega_{lmpq}) \right]
 ~~_{\textstyle{_{\textstyle .}}}~~~
  \ea
 where $\,G\,=\,6.674\,\times\,10^{-11}$ m$^3$ kg$^{-1}$ s$^{-2}\,$ is Newton's gravity constant.  As ever, the orbital elements with and without asterisk pertain to $\,\erbold^*\,$ and $\,\erbold\,$, correspondingly, while
 \ba
 \nonumber
 v_{lmpq}^{\,*}\,\equiv\,(l-2p)\,\omega^{\,*}+\,(l-2p+q)\,{\cal M}^{\,*}+\,m\,\Omega^{\,*}
 ~,\\
                                   \label{22}
                                   \label{A22}\\
 \nonumber
  v_{lmhj}\;\equiv\;(l-2h)\,\omega\,+\,(l-2h+j)\,{\cal M}\,+\,m\,\Omega\;\;.\quad
 \ea

 In the expressions (\ref{21} - \ref{22}), we assume that, generally, the perturber located at $\,\erbold^{\,*}\,$ does not coincide with the secondary residing in $\,\erbold\,$. In the special case, when they are the same body, we must first carry out the differentiation over $\,\erbold\,$ and only then set $\,\erbold^{\,*}\,=\,\erbold\,$.

 Both the dynamical Love numbers $\,k_l\,$ and the phase lags $\,\epsilon_l\,$ are functions of the tidal Fourier  modes
 $\,\omega_{lmpq}\,=\,(l-2p)\,\dot{\omega}^{\,*}\,+\,(l-2p+q)\,n^{\,*}\,+\,m\,(\dot{\Omega}^{\,*}-\,\dot{\theta}^{\,*}\,)\,$. After the
 secondary and the fictitious perturber are set to be the same body, and $\erbold^{\,*}$ is set equal to $\erbold$, the modes become
 \footnote{~While in Section \ref{label} we were using the osculating mean motion (defined in a standard way on the line between equations 71 and 72), here and hereafter we are using the anomalistic mean motion defined as in Table 1. We assume that the two are close, and therefore we interchangeably use the same notation n for both. The legitimacy of this is discussed in Efroimsky \& Makarov (2014, Appendix B).}
 \ba
 \omega_{lmpq}\;=\;(l-2p)\,\dot{\omega}\,+\,
 (l-2p+q)\,n\,+\,m\,(\dot{\Omega}\,-\,\dot{\theta}\,)\;\approx\;
 (l-2p+q)\,n-\,m\,\dot\theta~~.
 \label{222}
 \ea
 Interestingly, Kaula himself never addressed the Fourier modes in his works, probably (mis-)assuming that both the dynamical Love numbers $\,k_l\,$ and the phase lags $\,\epsilon_{l}\,$ are frequency-independent. The later development of geophysics demonstrated that the forms of the frequency-dependencies of $\,k_l\,$ and $\,\epsilon_{lmpq}\,$ play an important role in many situations. Hence the necessity to introduce the Fourier modes $\,\omega_{lmpq}\,$ (Efroimsky 2012).

 For a reader-friendly introduction to the Kaula theory, see Efroimsky \& Makarov (2013). It can be understood from equation (15) in that paper, that the tidal modes' absolute values,
 \ba
 \chi_{lmpq}\,\equiv\,|\,\omega_{lmpq}\,|\,\;,
 \ea
 are the physical forcing frequencies excited in the tidally deformed body.


 \section{Tidal evolution of the semimajor axis}

  \subsection{The general formula}

 In the Lagrange-type planetary equation for the semimajor axis rate (\ref{eq.dadt}),
 we should insert formula (\ref{eq.R}), and should perform differentiation over the mean motion. We further average the result over
 the mean anomaly and over the argument of pericentre as in Kaula (1964). This work, carried out in Appendix~\ref{AppA}, leads to:
 \ba
 \nonumber
 \frac{da}{dt}\;=\;-\;2\,a\,n\;\sum_{l=2}^{\infty}\sum_{m=0}^{l}\,\frac{(l - m)!}{(l + m)!}\;\left(2- \delta_{0m}\right)\sum_{p=0}^{l}\sum_{q=-\infty}^{\infty}G^{\,2}_{lpq}(e)\;(l-2p+q)
 \qquad\qquad\qquad\qquad\\
 \nonumber\\
 \nonumber\\
 \left[\,\left(\frac{R}{a}\right)^{\textstyle{^{2l+1}}}\frac{\;M^{\,\prime}}{M}\;
 F^{\,2}_{lmp}(i)\;K_l(\omega_{lmpq})\,+\;\left(\frac{R^{\,\prime}}{a\,}\right)^{\textstyle{^{2l+1}}}
 \frac{M\,}{M^{\,\prime}}\;F^{\,2}_{lmp}(i^{\,\prime})\;K^{\,\prime}_l(\omega_{lmpq}^{\,\prime})
  \right]
 ~~_{\textstyle{_{\textstyle ,}}}\qquad
 \label{28c}
 \label{A28c}
 \label{28}
 \label{A28}
 \ea
 where we employed a shortened notation for the quality functions of the primary:
 \ba
 K_l(\omega_{lmpq})\,\equiv\,k_l(\omega_{lmpq})\;\sin\epsilon_l(\omega_{lmpq})\,\;,
 \label{notation}
 \ea
 the Fourier tidal modes excited in the primary being
 \ba
 \omega_{lmpq}\;\equiv\;(l-2p)\dot{\omega}\,+\,(l-2p+q)n\,+\,m\,(\dot{\Omega}\,-\,\dot{\theta}\,)~\approx~(l-2p+q)n\,-\,m\,\dot{\theta}~~.
 \label{omega}
 \ea
 Likewise, for the quality functions of the secondary we introduced the notation
 \ba
 K^{\,\prime}_l(\omega_{lmpq}^{\,\prime})\,\equiv\,k^{\,\prime}_l(\omega^{\,\prime}_{lmpq})\;\sin\epsilon^{\,\prime}_l(\omega^{\,\prime}_{lmpq})\,\;,
 \label{notationprime}
 \ea
 the Fourier tidal modes excited in the secondary being
 \ba
 \omega^{\,\prime}_{lmpq}\;\equiv\;(l-2p)\dot{\omega}^{\,\prime}\,+\,(l-2p+q)n\,+\,m\,(\dot{\Omega}^{\,\prime}\,-\,\dot{\theta}^{\,\prime}\,)~\approx~(l-2p+q)n\,-\,m\,\dot{\theta}^{\,\prime}~~.
 \label{omegaprime}
 \ea
 Here $\,\Omega\,$, $\,i\,$, $\,\omega\,$ are the Euler angles of the orbit on the primary's equator, while $\,\Omega^{\,\prime}\,$, $\,i^{\,\prime}\,$, $\,\omega^{\,\prime}\,$ are those on the secondary's. The rotation rates of the primary and secondary are $\,\dot{\theta}\,$ and $\,\dot{\theta}^{\,\prime}\,$.

 Our expression (\ref{A28}) differs from its counterpart in Kaula (1964) by the factor of $\,(M+M^{\,\prime})/M\,$. The reason for this is explained above, in Section \ref{sec.Kaula2}.

 Finally, we would mention that our expression (\ref{A28}) behaves well when $\,M^{\,\prime}\rightarrow 0\,$ or $\,M\rightarrow 0\,$, because $\,K_l\,=\,O(M^{\,2})\,$. This can be proven via formulae (31), (40b) and (42)
 from Efroimsky (2015).

 \subsection{The leading inputs}

 By the formulae derived in Appendix \ref{AppendixA}, the quadrupole part of the major semiaxis' rate is
 \ba
 \nonumber
 \left(\frac{da}{dt}\right)_{l=2}=&-&3\,a\,n\left(1\,-\,5\,e^2\,+\,\frac{63}{8}\,e^4\right)
 \left[\frac{\;M^{\,\prime}}{M}\,\left(\frac{R}{a}\right)^5\,K_2(2n-2\dot{\theta})\,+\,
 \frac{M}{M^{\,\prime}}\,\left(\frac{R^{\,{\prime}}}{a}\right)^5\,K_2^{\,\prime}(2n-2\dot{\theta}^{\,\prime})
 \right]
 ~\\
 \nonumber\\
 &-&\frac{9}{4}\,a\,n\,e^2\,\left(1\,+\,\frac{9}{4}\,e^2\right)
 \left[\frac{\;M^{\,\prime}}{M}\,\left(\frac{R}{a}\right)^5\,K_2(n)\,+\,
 \frac{M}{M^{\,\prime}}\,\left(\frac{R^{\,{\prime}}}{a}\right)^5\,K_2^{\,\prime}(n)\right]
 \label{long}
 ~\\
 \nonumber\\
 \nonumber
 &-&\frac{3}{8}\,a\,n\,e^2\,\left(1\,-\,\frac{1}{4}\,e^2\right)
 \left[\frac{\;M^{\,\prime}}{M}\,\left(\frac{R}{a}\right)^5\,K_2(n-2\dot{\theta})\,+\,
 \frac{M}{M^{\,\prime}}\,\left(\frac{R^{\,{\prime}}}{a}\right)^5\,K_2^{\,\prime}(n-2\dot{\theta^{\,\prime}})\right]
 ~\\
 \nonumber\\
 \nonumber
 &-&\frac{441}{8}\,a\,n\,e^2\,\left(1\,-\,\frac{123}{28}\,e^2\right)
 \left[\frac{\;M^{\,\prime}}{M}\,\left(\frac{R}{a}\right)^5\,K_2(3n-2\dot{\theta})\,+\,
 \frac{M}{M^{\,\prime}}\,\left(\frac{R^{\,{\prime}}}{a}\right)^5\,K_2^{\,\prime}(3n-2\dot{\theta^{\,\prime}})\right]
 ~\\
 \nonumber\\
 \nonumber
 &-&\frac{867}{2}\,a\,n\,e^4\,
 \left[\frac{\;M^{\,\prime}}{M}\,\left(\frac{R}{a}\right)^5\,K_2(4n-2\dot{\theta})\,+\,
 \frac{M}{M^{\,\prime}}\,\left(\frac{R^{\,{\prime}}}{a}\right)^5\,K_2^{\,\prime}(4n-2\dot{\theta}^{\,\prime})
 \right]
 \nonumber\\
 \nonumber
 &-&\frac{81}{8}\,a\,n\,e^4\,
 \left[\frac{\;M^{\,\prime}}{M}\,\left(\frac{R}{a}\right)^5\,K_2(2n)\,+\,
 \frac{M}{M^{\,\prime}}\,\left(\frac{R^{\,{\prime}}}{a}\right)^5\,K_2^{\,\prime}(2n)\right]
\,+\,O(i^{\,2})\,+\,O({i^{\,\prime}}^{\,2})\,+\,O(e^{\,6})\;\,.\;\quad
 \ea
  This long formula can obviously be split into two parts:
 \ba
 \left(\frac{da}{dt}\right)_{l=2}=\; \left(\frac{da}{dt}\right)_{l=2}^{(prim)}+\;\left(\frac{da}{dt}\right)_{l=2}^{(sec)}\,\;.
 \ea
 where the first part is due to the tides in the primary and comprises the terms with $\,K_2(\omega_{lmpq})\,$. The second part is due to the tides in the secondary and comprises the terms with $\,K^{\,\prime}_2(\omega^{\,\prime}_{lmpq})\,$.

 \subsection{The case when the spin of neither partner is synchronised\label{case1}}

  If none of the partners is synchronised and both $\,i\,$ and $\,i^{\,\prime}\,$ are small, the leading terms are semidiurnal, i.e., those with   $\,\{lmpq\}\,=\,\{2200\}\,$. Approximated with these terms, the major semiaxis' rate is:
 \ba
 \left(\frac{da}{dt}\right)_{l=2}=\;-\;3\;n\,a\,
 \left[\left(\frac{R}{a}\right)^{\textstyle{^5}}\;\frac{\;M^{\,\prime}}{M}
 \;K_{2}(\omega_{2200})
  \,+\,\left(\frac{R^{\,{\prime}}}{a}\right)^{\textstyle{^5}}
 \frac{M\,}{M^{\,\prime}}\;K^{\,\prime}_{2}(\omega^{\,\prime}_{2200})
  \right]+O(i^{\,2})+O({i^{\,\prime}}^{\,2})+O(e^{\,2})
 ~_{\textstyle{_{\textstyle .}}}\quad
 \label{45a}
 \ea
 To compare the inputs, write the above as
 \bs
 \ba
 \frac{da}{dt}\;\approx\;-\;3\;n\,a\;\frac{M^{\,\prime}}{M\;}\,\left(\frac{R}{a}\right)^{\textstyle{^5}}\,\left[K_{2}(\omega_{2200})\,+\,\left(\frac{R^{\,{\prime}}}{R}\right)^5
 \left(\frac{M}{M^{\,\prime}}\right)^2 K^{\,\prime}_{2}(\omega^{\,\prime}_{2200})\right]
 \label{45b}
 \ea
 \ba
 =\;-\;3\;n\,a\;\frac{M^{\,\prime}}{M\;}\,\left(\frac{R}{a}\right)^{\textstyle{^5}}\,K_{2}(\omega_{2200})\;
 \left[\,1\,+\,\frac{{\rho^{\,\prime}}^{\,-2}\,{R^{\,{\prime}}}^{\,-1}}{{\rho}^{\,-2}\;\,{R}^{\,-1}}\;
 \frac{K^{\,\prime}_{2}(\omega^{\,\prime}_{2200})}{K_{2}(\omega_{2200})}\,\right]
 ~~_{\textstyle{_{\textstyle ,}}}\quad
 \label{45c}
  \label{nons}
 \ea
 \label{45}
 \es
 $\rho\,$ and $\,\rho^{\,\prime}\,$ being the mean densities of the primary and the secondary, correspondingly.

 When the role of the secondary is negligible, we are left with
 \ba
 \frac{da}{dt}\;=\;-\;3\;n\,a\;\frac{\;M^{\,\prime}}{M}\,
 \left(\frac{R}{a}\right)^{\textstyle{^{5}}}
 K_{2}(\omega_{2200})\;+\;O(i^{\,2})\;+\;O(e^{\,2})
  ~~_{\textstyle{_{\textstyle .}}}\;
  \label{ina}
  \label{46}
 \ea

 In the case when the spin is faster than orbiting, the Fourier mode $\,\omega_{2200}\,=\,2\,(n-\dot{\theta})\,$ is negative, and so is the phase lag
 \footnote{~Recall that the lag $\,\epsilon_l\,$ is an odd function of the Fourier mode $\,\omega_{lmpq}\,$.}
 $\,\epsilon_{2200}\,\equiv\,\epsilon_2(\omega_{2200})\,$. Then the above expression becomes:
 \ba
  \frac{da}{dt}\;\approx\;3\;n\,a\;\frac{M^{\,\prime}}{M\;}\,
 \left(\frac{R}{a}\right)^{\textstyle{^{5}}}
 \frac{k_{2}}{Q_2}
  ~~_{\textstyle{_{\textstyle ,}}}\;
  \label{final}
  \label{47}
 \ea
 where $\,Q_2\,$ is the quadrupole tidal quality factor defined through
 \ba
 \frac{1}{Q_2}\;=\;|\,\sin\epsilon_2(\omega_{2200})\,|\;\,.
 \label{48}
 \ea
 We reiterate that in expression (\ref{final}) the quality function $\,k_2/Q_2\,$, mass $\,M\,$, and radius $\,R\,$ are those of the partner tides wherein are dominant (the primary). The mass $\,M^{\,\prime}\,$ is that of the secondary (the tides wherein we neglected in equations \ref{ina} - \ref{final}).

 Approximation (\ref{45} - \ref{46}) contains only one tidal mode, the semidiurnal one. So this approximation looks the same, no matter what the frequency-dependence $\,K_2(\omega_{2mpq})\,$. For this reason, our equation (\ref{46}) agrees with the corresponding formulae by both Hut (1981, eqn 9)
 and Emelyanov (2018, eqn 18) who used the CTL (constant time lag) model. It also coincides with equation (A1) from Lainey et al. (2012) who relied on the CPL (constant phase lag) tidal model.

 \subsection{The case when the primary is not synchronised,\\ while the secondary is\label{case2}}

 If the primary is not synchronised ($\,\dot{\theta}\neq n\,$), the part $\;\left({da}/{dt}\right)_{l=2}^{(prim)}\;$ is approximated with formulae (\ref{ina} - \ref{final}).
 If the secondary is synchronised ($\,\dot{\theta}^{\,\prime}=n\,$), the terms with $\,K^{\,\prime}_2(2n-2\dot{\theta}^{\,\prime})\,$ and $\,K^{\,\prime}_2(n-\dot{\theta}^{\,\prime})\,$ in equation (\ref{long}) get nullified. In their absence, we are left with
 \ba
 \left(\frac{da}{dt}\right)_{l=2}^{(sec)}=\;
 -\;57\;a\,n\,e^2\,\left(\frac{R^{\,{\prime}}}{a}\right)^{\textstyle{^{5}}}\frac{\;M\;}{M^{\,{\prime}}}\;K_2^{\,\prime}(n)\;+\;O({i^{\,\prime}}^2)\;+\;O(e^4)
 \,\;,
 \label{49}
 \ea
 where we took into account that $\,K^{\,\prime}_2\,$ is an odd function.

 As expression (\ref{49}) contains only one tidal mode, $\,n\,$, the form of this expression is independent of the shape of the frequency-dependence $\,K_2^{\,\prime}(\omega_{2mpq})\,$. So our answer coincides with equation (28) from Emelyanov (2018) and also with equation (9) from Hut (1981) if we set $\,\Omega=n\,$ in Hut's equation. Our answer, however, differs from the first equation (A2) in the paper by Lainey et al. (2012), which contains an erroneous factor of $\;-\,21\,$ instead of $\;-\;57\,$. The same oversight is contained in equation (1) in Barnes et al. (2008) and in the first equation (40) in Shoji \& Kurita (2014).

 Together, the tides in both the primary and the secondary generate the rate $\,da/dt\,$ obtained by summing up the rates (\ref{46}) and (\ref{49}):
  \bs
 \ba
 \frac{da}{dt}\;\approx\;-\;3\;n\,a\;\frac{M^{\,\prime}}{M\;}\,\left(\frac{R}{a}\right)^{\textstyle{^5}}\,\left[K_{2}(\omega_{2200})\,+\,19\,e^2\,\left(\frac{R^{\,{\prime}}}{R}\right)^5
 \left(\frac{M}{M^{\,\prime}}\right)^2 K^{\,\prime}_{2}(n)\right]
 \ea
 \ba
 =\;-\;3\;n\,a\;\frac{M^{\,\prime}}{M\;}\,\left(\frac{R}{a}\right)^{\textstyle{^5}}\,K_{2}(\omega_{2200})\;
 \left[\,1\,+\,19\,e^2\,\frac{{\rho^{\,\prime}}^{\,-2}\,{R^{\,{\prime}}}^{\,-1}}{{\rho}^{\,-2}\;\,{R}^{\,-1}}\;
 \frac{K^{\,\prime}_{2}(n)}{K_{2}(\omega_{2200})}\,\right]
 ~~_{\textstyle{_{\textstyle .}}}\quad
 \ea
 \label{151}
 \es

 The dissipation rate in a synchronised satellite (and the corresponding input in the elements' rates) may be considerably amplified
 by longitudinal librations, when the satellite has a pronounced dynamical triaxiality (Frouard \& Efroimsky 2017a, Efroimsky 2018).

 \subsection{Beyond quadrupole\label{beyond}}

 Bills et al. (2005) argued that to attain a high precision in the modeling of Phobos' tidal dynamics the knowledge of $\,k_3\,$ and perhaps even $\,k_4\,$ may be needed. Later, Taylor \& Margot (2010) suggested that for very close asteroidal binaries the degrees $\,l\,$ up to $\,6\,$ may matter.

 In the quadrupole ($\,l=2\,$) approximation (\ref{long}), the smallest terms taken into account are of order $\,(R/a)^5\,e^4\,$. Now, if we choose to go beyond the quadrupole approximation and take into account the $\,l = 3\,$ terms, the largest of those will be of order $\;(R/a)^7\,e^0\,i^0\,=\,(R/a)^7\,$. Such are the terms with $\,\{lmpq\}\,=\,\{3300\}\,$ and $\,\{3110\}\,$. We may neglect them insofar as
 \ba
 (R/a)^7\,\ll\;(R/a)^5\,e^4\qquad\Longleftrightarrow\qquad R/a\,\ll\,e^2\,\;,
 \ea
 a somewhat stringent condition not necessarily obeyed by all close-in binaries.

 At the same time, had we kept in expression (\ref{long}) only the terms up to $\,(R/a)^5 e^2\,$, the neglect of the $\,l = 3\,$ would be justified under a more relaxed condition
 \ba
 (R/a)^7\,\ll\,(R/a)^5\,e^2\qquad\Longleftrightarrow\qquad R/a\,\ll\,e\,\;.
 \ea

 \subsection{Final caveat\label{caveat}}

 In both Sections \ref{case1} and \ref{case2} we passingly dropped the terms containing $\,e^4\,K_2(2n)\,$ and $\,e^4\,K^{\,\prime}_2(2n)\,$, see the last line in equation (\ref{long}). At first glance, this is legitimate when the eccentricity is not too high. In reality, the issue is subtle owing to the extremely sharp shapes of the frequency-dependencies of both $\,K_2(2n)\,$ and $\,K^{\,\prime}_2(2n)\,$. When the peak frequency happens to be equal or very close to $\,2n\,$, these terms may become prominent, even for modest values of $\,e\,$.

 \section{Tidal evolution of the eccentricity}

 The planetary equation for the eccentricity evolution is given in equation (\ref{eq.dedt}).
 The insertion  of expressions (\ref{eq.R}), (\ref{A27}) and (\ref{tacit}) in this equation and the
 removal of the short and long period oscillating terms gives us
 \ba
 \frac{de}{dt}\,=\;-\;\frac{1-e^2}{e}\;n\;\frac{1}{M\;M^{\,\prime}}\;
 \sum_{l=2}^{\infty}\sum_{m=0}^{l}\,\frac{(l - m)!}{(l + m)!}\;\left(2- \delta_{0m}\right)\sum_{p=0}^{l}\sum_{q=-\infty}^{\infty}G^{\,2}_{lpq}(e)\;(l-2p+q)\;
 \qquad
 \nonumber\\
 \nonumber\\
 \nonumber\\
 \nonumber
 \left[\,\left(\frac{R}{a}\right)^{\textstyle{^{2l+1}}}
 {M^{\,\prime}}^{\,2}\,F^{\,2}_{lmp}(i)\;K_l(\omega_{lmpq})\,+\;\left(\frac{R^{\,\prime}}{a}\right)^{\textstyle{^{2l+1}}}
 M^2\,F^{\,2}_{lmp}(i^{\,\prime})\;K^{\,\prime}_{l}(\omega^{\,\prime}_{lmpq})
 \right]\\
 \nonumber\\
 \nonumber\\
 \nonumber
 +\;\frac{(1\,-\,e^2)^{1/2}}{e}\;n\;\frac{1}{M\;M^{\,\prime}}\;
 \sum_{l=2}^{\infty}\sum_{m=0}^{l}\,\frac{(l - m)!}{(l + m)!}\;\left(2- \delta_{0m}\right)\sum_{p=0}^{l}\sum_{q=-\infty}^{\infty}G^{\,2}_{lpq}(e)\;(l-2p)\,\qquad\;\quad
 \qquad\\
 \nonumber\\
 \nonumber\\
 \left[\,\left(\frac{R}{a}\right)^{\textstyle{^{2l+1}}}
 {M^{\,\prime}}^{\,2}\,F^{\,2}_{lmp}(i)\;K_{l}(\omega_{lmpq})\,+\;\left(\frac{R^{\,\prime}}{a}\right)^{\textstyle{^{2l+1}}}
 M^2\,F^{\,2}_{lmp}(i^{\,\prime})\;K^{\,\prime}_{l}(\omega^{\,\prime}_{lmpq})
 \right]
  ~~_{\textstyle{_{\textstyle ,}}}
 \qquad\;\qquad
\label{30}
 \ea
 where we used the shortened notation (\ref{notation}) and (\ref{notationprime}).

  As expected (see Section \ref{sec.Kaula2} above), our expression (\ref{30}) differs from the corresponding formula in Kaula (1964, eqn 38) by a factor of $\,(M+M^{\,\prime})/M\,$.

 The quadrupole part of expression (\ref{30}) reads (see Appendix \ref{AppendixB}):
 \begin{eqnarray}
 \nonumber
 \left(\frac{de}{dt}\right)_{l=2}\;=\;\left(\frac{de}{dt}\right)_{l=2}^{(prim)}+\;\left(\frac{de}{dt}\right)_{l=2}^{(sec)} \qquad\qquad\qquad\qquad\qquad\qquad\qquad\qquad\qquad\qquad\qquad\qquad\qquad\;\;
 ~\\
 \nonumber\\
 \nonumber
 =\;-\;n\,e\;\frac{\,M^{\,\prime}}{M\,}\,\left(\frac{R}{a}\right)^{\textstyle{^{5}}}\,
 \left[\,
 -\,\left(1\,-\,\frac{e^2}{4}\right)\;\frac{3}{16}\;K_2(n-2\dot{\theta})\;-\;\frac{3}{4}\,\left(1\,-\,\frac{21}{4}\,e^2\right)\;K_2(2n-2\dot{\theta})\;+
 \right.  \;\;\;\;\\
 \nonumber\\
 \nonumber\\
 \nonumber
 \left.
 \frac{147}{16}\,\left(1\,-\,\frac{\textstyle 179}{\textstyle 28}\,e^2\right)\,K_2(3n-2\dot{\theta})
 \;+\;\frac{867}{8}\,e^2\,K_2(4n-2\dot{\theta})
 \;+\;\frac{9}{8}\,\left(1\,+\,\frac{5}{4}\,e^2\right)\,K_2(n)\;+\;\frac{81}{16}\;e^2\;K_2(2n)
 \,\right]\;\;\\
 \nonumber\\
  \label{38}\\
  \nonumber
 \;-\;n\,e\;\frac{M\,}{\,M^{\,\prime}}\,\left(\frac{R^{\;\prime}}{a}\right)^{\textstyle{^{5}}}\,
 \left[\,
 -\,\left(1\,-\,\frac{e^2}{4}\right)\;\frac{3}{16}\;K_2^{\,\prime}(n-2\dot{\theta}^{\,\prime})\;-\;\frac{3}{4}\,\left(1\,-\,\frac{21}{4}\,e^2\right)\;K_2^{\,\prime}(2n-2\dot{\theta}^{\,\prime})\;+
 \right.\;\;\;\;\\
 \nonumber\\
 \nonumber\\
 \nonumber
 \left.
 \frac{147}{16}\,\left(1\,-\,\frac{\textstyle 179}{\textstyle 28}\,e^2\right)\,K_2^{\,\prime}(3n-2\dot{\theta}^{\,\prime})
 \,+\;\frac{867}{8}\,e^2\,K_2^{\,\prime}(4n-2\dot{\theta}^{\,\prime})
 \,+\;\frac{9}{8}\,\left(1\,+\,\frac{5}{4}\,e^2\right)\,K_2^{\,\prime}(n)\,+\;\frac{81}{16}\;e^2\,K_2^{\,\prime}(2n)
 \,\right]\\
 \nonumber\\
 \nonumber\\
 \nonumber
  \;+\;O(e^5)\;+\;O(i^2)\,+\,O({i^{\,\prime\;}}^2)~~_{\textstyle{_{\textstyle ,}}}\qquad
 \end{eqnarray}
 where the contribution $\;(de/dt)^{(prim)}\;$ comprises all the primary-generated terms (those with $\,K_2\,$), while $\;(de/dt)^{(sec)}\;$ comprises the secondary-generated terms (those with $\,K_2^{\,\prime}\,$).

 \subsection{The case when neither partner is synchronised}

 When the spin of neither body is synchronised, while both inclinations are small, the leading terms in the above equation are those linear in $\,e\;$:
 \ba
 \nonumber
 \left(\frac{de}{dt}\right)_{l=2}
 =\,-\,n\,e\,\frac{\,M^{\,\prime}}{M\,}\,\left(\frac{R}{a}\right)^{\textstyle{^{5}}}\,
 \left[
 -\,\frac{3}{16}\,K_2(n-2\dot{\theta})\,-\,\frac{3}{4}\,K_2(2n-2\dot{\theta})\,+
 \,
 \frac{147}{16}\,\,K_2(3n-2\dot{\theta})
 \,+\,\frac{9}{8}\,\,K_2(n)
 \right]\;\;\;\;\\
 \nonumber\\
 \nonumber
 -\,n\,e\,\frac{M\,}{\,M^{\,\prime}}\,\left(\frac{R^{\;\prime}}{a}\right)^{\textstyle{^{5}}}\,
 \left[
 -\,\frac{3}{16}\,K_2^{\,\prime}(n-2\dot{\theta}^{\,\prime})\,-\,\frac{3}{4}\,K_2^{\,\prime}(2n-2\dot{\theta}^{\,\prime})\,+
  \,\frac{147}{16}\,K_2^{\,\prime}(3n-2\dot{\theta}^{\,\prime})
  \,+\,\frac{9}{8}\,K_2^{\,\prime}(n)
 \right]\\
 \nonumber\\
 \;+\;O(e^2)\;+\;O(i^2)\,+\,O({i^{\,\prime\;}}^2)~~_{\textstyle{_{\textstyle ,}}}\quad
 \label{53}
 \ea
 Specifically, when both partners satisfy the Constant Time Lag (CTL) model (i.e., when both $\,K_2\,$ and $\,K_2^{\,\prime}\,$ are linear in the tidal mode), the above expression becomes

 \ba
 \nonumber
 ^{\textstyle{^{(CTL)}}}\left(\frac{de}{dt}\right)_{l=2}
 =\qquad\qquad\qquad\qquad\qquad\qquad\qquad\qquad\qquad\qquad\qquad\qquad\qquad\qquad\qquad\qquad\qquad\\
 \nonumber\\
 \nonumber\\
 \nonumber
 \frac{3}{4}\,n\,e\,\left[
 \frac{\,M^{\,\prime}}{M\,}\,\left(\frac{R}{a}\right)^{\textstyle{^{5}}}\,\frac{11\;\dot{\theta}\;-\;18\;n}{\dot{\theta}\;-\;n}
 \;K_2(2n-2\dot{\theta})
 \;+\;
 \frac{M\,}{\,M^{\,\prime}}\,\left(\frac{R^{\;\prime}}{a}\right)^{\textstyle{^{5}}}\,\frac{11\;\dot{\theta}^{\,\prime}\;-\;18\;n}{\dot{\theta}^{\,\prime}\;-\;n}
 \;K_2^{\,\prime}(2n-2\dot{\theta}^{\,\prime})\right]\\
 \nonumber\\
 \;+\;O(e^2)\;+\;O(i^2)\,+\,O({i^{\,\prime\;}}^2)~~_{\textstyle{_{\textstyle .}}}\quad
 \ea
 This agrees with equations (10) from Hut (1981) and (19) from Emelyanov (2018). Apart from the afore-mentioned factor of $\,(M+M^{\,\prime})/M\,$, this also agrees with the first line of equation (46) from Kaula (1964). (On the second line, Kaula lost the factor of 4 in the denominator.)

 When both partners satisfy the Constant Phase Lag (CPL) model (so both $\,K_2\,$ and $\,K_2^{\,\prime}\,$ are constants) and  both $\,\dot{\theta}\,$ and $\,\dot{\theta}^{\,\prime}\,$ exceed $\,3n/2\,$, we have:
 $$K_2(n)\,=\,-\,K_2(n-2\dot{\theta})\,=\,-\,K_2(2n-2\dot{\theta})\,=\,-\,K_2(3n-2\dot{\theta})\,=\,k_2/Q\,$$
 and
 $$K^{\,\prime}_2(n)\,=\,-\,K_2^{\,\prime}(n-2\dot{\theta}^{\,\prime})\,=\,-\,K_2^{\,\prime}(2n-2\dot{\theta}^{\,\prime})\,=\,-\,K_2^{\,\prime}(3n-2\dot{\theta}^{\,\prime})\,=\,k_2^{\,\prime}/Q^{\,\prime}\,\;,$$
 wherefrom
 \ba
 ^{\textstyle{^{(CPL)}}}\left(\frac{de}{dt}\right)_{l=2}
 =\,\frac{57}{8}\,n\,e\,\left[
 \frac{\,M^{\,\prime}}{M\,}\,\left(\frac{R}{a}\right)^{\textstyle{^{5}}}\,\frac{k_2}{Q}
 \;+\;
 \frac{M\,}{\,M^{\,\prime}}\,\left(\frac{R^{\;\prime}}{a}\right)^{\textstyle{^{5}}}\,\frac{k_2^{\,\prime}}{Q^{\,\prime}}
 \right]
 \;+\;O(e^2)\;+\;O(i^2)\,+\,O({i^{\,\prime\;}}^2)~~_{\textstyle{_{\textstyle .}}}\quad
 \ea
 This is in agreement with the second expression in equation (A1) from Lainey et al. (2012), but differs from the corresponding formulae in
 some other works.

 \subsection{The case when the primary is not synchronised,\\ while the secondary is}

 If the primary is not synchronised ($\dot{\theta}\neq n$), the part $\,\left({de}/{dt}\right)_{l=2}^{(prim)}$ is still approximated by the no-asterisk terms from the expressions above. If at the same time the secondary is synchronised ($\dot{\theta}^{\,\prime}=n$), then in formula (\ref{53}) the term with $\,K^{\,\prime}_2(2n-2\dot{\theta}^{\,\prime})\,$ must be set zero. Thence,
 \ba
 \nonumber
 \left(\frac{de}{dt}\right)_{l=2}^{(sec)}=\;
 -\,n\,e\,\frac{M\,}{\,M^{\,\prime}}\,\left(\frac{R^{\;\prime}}{a}\right)^{\textstyle{^{5}}}\,
 \left[
 -\,\frac{3}{16}\,K_2^{\,\prime}(-n)\,+
  \,\frac{147}{16}\,K_2^{\,\prime}(n)
  \,+\,\frac{9}{8}\,K_2^{\,\prime}(n)
 \right]
 \;+\;O(e^2)\;+\,O({i^{\,\prime\;}}^2)\qquad \\
 \label{56}\\
 \nonumber
 =\;
 -\,\frac{21}{2}\,n\,e\,\frac{M\,}{\,M^{\,\prime}}\,\left(\frac{R^{\;\prime}}{a}\right)^{\textstyle{^{5}}}\,
 K_2^{\,\prime}(n)
 \;+\;O(e^2)\;+\,O({i^{\,\prime\;}}^2)~~_{\textstyle{_{\textstyle ,}}}\qquad
 \ea
 where we took into consideration that $\,K^{\,\prime}_2\,$ is an odd function.

 Containing only one frequency, expression (\ref{56}) bears no dependence on the shape of the frequency-dependence of $\,K_2^{\,\prime}\,$.
 So it coincides with the corresponding expressions from Emelyanov (2018, eqn 29) and Hut (1981, eqn 10) both of whom employed the CTL model. It also is in agreement with equation (A2) by Lainey et al. (2012) who used the CPL model.

  \subsection{Beyond quadrupole}

 Under what condition may the degree-$3\,$ terms be ignored in the expression for $\,de/dt\;$?

 In the quadrupole ($\,l=2\,$) approximation (\ref{38}), the smallest terms taken into account are of order $\,(R/a)^5\,e^3\,$. Had it been our intention to include there also the $\,l = 3\,$ terms, the largest of those would be the ones with $\,\{lmpq\}\,=\,3300\,$ and $\,3110\,$. Being of order $\;(R/a)^7\,e^1\,i^0\,=\,(R/a)^7\,e\,$, they may be ignored if
 \ba
 (R/a)^7\,e\;\ll\;(R/a)^5\,e^3\qquad\Longleftrightarrow\qquad R/a\,\ll\,e\,\;.
 \ea

 Had we kept in expression (\ref{38}) only the terms up to $\,(R/a)^5 e\,$, the neglect of the $\,l = 3\,$ would be justified in all realistic situations:
 \ba
 (R/a)^7\,e\,\ll\,(R/a)^5\,e\qquad\Longleftrightarrow\qquad R/a\,\ll\,1\,\;.
 \ea

 \section{Tidal evolution of the inclination\label{inc}}

 In the Lagrange-type equation for the inclination rate (\ref{eq.didt}),
 we should insert expression (\ref{eq.R}), perform the differentiation over $\,\omega\,$, $\,\omega^{\,\prime}\,$, $\,\Omega\,$, $\,\Omega^{\,\prime}\,$ and $i^{\,\prime}$ and then extract the secular terms by averaging over ${\cal M}$, $\omega$ and $\omega^{\,\prime}$ to be consistent with Kaula (1964) who also removed the oscillating part. This work is carried out in Appendices \ref{AppB} and \ref{AppC}, the result being

 \ba
 \frac{di}{dt}\;=\;\qquad\qquad\qquad\qquad\qquad\qquad\qquad\qquad\qquad\qquad\qquad\qquad\qquad\qquad\qquad\qquad\qquad\qquad
 \nonumber
 \ea
 \ba
 \nonumber
  -\;\frac{n}{\sqrt{1-e^2}}\;\frac{{M^{\,\prime}}}{M\;}
 \sum_{l=2}^{\infty}\left(\frac{R}{a}\right)^{\textstyle{^{2l+1}}}\sum_{m=0}^{l}\frac{(l - m)!}{(l + m)!}\left(2- \delta_{0m}\right)
 \sum_{p=0}^{l}
 \frac{(l-2p)\cos i-m}{\sin i}\,F^{\,2}_{lmp}(i)\sum_{q=-\infty}^{\infty}G^{\,2}_{lpq}(e)\,K_{l}(\omega_{lmpq})\qquad\\
 \label{eq.explicitdidt}\\
 \nonumber
 +\;\frac{\beta\,n^2 a^2}{C\dot\theta}\;\frac{M^{\,\prime}}{M\,}
 \sum_{l=2}^\infty \left(\frac{R}{a}\right)^{\textstyle{^{2l+1}}}\sum_{m=0}^{l}\frac{(l-m)!}{(l+m)!}\left(2-\delta_{0m}\right)
 \sum_{p=0}^{l} \frac{m\cos i-(l-2p)}{\sin i}\,F^{\,2}_{lmp}(i)
 \sum_{q=-\infty}^{\infty}G^{\,2}_{lpq}(e)\,K_{l}(\omega_{lmpq})\;.\qquad
 \ea

 Apart from the multiplier of $\,(M+M^{\,\prime})/M\,$ discussed previously (see Section
\ref{sec.Kaula2}), the first term of the above expression coincides with the first term of the
expression (38) in Kaula (1964), while the second term in our formula differs from that in Kaula
(1964, eqn 38) considerably. The second term in Kaula (1964, eqn 38) is the equivalent of the first
with primed Keplerian elements. However, as explained in Section~\ref{sec.Kaula2}, the rotation
matrix $\trans{\mat R}\mat R^{\,\prime}$ from the secondary's frame to the primary's frame breaks
the symmetry between primed and unprimed Keplerian elements in $di/dt$. According to
equation~(\ref{eq.didt}), the primed equivalent of the first line of
equation~(\ref{eq.explicitdidt}) is multiplied by the slow oscillating
$\cos(\omega-\omega^{\,\prime})$ and vanishes once averaged over the argument of pericentres
$\omega$ and $\omega^{\,\prime}$ (see the non-averaged expression in
Appendix~\ref{sec.nonaveragedidt}). Moreover, in our derivation of the inclination rate, the second
line of equation~(\ref{eq.explicitdidt}) is an inertial force due to the non-Galilean nature of the
coprecessing frame of the primary. More explicitly, this term expresses the variation of the orbital
inclination with respect to the primary's equator induced by the motion of the primary's spin axis.

 We would emphasize that the apparent lack of symmetry between the two components in the expression
of $di/dt$ is due to the fact that the inclination $i$ is reckoned from the primary's equatorial
plane. Unlike $a$ and $e$ which have the same definition in both body frames and whose rates are
symmetric in primed and unprimed variables, here the symmetry is recovered by writing the time
derivatives of the orbital inclination with respect to the secondary's equatorial plane, namely,
 \ba
 \frac{di^{\,\prime}}{dt}\;=\;\qquad\qquad\qquad\qquad\qquad\qquad\qquad\qquad\qquad\qquad\qquad\qquad\qquad\qquad\qquad\qquad\qquad\qquad
 \nonumber
 \ea
 \ba
 \nonumber
  -\;\frac{n}{\sqrt{1-e^2}}\;\frac{M\;}{{M^{\,\prime}}}
 \sum_{l=2}^{\infty}\left(\frac{R^{\,\prime}}{a}\right)^{\textstyle{^{2l+1}}}\sum_{m=0}^{l}\frac{(l - m)!}{(l + m)!}\left(2- \delta_{0m}\right)
 \sum_{p=0}^{l}
 \frac{(l-2p)\cos i^{\,\prime}-m}{\sin i^{\,\prime}}\,F^{\,2}_{lmp}(i^{\,\prime})\sum_{q=-\infty}^{\infty}G^{\,2}_{lpq}(e)\,K^{\,\prime}_{l}(\omega^{\,\prime}_{lmpq})\qquad\\
 \label{eq.explicitdidtprime}\\
 \nonumber
 +\;\frac{\beta\,n^2 a^2}{C^{\,\prime}\dot\theta^{\,\prime}}\;\frac{M\,}{M^{\,\prime}}
 \sum_{l=2}^\infty \left(\frac{R^{\,\prime}}{a}\right)^{\textstyle{^{2l+1}}}\sum_{m=0}^{l}\frac{(l-m)!}{(l+m)!}\left(2-\delta_{0m}\right)
 \sum_{p=0}^{l} \frac{m\cos i^{\,\prime}-(l-2p)}{\sin i^{\,\prime}}\,F^{\,2}_{lmp}(i^{\,\prime})
 \sum_{q=-\infty}^{\infty}G^{\,2}_{lpq}(e)\,K^{\,\prime}_{l}(\omega^{\,\prime}_{lmpq})\;.\qquad
 \ea
 In addition, it shall be reminded that neither the expression (\ref{eq.explicitdidt}) for $\,di/dt\,$ nor the expression (\ref{eq.explicitdidtprime}) for $\,di^{\,\prime}/dt\,$ allows, by itself, to deduce the motion of the orbital plane with respect to the inertial frame. For example, if the inclination rate $\,di/dt\,$ happens to be zero, the orbit is fastened to the primary's equator, and its orientation in the inertial frame is then governed by its longitude of the ascending node $\,\Omega\,$ and the primary's Euler angles $\,(\psi, \varepsilon)\,$. Reciprocally, when $\,di/dt\,$ is non-zero, the orbit can still remain at rest in the inertial frame, in which case the apparent inclination evolution is only due the motion of the primary's spin axis. This would have precisely been the situation if most of the total angular momentum of the system were associated with the orbital motion. Nevertheless, as in Kaula (1964), we are not interested in the motion of the orbit plane relative to the inertial frame. This is why the orientation of the orbit is measured either in the primary's or in the secondary's frame. The presence of two inclinations to represent the orientation of a single orbit plane might seem odd at first sight; however, the angles $\,i\,$ and $\,i^{\,\prime}\,$ can also be interpreted as
 the obliquities of the bodies relative to the orbit.

 In a situation where the orbital angular momentum is much lower than the angular momentum of the primary's spin, and where the dissipation in the perturber can be neglected, our result coincides with that of Kaula (1964, eqn 38), up to the afore-mentioned factor of $\,(M+M^{\,\prime})/M\,$.

 The quadrupole inputs of expression (\ref{eq.explicitdidt}) reads:
\begin{equation}
\begin{split}
& \left(\frac{di}{dt}\right)_{l=2} =
n\;\sin i\;\frac{M^{\,\prime}}{M}\left(\frac{R}{a}\right)^5 \Bigg[
\\ &
     \frac{243}{64} (1+\varrho) e^4 K_2(-2n-\dot\theta)
   + \frac{27}{16} e^2 \left[1+\varrho+\left(\frac{11}{4}+\frac{9}{4} \varrho\right) e^2\right] K_2(-n-\dot\theta)
\\[0.3em] &
   + \frac{3}{4} \left[1+\varrho+\left(\frac{7}{2}+3 \varrho\right) e^2+\left(\frac{63}{8}+6 \varrho\right) e^4\right] K_2(-\dot\theta)
   + \frac{3}{16} e^2 \left[1-\varrho+\frac{1}{4} \left(1+\varrho\right) e^2\right] K_2(n-2\dot\theta)
\\[0.3em] &
   + \frac{3}{2} e^2 \left[1+\varrho+\left(\frac{49}{16}+\frac{41}{16} \varrho\right) e^2\right] K_2(n-\dot\theta)
   + \frac{3}{4} \left[1-\varrho-\left(\frac{9}{2}-5 \varrho\right) e^2+\left(\frac{23}{4}-\frac{63}{8} \varrho\right) e^4\right] K_2(2n-2\dot\theta)
\\[0.3em] &
   - \frac{3}{4} \left[1+\varrho-\left(\frac{9}{2}+5 \varrho\right) e^2+\left(\frac{11}{16}+\frac{45}{16} \varrho\right) e^4\right] K_2(2n-\dot\theta)
   + \frac{147}{16} e^2 \left[1-\varrho-\left(\frac{109}{28}-\frac{123}{28}\varrho\right) e^2\right] K_2(3n-2\dot\theta)
\\[0.3em] &
   - \frac{147}{16} e^2 \left[1+\varrho-\left(\frac{109}{28}+\frac{123}{28}\varrho\right] e^2\right) K_2(3n-\dot\theta)
   + \frac{867}{16} \left(1-\varrho\right) e^4 K_2(4n-2\dot\theta)
   - \frac{867}{16} \left(1+\varrho\right) e^4 K_2(4n-\dot\theta)
 \\&
 \Bigg]
 +\;O(i^3)\;+\;O(e^6)\,\;,
 \label{single}
 \end{split}
 \end{equation}
 with $\,\varrho\,=\,\beta\,n\,a^2 / (C\,\dot\theta)\,$. For a very small eccentricity, a cruder approximation is acceptable
 \ba
 \nonumber
  \left(\frac{\textstyle di}{\textstyle dt}\right)_{l=2}\,=\qquad\qquad\qquad\qquad\qquad \qquad\qquad\qquad\qquad\qquad \qquad\qquad\qquad\qquad\qquad \qquad \qquad\qquad
 \\
 \label{both}\\
 \frac{3}{4}\,n\;\sin i\;\frac{M^{\,\prime}}{M}\left(\frac{R}{a}\right)^5 \Bigg[
     (1+\varrho) K_2(-\dot\theta)
   + (1-\varrho) K_2(2n-2\dot\theta)
   - (1+\varrho) K_2(2n-\dot\theta)
 \Bigg]
 +\;O(i^3)\;+\;O(e^2)\,\;.\qquad
 \nonumber
 \ea
 Without loss of precision, $\;\sin i\;$ may be changed to $\;i\;$ in both equations (\ref{single}) and (\ref{both}).

 In many realistic situations, the inclination is stabilised in the sense that $\,di/dt \propto -i\,$. Specifically, it can be seen from equation (\ref{both}) that at small eccentricities stabilisation is taking place for a synchronous orbit ($\,\dot\theta = n\,$), for the \,2:1\, spin-orbit resonance ($\,\dot\theta = 2n\,$), and for fast prograde rotation ($\,\dot\theta \gg n\,$). For other values of the angular velocity, however, it is not possible to determine the sign of $\,di/dt\,$ which, generally, depends on the rheology and on the value of the parameter $\,\varrho\,$.

 \section{Tidal evolution of $\,\omega\,$, $\Omega$ and $\,{\cal{M}}_0\,$ }

 Tidal evolution of $\,\omega\,$, $\,\Omega$ and $\,{\cal{M}}_0\,$ can be described by the same tools. 
 Be mindful, though, that the  rate of these angles contains an input due to the oblateness of the primary (see, e.g., Efroimsky 2005a). Therefore, even if a total rate is measured, it will not be easy to single out the tidal input in it. Also, for $\,d\omega/dt\,$ of a very close-in planet, the relativistic correction may supersede the tidal effect, like in the case of Mercury.

 \section{Conclusions}

 In this paper, we have revisited the theory of Kaula (1964) basing our calculation on a non-canonical Hamiltonian
formalism with constraint. We
have written down the rates $\,da/dt\,$, $\,de/dt\,$, and $\,di/dt\,$ to order $\,e^4\,$, inclusively.
 They differ from Kaula's expressions which contain a redundant factor of $\,M/(M+M^{\,\prime})\,$, with $\,M\,$ and $\,M^{\,\prime}\,$ being the masses of the primary and the secondary. Since Kaula was interested in the Earth-Moon system, this redundant factor was close to unity and was unimportant. This omission, however, must be corrected when Kaula's theory is applied to a binary composed of partners of comparable masses.

 We have pointed out that, while it is legitimate to simply sum the primary's and secondary's inputs in $\,da/dt\,$ or $\,de/dt\,$, this is not the case for $\,di/dt\,$, so our expression for the inclination rate $\,di/dt\,$ differs from that of Kaula in two regards. First, in the expression for the primary's $\,di/dt\,$ the contribution due to the dissipation in the secondary averages out completely, provided the apsidal precession is uniform. Second, we have an additional term which emerges owing to the conservation of the angular momentum: a change in the inclination of the orbit causes a change of the primary's plane of equator.

 We have carried out our developments in the gyroscopic approximation (which implies that the spin of a body is much faster than the evolution of the spin axis' orientation). As a by-product, our work also provides a full set of equations of motion as it reads before the this approximation is applied.


 \section*{Acknowledgements}

 \noindent
 One of the authors (ME) would like to thank Valeri V. Makarov and Dimitri Veras for numerous stimulating discussions.

  ~\\
 {\bf{{Conflict of interest}}} The authors hereby state that they have no conflict of interest to declare.\\


 \appendix
 \begin{center}
  {\underline{\Large{\bf{Appendix}}}}
 \end{center}

 \section{Differential of the rotation matrix $\,\mat F = \trans{\mat R}\mat R'\,$\label{sec.dX}}
 Let $\mat R_1(\varphi)$ and $\mat R_3(\varphi)$ be the rotation matrices of angle $\varphi$ around the first axis $\vec i =
\trans{(1, 0, 0)}$ and around the third one $\vec k = \trans{(0, 0, 1)}$, respectively:
 \begin{equation}
 \mat R_1(\varphi) = \begin{bmatrix}
1 & 0 & 0 \\
0 & \cos\varphi & -\sin\varphi \\
0 & \sin\varphi & \cos\varphi
\end{bmatrix}\;,\qquad
 \mat R_3(\varphi) = \begin{bmatrix}
\cos\varphi & -\sin\varphi & 0 \\
\sin\varphi & \cos\varphi & 0  \\
0 & 0 & 1
\end{bmatrix}\;.
 \end{equation}
 The derivatives of these matrices read
 \begin{equation}
 d\mat R_1(\varphi) = \begin{bmatrix}
0 & 0 & 0 \\
0 & -\sin\varphi & -\cos\varphi \\
0 & \cos\varphi & -\sin\varphi
\end{bmatrix}\,d\varphi\;,\qquad
 d\mat R_3(\varphi) = \begin{bmatrix}
-\sin\varphi & -\cos\varphi & 0 \\
\cos\varphi & -\sin\varphi & 0 \\
0 & 0 & 0
\end{bmatrix}\,d\varphi\;.
 \end{equation}
 A direct calculation gives
\begin{equation}
 d\mat R_1(\varphi)\,\trans{\mat R}_1(\varphi) = \begin{bmatrix}
 0 & 0 & 0 \\
 0 & 0 & -1 \\
 0 & 1 & 0
\end{bmatrix}\,d\varphi\;,\qquad
 d\mat R_3(\varphi)\,\trans{\mat R}_3(\varphi) = \begin{bmatrix}
 0 & -1 & 0 \\
 1 & 0 & 0 \\
 0 & 0 & 0
\end{bmatrix}\,d\varphi\;.
\label{eq.dR1dR3}
\end{equation}
 We introduce the hat operator which associate to any vector $\vec v = \trans{(x, y, z)} \in \mathbb{R}^3$ the
skew-symmetrix matrix $\hvec v$ defined as
\begin{equation}
\hvec v = \begin{bmatrix}
0 & -z & y \\
z & 0 & -x \\
-y & x & 0
\end{bmatrix}\;.
\end{equation}
This skew-symmetric matrix is such that for any two vectors $\vec a, \vec b \in \mathbb{R}^3$, their
vector product $\vec a \times\vec b$ reads $\hvec a \vec b$. From Eq~(\ref{eq.dR1dR3}), one notices
that
\begin{equation}
d\mat R_1(\varphi)\,\trans{\mat R}_1(\varphi) = \hvec i \,d\varphi\,,\qquad
d\mat R_3(\varphi)\,\trans{\mat R}_3(\varphi) = \hvec k \,d\varphi\,.
\label{eq.dR1R1dR3R3}
\end{equation}
Let us recall the definition of the rotation matrix $\mat F$:
\begin{equation}
\mat F = \mat R_3(\bar\Omega)\mat R_1(i)\mat R_3(\omega-\omega^{\,\prime}) \mat R_1(-i^{\,\prime}) \mat R_3(-\bar\Omega^{\,\prime}) \;.
\end{equation}
Applying the rule (\ref{eq.dR1R1dR3R3}), we get
\begin{equation}
\begin{split}
d\mat F = &
\hvec k\;\mat R_3(\bar\Omega) \mat R_1(i) \mat R_3(\omega-\omega')\mat R_1(-i')\mat R-3(-\bar\Omega')\,d\bar\Omega
\\ &
+ \mat R_3(\bar \Omega)\;\hvec i\;\mat R_1(i) \mat R_3(\omega-\omega') \mat R_1(-i') \mat R_3(-\bar\Omega')\,di
\\ &
+ \mat R_3(\bar \Omega) \mat R_1(i)\;\hvec k\;\mat R_3(\omega-\omega') \mat R_1(-i') \mat R_3(-\bar\Omega')\,d(\omega-\omega')
\\ &
- \mat R_3(\bar \Omega) \mat R_1(i) \mat R_3(\omega-\omega')\;\hvec i\;\mat R_1(-i') \mat R_3(-\bar\Omega')\,di'
\\ &
- \mat R_3(\bar \Omega) \mat R_1(i) \mat R_3(\omega-\omega') \mat R_1(-i')\;\hvec k\;\mat R_3(-\bar\Omega')\,d\bar\Omega'\,.
\end{split}
\end{equation}
To simplify the result, we introduce the vectors $\vec i'$, $\vec K$, $\vec K'$ defined as
\begin{equation}
\vec i' = \mat R_3(\omega-\omega')\,\vec i\,,\quad
\vec K  = \mat R_1(-i)\,\vec k\,,\quad
\vec K' = \mat R_3(\omega-\omega')\,\mat R_1(-i')\vec k\,.
\end{equation}
The associated skew-symmetric matrices $\hvec i'$, $\hvec K$ and $\hvec K'$ are given by
\begin{equation}
\begin{split}
&
\hvec i' = \mat R_3(\omega-\omega')\,\hvec i\,\trans{\mat R}_3(\omega-\omega')\,,\qquad
\hvec K  = \mat R_1(-i)\,\hvec k\,\trans{\mat R}_1(-i)\,,
\\ &
\hvec K' = \mat R_3(\omega-\omega')\,\mat R_1(-i')\hvec k\,\trans{\mat R_1}(-i')\trans{\mat R}_3(\omega-\omega')\,.
\end{split}
\end{equation}
A direct calculation shows that $d\mat X$, defined as $\mat R_1(-i) \mat R_3(-\bar\Omega)\;(d\mat
F\,\trans{\mat F})\;\mat R_3(\bar\Omega) \mat R_1(i)$, is equal to
\begin{equation}
d\mat X = \hvec K d\bar\Omega + \hvec i di + \hvec k d(\omega-\omega') - \hvec i' di' - \hvec K'
d\bar\Omega'\,.
\end{equation}

 \section{Equations of motion in terms of the Keplerian elements
 \label{sec.demoEq56}}

 Let $\,\mat M_{\mathrm{C}/\mathrm{C^{\,\prime}}}\,$ be the Jacobian $\,\partial(\vec p,\,\vec r)/\partial(\vec
 p^{\,\prime},\,\vec r^{\,\prime})\,$ describing the transition between the two cartesian coordinate systems and given by
\begin{equation}
\mat M_{\mathrm{C}/\mathrm{C^{\,\prime}}} = \begin{bmatrix}
\trans{\mat R}\mat R^{\,\prime} & \vec 0 \\
\vec 0 & \trans{\mat R}\mat R^{\,\prime}
\end{bmatrix}\;.
\end{equation}
As shown by equations (\ref{eq.dtr}) and (\ref{eq.drt}) in the main text, the Hamiltonian equations for $\,(\vec p,\,\vec r)\,$ are
\begin{equation}
\frac{d}{dt}\begin{pmatrix} \vec p \\ \vec r \end{pmatrix} =
\begin{bmatrix}
\vec 0 & -\Id \\
\Id & \vec 0
\end{bmatrix}
\begin{pmatrix}
\Frac{\partial \cal H}{\partial \vec p} \\[1.em]
\Frac{\partial \cal H}{\partial \vec r}
\end{pmatrix}
+ \mat M_{\mathrm{C}/\mathrm{C^{\,\prime}}}
\begin{bmatrix}
\vec 0 & -\Id \\
\Id & \vec 0
\end{bmatrix}
\begin{pmatrix}
\Frac{\partial \cal H}{\partial \vec p^{\,\prime}} \\[1em]
\Frac{\partial \cal H}{\partial \vec r^{\,\prime}}
\end{pmatrix}\,.
\end{equation}
We shall precise that in our problem ${\cal H}$ bears not dependence on $\,\vec p^{\,\prime}\,$, wherefore $\,\partial
{\cal H}/\partial \vec p^{\,\prime} = \vec 0\,$. Now, let $\,\mat M_{\mathrm{K}/\mathrm{C}}\,$ and $\,\mat M_{\mathrm{K}^{\,\prime}/\mathrm{C}^{\,\prime}}\,$ be the Jacobian matrices $\,\partial (\vec Y, \vec y)/\partial (\vec p, \vec r)\,$ and $\,\partial (\vec Y^{\,\prime}, \vec y^{\,\prime})/\partial (\vec p^{\,\prime}, \vec r^{\,\prime})\,$, respectively, where $\,\vec Y = (a, e, i)\,$ and $\,\vec y = ({\cal M}, \omega, \bar \Omega)\,$. These matrices describe transitions to Keplerian (K) variables from the Cartesian (C) ones, hence the notation. Applying the chain rule, we have
\begin{equation}
\begin{split}
\frac{d}{dt}\begin{pmatrix}
\vec Y \\ \vec y
\end{pmatrix} &= \mat M_{\mathrm{K}/\mathrm{C}} \frac{d}{dt}\begin{pmatrix}
\vec p \\ \vec r
\end{pmatrix}
\\ &
= \mat M_{\mathrm{K}/\mathrm{C}}
\begin{bmatrix}
\vec 0 & -\Id \\
\Id & \vec 0
\end{bmatrix}
\trans{\mat M_{\mathrm{K}/\mathrm{C}}}
\begin{pmatrix}
\Frac{\partial \cal H}{\partial \vec Y} \\[1.em]
\Frac{\partial \cal H}{\partial \vec y}
\end{pmatrix}
+\mat M_{\mathrm{K}/\mathrm{C}} \mat M_{\mathrm{C}/\mathrm{C^{\,\prime}}}
\begin{bmatrix}
\vec 0 & -\Id \\
\Id & \vec 0
\end{bmatrix}
\trans{\mat M_{\mathrm{K}^{\,\prime}/\mathrm{C}^{\,\prime}}}
\begin{pmatrix}
\Frac{\partial \cal H}{\partial \vec Y^{\,\prime}} \\[1.em]
\Frac{\partial \cal H}{\partial \vec y^{\,\prime}}
\end{pmatrix}\;.
\end{split}
\label{eq.dK}
\end{equation}
Then, we define the Poisson matrices $\mat B$ and $\mat B^{\,\prime}$ as
\begin{equation}
\mat B \equiv \mat M_{\mathrm{K}/\mathrm{C}}
\begin{bmatrix}
\vec 0 & -\Id \\
\Id & \vec 0
\end{bmatrix}
\trans{\mat M_{\mathrm{K}/\mathrm{C}}}\;,\qquad
\mat B^{\,\prime} \equiv \mat M_{\mathrm{K}^{\,\prime}/\mathrm{C}^{\,\prime}}
\begin{bmatrix}
\vec 0 & -\Id \\
\Id & \vec 0
\end{bmatrix}
\trans{\mat M_{\mathrm{K}^{\,\prime}/\mathrm{C}^{\,\prime}}}\;,
\label{eq.Poissons}
\end{equation}
 and introduce the total Jacobian matrix
 \begin{equation}
 \mat A\;\equiv\;\frac{\partial (\vec Y,\vec y)}{\partial (\vec Y^{\,\prime}, \vec y^{\,\prime})}\;=\;
 \frac{\partial (\vec Y,\vec y)}{\partial (\vec p, \vec r)}
 \frac{\partial (\vec p,\vec r)}{\partial (\vec p^{\,\prime}, \vec r^{\,\prime})}
 \frac{\partial (\vec p^{\,\prime},\vec r^{\,\prime})}{\partial (\vec Y^{\,\prime}, \vec y^{\,\prime})}\;=\;\mat M_{\mathrm{K}/\mathrm{C}}\,\mat M_{\mathrm{C}/\mathrm{C^{\,\prime}}}\,(\mat M_{\mathrm{K}^{\,\prime}/\mathrm{C}^{\,\prime}})^{-1}\;\,.
\label{eq.matA}
\end{equation}
 Combined, equations (\ref{eq.dK}-\ref{eq.matA}) render us
\begin{equation}
\frac{d}{dt}\begin{pmatrix}
\vec Y \\ \vec y
\end{pmatrix} = \mat B
\begin{pmatrix}
\Frac{\partial \cal H}{\partial \vec Y} \\[1.em]
\Frac{\partial \cal H}{\partial \vec y}
\end{pmatrix}
+\mat A \mat B^{\,\prime}
\begin{pmatrix}
\Frac{\partial \cal H}{\partial \vec Y^{\,\prime}} \\[1.em]
\Frac{\partial \cal H}{\partial \vec y^{\,\prime}}
\end{pmatrix}\;\;.
\end{equation}

 \section{Differentiation of $\,{\cal{R}}\,$ with respect to the mean motion\label{AppA}}

 First, we differentiate $\,U\,$ with respect to the mean motion $\,{\cal{M}}\,$:
 \ba
 \nonumber
 \frac{\partial\;}{\partial{\cal{M}}}\,U(\erbold,\,\erbold^*) \,=\;
  %
-\;\sum_{l=2}^{\infty}\,\left(\frac{R}{a}\right)^{\textstyle{^{l+1}}}\frac{{\cal{G}}\,M^{\,\prime}}{a^*}\,\left(\frac{R}{a^*}\right)^{\textstyle{^l}}\sum_{m=0}^{l}\,\frac{(l - m)!}{(l + m)!}\;\left(2- \delta_{0m}\right)\sum_{p=0}^{l}F_{lmp}(\inc^*)\sum_{q=-\infty}^{\infty}G_{lpq}(e^*)~\quad~
 ~\\
                                   \label{24}
                                   \label{A24}\\
                                    \nonumber
 \sum_{h=0}^{\it l}F_{lmh}(\inc)\sum_{j=-\infty}^{\infty}
 G_{{\it l}hj}(e)\;(l-2h+j)\;k_{l}\,\sin\left[
 \left(v_{lmpq}^*-m\theta^*\right)-
 \left(v_{lmhj}-m\theta\right)-
 \epsilon_{lmpq} \right]
 ~~_{\textstyle{_{\textstyle .}}}~~~
 \ea
 After we set $\,\erbold\,$ and $\,\erbold^*\,$ equal to one another (and drop the now-redundant asterisks), the \,{\it{secular part}}\, of the above derivative will become
 \ba
 \frac{\partial U}{\partial{\cal{M}}}\;=\;
                                   \label{25}
                                   \label{A25}
 \sum_{l=2}^{\infty}\,\left(\frac{R}{a}\right)^{\textstyle{^{2l+1}}}\frac{{\cal{G}}\,M^{\,\prime}}{a}\,\sum_{m=0}^{l}\,\frac{(l - m)!}{(l + m)!}\;\left(2- \delta_{0m}\right)\sum_{p=0}^{l}F^{\,2}_{lmp}(\inc)\sum_{q=-\infty}^{\infty}G^{\,2}_{lpq}(e)\;(l-2p+q)\;k_{l}\,\sin\epsilon_{lmpq}
 ~~_{\textstyle{_{\textstyle .}}}~~~\quad
 \ea
 Similarly, the secular part of the derivative of the potential created by the tidally deformed secondary, at the place where the primary resides, will read:
 \ba
 \frac{\,\partial U^{\,\prime}}{\partial{\cal{M}}}\;=\;
                                   \label{26}
                                   \label{A26}
 \sum_{l=2}^{\infty}\,\left(\frac{R^{\,\prime}}{a}\right)^{\textstyle{^{2l+1}}}\frac{{\cal{G}}\,M}{a}\,\sum_{m=0}^{l}\,\frac{(l - m)!}{(l + m)!}\;\left(2- \delta_{0m}\right)\sum_{p=0}^{l}F^{\,2}_{lmp}(i^{\,\prime})\sum_{q=-\infty}^{\infty}G^{\,2}_{lpq}(e)\;(l-2p+q)\;k^{\,\prime}_{l}\,\sin\epsilon^{\,\prime}_{lmpq}
 ~~_{\textstyle{_{\textstyle .}}}~~~\quad
 \ea
 While in equation (\ref{A25}) $\,i\,$ is standing for the secondary's inclination on the planetary equator, in equation (\ref{A26}) $\,i^{\,\prime}\,$ denotes the inclination of the primary's apparent orbit on the secondary's equator. Likewise, while $\,k_l\,\sin\epsilon_{lmpq}\,$ is the quality function of the primary, $\,k_l^{\,\prime}\,\sin\epsilon^{\,\prime}_{lmpq}\,$ is that of the secondary.

 According to formula (\ref{eq.R}), the sum of the primary's and secondary's inputs in the derivative of the disturbing function over $\,{\cal{M}}\,$ will be

 \bs
 \ba
 \nonumber
 \frac{\partial\cal R}{\partial{\cal{M}}}\;=\;\frac{M+M^{\,\prime}}{M\;M^{\,\prime}}\;
 \frac{\partial\;}{\partial{\cal{M}}}\,\left(\,-\,M^{\,\prime}\,U\,-\,M\,U^{\,\prime}\right)\;=\;\quad\;\qquad\;\qquad\;\qquad\;\qquad\;\qquad\;\qquad
 ~\\
                                  \nonumber\\
                                    \nonumber\\
                                    \nonumber
 -\;\frac{M+M^{\,\prime}}{M\;M^{\,\prime}}\;
 \sum_{l=2}^{\infty}\sum_{m=0}^{l}\,\frac{(l - m)!}{(l + m)!}\;\left(2- \delta_{0m}\right)\sum_{q=-\infty}^{\infty}G^{\,2}_{lpq}(e)\;(l-2p+q)\,\qquad\;\quad
 \qquad\qquad\qquad\\
 \\
 \nonumber\\
 \nonumber
 \left[\,\left(\frac{R}{a}\right)^{\textstyle{^{2l+1}}}
 \frac{{\cal{G}}\,{M^{\,\prime}}^{\,2}}{a}\,F^{\,2}_{lmp}(i)\;k_{l}\,\sin\epsilon_{lmpq}\,+\;\left(\frac{R^{\,\prime}}{a}\right)^{\textstyle{^{2l+1}}}
 \frac{{\cal{G}}\,M^2}{a}\,F^{\,2}_{lmp}(i^{\,\prime})\;k^{\,\prime}_{l}\,\sin\epsilon^{\,\prime}_{lmpq}
 \right]
 \qquad\qquad
 \ea
 \ba
 \nonumber
 =\;-\;n^2\,a^2\;
 \sum_{l=2}^{\infty}\sum_{m=0}^{l}\,\frac{(l - m)!}{(l + m)!}\;\left(2- \delta_{0m}\right)\sum_{q=-\infty}^{\infty}G^{\,2}_{lpq}(e)\;(l-2p+q)\,\qquad\;\quad
 \qquad\\
 \label{A27}\\
 \nonumber\\
 \nonumber
 \left[\,\left(\frac{R}{a}\right)^{\textstyle{^{2l+1}}}
 \frac{M^{\,\prime}}{M}\,F^{\,2}_{lmp}(i)\;k_{l}\,\sin\epsilon_{lmpq}\,+\;\left(\frac{R^{\,\prime}}{a}\right)^{\textstyle{^{2l+1}}}
 \frac{M}{M^{\,\prime}}\,F^{\,2}_{lmp}(i^{\,\prime})\;k^{\,\prime}_{l}\,\sin\epsilon^{\,\prime}_{lmpq}
 \right]
 ~~_{\textstyle{_{\textstyle .}}}
 \ea
 \es

 \section{Differentiation of $\,{\cal{R}}\,$ with respect to the argument of the pericentre\label{AppB}}

 Differentiation of $\,U\,$ over $\,\omega\,$ renders us
 \ba
 \nonumber
 \frac{\partial\;}{\partial\omega}\,U(\erbold,\,\erbold^*) \,=\;
  %
-\;\sum_{l=2}^{\infty}\,\left(\frac{R}{a}\right)^{\textstyle{^{l+1}}}\frac{{\cal{G}}\,M^{\,\prime}}{a^*}\,\left(\frac{R}{a^*}\right)^{\textstyle{^l}}\sum_{m=0}^{l}\,\frac{(l - m)!}{(l + m)!}\;\left(2- \delta_{0m}\right)\sum_{p=0}^{l}F_{lmp}(\inc^*)\sum_{q=-\infty}^{\infty}G_{lpq}(e^*)~\quad~
 ~\\
                                   \label{240}
                                   \label{A240}\\
                                    \nonumber
 \sum_{h=0}^{\it l}F_{lmh}(\inc)\sum_{j=-\infty}^{\infty}
 G_{{\it l}hj}(e)\;(l-2h)\;k_{l}\,\sin\left[
 \left(v_{lmpq}^*-m\theta^*\right)-
 \left(v_{lmhj}-m\theta\right)-
 \epsilon_{lmpq} \right]
 ~~_{\textstyle{_{\textstyle .}}}~~~
 \ea
 For $\,\erbold\,=\,\erbold^*\,$, the \,{\it{secular part}}\, of the derivative reduces to
 \ba
 \nonumber
 \frac{\partial U}{\partial\omega}\;=\;
 \sum_{l=2}^{\infty}\,\left(\frac{R}{a}\right)^{\textstyle{^{2l+1}}}\frac{{\cal{G}}\,M^{\,\prime}}{a}\,\sum_{m=0}^{l}\,\frac{(l - m)!}{(l + m)!}\;\left(2- \delta_{0m}\right)\sum_{p=0}^{l}F^{\,2}_{lmp}(\inc)\sum_{q=-\infty}^{\infty}G^{\,2}_{lpq}(e)\;(l-2p)\;k_{l}\,\sin\epsilon_{lmpq}
 ~~_{\textstyle{_{\textstyle .}}}~~~\quad
                                   \label{250}
                                   \label{A250}
 \ea
 Equivalently, the secular part of the derivative of the potential created by the tidally deformed
secondary and acting on the primary is:
 \ba
 \nonumber
 \frac{\partial U^{\,\prime}}{\partial\omega^{\,\prime}}\;=\;
 \sum_{l=2}^{\infty}\,\left(\frac{R^{\,\prime}}{a}\right)^{\textstyle{^{2l+1}}}\frac{{\cal{G}}\,M^{\,\prime}}{a}\,\sum_{m=0}^{l}\,\frac{(l - m)!}{(l + m)!}\;\left(2- \delta_{0m}\right)\sum_{p=0}^{l}F^{\,2}_{lmp}(\inc^{\,\prime})\sum_{q=-\infty}^{\infty}G^{\,2}_{lpq}(e)\;(l-2p)\;k^{\,\prime}_{l}\,\sin\epsilon^{\,\prime}_{lmpq}
 ~~_{\textstyle{_{\textstyle ,}}}~~~\quad
                                   \label{260}
                                   \label{A260}
 \ea
 where $\,i^{\,\prime}\,$, $\,\omega^{\,\prime}\,$, and $\,\Omega^{\,\prime}\,$ denote the inclination, the argument of the pericentre, and the longitude of the node of the planet's apparent orbit as seen from the perturber.

 Combining the last two equations with expression (\ref{eq.R}) for $\,{\cal{R}}\,$ as a function of $\,U\,$ and $\,U^{\,\prime}$, we obtain:
 \bs
 \ba
 \nonumber
 \frac{\partial\cal R}{\partial\omega}\;+\;\frac{\partial\cal R}{\partial\omega^{\,\prime}}\;=\;\frac{M+M^{\,\prime}}{M\;M^{\,\prime}}\;\left(
 \frac{\partial\;}{\partial\omega}\,\left(\,-\,M^{\,\prime}\,U\right)\;+\;\frac{\partial\;}{\partial\omega^{\,\prime}}\,\left(\,-\,M\,U^{\,\prime\,}\right)\right)\;\quad\;\qquad\;\qquad\;\qquad\;\qquad\;\qquad\;\qquad
\nonumber
\ea
\ba
 =-\;\frac{M+M^{\,\prime}}{M\;M^{\,\prime}}\;
 \sum_{l=2}^{\infty}\sum_{m=0}^{l}\,\frac{(l - m)!}{(l + m)!}\;\left(2- \delta_{0m}\right)\sum_{p=0}^{l}\sum_{q=-\infty}^{\infty}G^{\,2}_{lpq}(e)\;(l-2p)\,\qquad\;\quad
 \qquad
 \\
 \nonumber\\
 \nonumber
 \left[\,\left(\frac{R}{a}\right)^{\textstyle{^{2l+1}}}
 \frac{{\cal{G}}\,{M^{\,\prime}}^{\,2}}{a}\,F^{\,2}_{lmp}(i)\;k_{l}\,\sin\epsilon_{lmpq}\,+\;\left(\frac{R^{\,\prime}}{a}\right)^{\textstyle{^{2l+1}}}
 \frac{{\cal{G}}\,M^2}{a}\,F^{\,2}_{lmp}(i^{\,\prime})\;k^{\,\prime}_{l}\,\sin\epsilon^{\,\prime}_{lmpq}
 \right]
 \ea
 \ba
 =\;-\;n^2\;a^2\;
 \sum_{l=2}^{\infty}\sum_{m=0}^{l}\,\frac{(l - m)!}{(l + m)!}\;\left(2- \delta_{0m}\right)\sum_{p=0}^{l}\sum_{q=-\infty}^{\infty}G^{\,2}_{lpq}(e)\;(l-2p)\,\qquad\;\qquad
 \qquad
 \label{tacit}\\
 \nonumber\\
 \nonumber
 \left[\,\left(\frac{R}{a}\right)^{\textstyle{^{2l+1}}}
 \frac{{M^{\,\prime}}}{M\;}\;F^{\,2}_{lmp}(i)\;k_{l}\,\sin\epsilon_{lmpq}\,+\;\left(\frac{R^{\,\prime}}{a}\right)^{\textstyle{^{2l+1}}}
 \frac{M\;}{M^{\,\prime}}\;F^{\,2}_{lmp}(i^{\,\prime})\;k^{\,\prime}_{l}\,\sin\epsilon^{\,\prime}_{lmpq}
 \right]
 ~~_{\textstyle{_{\textstyle .}}}
 \ea
 \es

 \section{Differentiation of $\,{\cal{R}}\,$ with respect to the longitude of the node\label{AppC}}

 Differentiation of $\,U\,$ over $\,\Omega\,$ gives us
 \ba
 \nonumber
 \frac{\partial\;}{\partial\Omega}\,U(\erbold,\,\erbold^*) \,=\;-\;\sum_{l=2}^{\infty}\,\left(\frac{R}{a}\right)^{\textstyle{^{l+1}}}\frac{{\cal{G}}\,M^{\,\prime}}{a^*}\,\left(\frac{R}{a^*}\right)^{\textstyle{^l}}\sum_{m=0}^{l}\,\frac{(l - m)!}{(l + m)!}\;\left(2- \delta_{0m}\right)\sum_{p=0}^{l}F_{lmp}(\inc^*)\sum_{q=-\infty}^{\infty}G_{lpq}(e^*)~\quad~
 ~\\
                                   \\
                                    \nonumber
 \sum_{h=0}^{\it l}F_{lmh}(\inc)\sum_{j=-\infty}^{\infty}
 G_{{\it l}hj}(e)\;m\;k_{l}\,\sin\left[
 \left(v_{lmpq}^*-m\theta^*\right)-
 \left(v_{lmhj}-m\theta\right)-
 \epsilon_{lmpq} \right]
 ~~_{\textstyle{_{\textstyle .}}}~~~
 \ea
 For $\,\erbold\,=\,\erbold^*\,$, the \,{\it{secular part}}\, of the above expression becomes
 \ba
 \frac{\partial U}{\partial\Omega}\;=\;
 \sum_{l=2}^{\infty}\,\left(\frac{R}{a}\right)^{\textstyle{^{2l+1}}}\frac{{\cal{G}}\,M^{\,\prime}}{a}\,\sum_{m=0}^{l}\,\frac{(l - m)!}{(l + m)!}\;\left(2- \delta_{0m}\right)\sum_{p=0}^{l}F^{\,2}_{lmp}(\inc)\sum_{q=-\infty}^{\infty}G^{\,2}_{lpq}(e)\;m\;k_{l}\,\sin\epsilon_{lmpq}
 ~~_{\textstyle{_{\textstyle .}}}~~~\quad
 \ea
 Combined with equation (\ref{eq.R}), the above formula yields:
 \ba
 \frac{\partial \cal R}{\partial\Omega}\;=\;
 \frac{M+M^{\,\prime}}{M\;M^{\,\prime}}\;\frac{\partial\;}{\partial\Omega}\,\left(\,-\,M^{\,\prime}\,U\,-\,M\,U^{\,\prime\,}\right)\;=\;
 -\;\frac{M+M^{\,\prime}}{M}\;\frac{\partial U}{\partial\Omega}\,\;.\quad
 \ea
 In this situation,
 \ba
 \nonumber
 \frac{\partial \cal R}{\partial\Omega}\;=
 -\;\frac{M+M^{\,\prime}}{M}\;\sum_{l=2}^{\infty}\,\left(\frac{R}{a}\right)^{\textstyle{^{2l+1}}}\frac{{\cal{G}}\,M^{\,\prime}}{a}\,\sum_{m=0}^{l}\,\frac{(l - m)!}{(l + m)!}\;\left(2- \delta_{0m}\right)\sum_{p=0}^{l}F^{\,2}_{lmp}(\inc)\sum_{q=-\infty}^{\infty}G^{\,2}_{lpq}(e)\;m\;k_{l}\,\sin\epsilon_{lmpq}\\
 \nonumber\\
 \label{tacitly}
  =\;-\;n^2\;a^2\;\frac{\,M^{\,\prime}}{M\,}\;\sum_{l=2}^{\infty}\,\left(\frac{R}{a}\right)^{\textstyle{^{2l+1}}}\sum_{m=0}^{l}\,\frac{(l - m)!}{(l + m)!}\;\left(2- \delta_{0m}\right)\sum_{p=0}^{l}F^{\,2}_{lmp}(\inc)\sum_{q=-\infty}^{\infty}G^{\,2}_{lpq}(e)\;m\;k_{l}\,\sin\epsilon_{lmpq}
 ~~_{\textstyle{_{\textstyle .}}} \quad
 \ea

 Similarly, differentiation over the longitude of the node $\,\Omega^{\,\prime}\,$ reckoned from the secondary's equator gives
 \ba
 \nonumber
 \frac{\partial \cal R}{\partial\Omega^{\,\prime}}\;=
 -\;\frac{M+M^{\,\prime}}{M^{\,\prime}}\;\sum_{l=2}^{\infty}\,\left(\frac{R^{\,\prime}}{a}\right)^{\textstyle{^{2l+1}}}\frac{{\cal{G}}\,M}{a}\,\sum_{m=0}^{l}\,\frac{(l - m)!}{(l + m)!}\;\left(2- \delta_{0m}\right)\sum_{p=0}^{l}F^{\,2}_{lmp}(\inc^{\,\prime})\sum_{q=-\infty}^{\infty}G^{\,2}_{lpq}(e)\;m\;k_{l}^{\,\prime}\,\sin\epsilon_{lmpq}^{\,\prime}\\
 \nonumber\\
 \label{tacitly}
  =\;-\;n^2\;a^2\;\frac{M\,}{\,M^{\,\prime}}\;\sum_{l=2}^{\infty}\,\left(\frac{R^{\,\prime}}{a}\right)^{\textstyle{^{2l+1}}}\sum_{m=0}^{l}\,\frac{(l - m)!}{(l + m)!}\;\left(2- \delta_{0m}\right)\sum_{p=0}^{l}F^{\,2}_{lmp}(\inc^{\,\prime})\sum_{q=-\infty}^{\infty}G^{\,2}_{lpq}(e)\;m\;k_{l}^{\,\prime}\,\sin\epsilon_{lmpq}^{\,\prime}
 ~~_{\textstyle{_{\textstyle .}}} \quad
 \ea

 \section{Differentiation of $\,{\cal{R}}\,$ with respect to the inclination\label{D}}

 The derivative of $\,U\,$ with respect to $\,i\,$ is
 \ba
 \nonumber
  \frac{\partial\;}{\partial i}\,U(\erbold,\,\erbold^*) \,=\,
     %
-\sum_{l=2}^{\infty}\,\left(\frac{R}{a}\right)^{\textstyle{^{l+1}}}\frac{{\cal{G}}\,M^{\,\prime}}{a^*}\,\left(\frac{R}{a^*}\right)^{\textstyle{^l}}\sum_{m=0}^{l}\,\frac{(l - m)!}{(l + m)!}\;\left(2- \delta_{0m}\right)\sum_{p=0}^{l}F_{lmp}(\inc^*)\sum_{q=-\infty}^{\infty}G_{lpq}(e^*)~\quad~\quad
 ~\\
 \\
 \nonumber
 \sum_{h=0}^{\it l}\frac{dF_{lmh}(i)}{di}\sum_{j=-\infty}^{\infty}
 G_{lhj}(e)\;k_{l}(\omega_{lmpq})\,\cos\left[
 \left(v_{lmpq}^*-m\theta^*\right)-
 \left(v_{lmhj}-m\theta\right)-
 \epsilon_l(\omega_{lmpq}) \right]
 ~~_{\textstyle{_{\textstyle ,}}}~~~\qquad
 \ea
 For $\,\erbold\,=\,\erbold^*\,$, the \,{\it{secular part}}\, of this expression takes the form of
 \ba
 \frac{\partial U}{\partial i}\,=\,
    %
-\sum_{l=2}^{\infty}\left(\frac{R}{a}\right)^{\textstyle{^{l+1}}}\frac{{\cal{G}}\,M^{\,\prime}}{a}\left(\frac{R}{a}\right)^{\textstyle{^l}}
 \sum_{m=0}^{l}\frac{(l - m)!}{(l + m)!}\;\frac{2- \delta_{0m}}{2}\sum_{p=0}^{l}\frac{dF^2_{lmp}(\inc)}{di}\sum_{q=-\infty}^{\infty}G^2_{lpq}(e)\;k_{l}(\omega_{lmpq})\,\cos\epsilon_l(\omega_{lmpq})
 ~~_{\textstyle{_{\textstyle .}}}\quad
 \ea
 It should be noted that in the differentiation of $U$ with respect to $i$, the effect of the primary's oblateness $J_2$ does not average out as it was the case in the differentiation over $\,\cal M\,$, $\,\omega\,$ or $\,\Omega\,$.
 Combining this with formula (\ref{eq.R}), we obtain:
  \ba
 \nonumber
 \frac{\partial \cal R}{\partial i}\;=\;
 \frac{M+M^{\,\prime}}{M\;M^{\,\prime}}\;\frac{\partial\;}{\partial i}\,\left(\,-\,M^{\,\prime}\,U\,-\,M\,U^{\,\prime\,}\right)\;=\;
-\;\frac{M+M^{\,\prime}}{M}\;\frac{\partial U}{\partial i}\,\;,\quad
 \ea
 which gives
 \ba
 \nonumber
 \frac{\partial \cal R}{\partial i}\;=\;
   %
\frac{M+M^{\,\prime}}{M}\;\sum_{l=2}^{\infty}\,\left(\frac{R}{a}\right)^{\textstyle{^{2l+1}}}\frac{{\cal{G}}\,M^{\,\prime}}{a}\,\sum_{m=0}^{l}\,\frac{(l - m)!}{(l + m)!}\;\frac{2- \delta_{0m}}{2}\sum_{p=0}^{l}\frac{dF^{\,2}_{lmp}(i)}{di}\sum_{q=-\infty}^{\infty}G^{\,2}_{lpq}(e)\;m\;k_{l}\,\cos\epsilon_{lmpq}\;\;\\
 \nonumber\\
 \label{ta}
  =\;
  %
  n^2\;a^2\;\frac{\,M^{\,\prime}}{M\,}\;\sum_{l=2}^{\infty}\,\left(\frac{R}{a}\right)^{\textstyle{^{2l+1}}}\sum_{m=0}^{l}\,\frac{(l - m)!}{(l + m)!}\;\frac{2- \delta_{0m}}{2}\sum_{p=0}^{l}\frac{dF^{\,2}_{lmp}(i)}{di}\sum_{q=-\infty}^{\infty}G^{\,2}_{lpq}(e)\;m\;k_{l}\,\cos\epsilon_{lmpq}
 ~~_{\textstyle{_{\textstyle .}}} \quad
 \ea
 Similarly, the secular part of the derivative of the potential created by the tidally deformed secondary and acting on the primary is:
 \ba
 \nonumber
 \frac{\partial \cal R}{\partial i^{\,\prime}}\;
  =\;
  n^2\;a^2\;\frac{M\,}{\,M^{\,\prime}}\;\sum_{l=2}^{\infty}\,\left(\frac{R^{\,\prime}}{a}\right)^{\textstyle{^{2l+1}}}\sum_{m=0}^{l}\,\frac{(l - m)!}{(l + m)!}\;\frac{2- \delta_{0m}}{2}\sum_{p=0}^{l}\frac{dF^{\,2}_{lmp}(i^{\,\prime})}{di^{\,\prime}}\sum_{q=-\infty}^{\infty}G^{\,2}_{lpq}(e)\;m\;k_{l}^{\,\prime}\,\cos\epsilon_{lmpq}^{\,\prime}
 ~~_{\textstyle{_{\textstyle .}}} \quad
 \ea

 \section{Details of the calculation of $\,da/dt\,$}\label{AppendixA}

 Writing $\,da/dt\,$ in the leading order over the inclination requires the knowledge of the squares of the inclination functions
 $\;F_{201}^2\,=\,\frac{\textstyle 1}{\textstyle 4}\,+\,O(i^2)\;$ and $\;F_{220}^2\,=\,9\,+\,O(i^2)\;$, the other $\,F^2_{lmp}(i)\,$ being of order $\,O(i^2)\,$ or higher. So we shall work with the sets of integers $\;(lmpq)\,=\,(201q)\;$ and $\;(lmpq)\,=\,(220q)\;$. The corresponding eccentricity functions are:
 \ba
 \nonumber
 G_{21,-2}(e)&=&G_{212}(e)\;=\;\frac{9}{4}\;e^2\;\left(1\;+\;\frac{7}{9}\;e^2\right)\;+\;O(e^6)\\
 \nonumber
 G_{21,-1}(e)&=&G_{211}(e)\,=\,\frac{\textstyle 3}{\textstyle 2}\,e\,\left(1\,+\,\frac{\textstyle 9}{\textstyle 8}\,e^2\right)\,+\,O(e^5)\;\;,
 \qquad G_{210}(e)\;=\;(1\,-\,e^2)^{-3/2}\;\;,\\
 \nonumber\\
 G_{20,-1}(e)&=&-\;\frac{\textstyle 1}{\textstyle 2}\,e\,\left(1\,-\,\frac{\textstyle 1}{\textstyle 8}\,e^2\right)\,+\,O(e^5)\;\,\;,\qquad
 G_{201}(e)\;=\;\frac{\textstyle 7}{\textstyle 2}\,e\,-\,\frac{\textstyle 123}{\textstyle 16}\,e^3\,+\,O(e^5)\;\;,\;\qquad
 \label{G}\\
 \nonumber
 G_{20,-2}(e)&=&0\;\;\;,\quad G_{202}(e)\;=\;\frac{17}{2}\;e^2\;+\;O(e^4)\;\;,\;\;\;
 G_{200}(e)=1\;-\;\frac{\textstyle 5}{\textstyle 2}\,e^2\,+\;\frac{\textstyle 13}{\textstyle 16}\,e^4\,+\,O(e^6)\;\,\;.
 \ea
 Also mind that for $\;(lmpq)\,=\,(2010)\;$ the expression $\;(2-2p+q)\;$ is zero~---~and so is the input $\,(da/dt)_{2010}\,$.
 Below is an inventory of the relevant inputs:
 \ba
 \nonumber
 \left(\frac{da}{dt}\right)_{201,-2}=\;\frac{81}{16}\;a\,n\;\frac{M^{\,\prime}}{M\;}\;e^4\,\left[
 \left(\frac{R}{a}\right)^5K_2(-2n)     \right. \qquad\qquad\;\qquad\;\qquad \qquad\;\qquad \\
 \\
 \nonumber
 \left.
 +\;\left(\frac{M}{M^{\,\prime}}\right)^2\,\left(\frac{R^{\,\prime}}{a}\right)^5K^{\,\prime}_2(-2n)
 \right]\,+\,O(i^2)\,+\,O(e^6)
 \;\,,
 \ea
 \ba
 \nonumber
 \left(\frac{da}{dt}\right)_{201,-1}=\;\frac{9}{8}\;a\,n\;\frac{M^{\,\prime}}{M\;}\;e^2\,\left(1\,+\,\frac{9}{4}\,e^2\right)\,\left[
 \left(\frac{R}{a}\right)^5K_2(-n)     \right. \qquad\qquad\;\qquad\;\qquad \\
 \\
 \nonumber
 \left.
 +\;\left(\frac{M}{M^{\,\prime}}\right)^2\,\left(\frac{R^{\,\prime}}{a}\right)^5K^{\,\prime}_2(-n)
 \right]\,+\,O(i^2)\,+\,O(e^6)
 \;\,,
 \ea
 \ba
 \left(\frac{da}{dt}\right)_{2010}\;=\;0\;\,,\qquad\qquad\qquad\qquad\qquad\qquad\qquad\qquad\qquad\qquad\qquad\qquad\qquad\qquad\qquad\;\;\;
 \ea
 \ba
 \nonumber
 \left(\frac{da}{dt}\right)_{2011}=\;-\;\frac{9}{8}\;a\,n\;\frac{M^{\,\prime}}{M\;}\;e^2\,\left(1\,+\,\frac{9}{4}\,e^2\right)\,\left[
 \left(\frac{R}{a}\right)^5K_2(n)  \right. \qquad\qquad\;\qquad\;\qquad\quad \\
 \\
 \nonumber
 \left.
 +\;\left(\frac{M}{M^{\,\prime}}\right)^2\,\left(\frac{R^{\,\prime}}{a}\right)^5K^{\,\prime}_2(n)
 \right]\,+\,O(i^2)\,+\,O(e^6)
 \;\,,
 \ea
  \ba
 \nonumber
 \left(\frac{da}{dt}\right)_{2012}=\;-\;\frac{81}{16}\;a\,n\;\frac{M^{\,\prime}}{M\;}\;e^4\,\left[
 \left(\frac{R}{a}\right)^5K_2(2n)     \right. \qquad\qquad\;\qquad\;\qquad \qquad\;\qquad \\
 \\
 \nonumber
 \left.
 +\;\left(\frac{M}{M^{\,\prime}}\right)^2\,\left(\frac{R^{\,\prime}}{a}\right)^5K^{\,\prime}_2(2n)
 \right]\,+\,O(i^2)\,+\,O(e^6)
 \;\,,
 \ea
 \ba
 \nonumber
 \left(\frac{da}{dt}\right)_{220,-1}=\;-\;\frac{3}{8}\;a\,n\;\frac{M^{\,\prime}}{M\;}\;e^2\,\left(1\,-\,\frac{1}{4}\,e^2\right)\,\left[
 \left(\frac{R}{a}\right)^5K_2(n-2\dot{\theta})     \right. \qquad\qquad \\
 \\
 \nonumber
 \left.
 +\;\left(\frac{M}{M^{\,\prime}}\right)^2\,\left(\frac{R^{\,\prime}}{a}\right)^5K^{\,\prime}_2(n-2\dot{\theta}^{\,\prime})
 \right]\,+\,O(i^2)\,+\,O(e^6)
 \;\,,
 \ea
 \ba
 \nonumber
 \left(\frac{da}{dt}\right)_{2200}=\;-\;3\;a\,n\;\frac{M^{\,\prime}}{M\;}\,\left(1\,-\,5\,e^2\,+\,\frac{63}{8}\,e^4\right)\,\left[
 \left(\frac{R}{a}\right)^5K_2(2n-2\dot{\theta})    \right. \qquad\qquad \\
 \\
 \nonumber
 \left.
 +\;\left(\frac{M}{M^{\,\prime}}\right)^2\,\left(\frac{R^{\,\prime}}{a}\right)^5K^{\,\prime}_2(2n-2\dot{\theta}^{\,\prime})
 \right]\,+\,O(i^2)\,+\,O(e^6)
 \;\,,
 \ea
 \ba
 \nonumber
 \left(\frac{da}{dt}\right)_{2201}=\;-\;\frac{441}{8}\;a\,n\;\frac{M^{\,\prime}}{M\;}\;e^2\,\left(1\,-\,\frac{123}{28}\,e^2\right)\,\left[
 \left(\frac{R}{a}\right)^5K_2(3n-2\dot{\theta})     \right. \qquad\qquad \\
 \\
 \nonumber
 \left.
 +\;\left(\frac{M}{M^{\,\prime}}\right)^2\,\left(\frac{R^{\,\prime}}{a}\right)^5K^{\,\prime}_2(3n-2\dot{\theta}^{\,\prime})
 \right]\,+\,O(i^2)\,+\,O(e^6)
 \;\,,
 \ea
  \ba
 \nonumber
 \left(\frac{da}{dt}\right)_{2202}=\;-\;\frac{867}{2}\;a\,n\;\frac{M^{\,\prime}}{M\;}\;e^4\,\left[
 \left(\frac{R}{a}\right)^5K_2(4n-2\dot{\theta})     \right. \qquad\qquad\qquad\qquad\quad\quad\;\, \\
 \\
 \nonumber
 \left.
 +\;\left(\frac{M}{M^{\,\prime}}\right)^2\,\left(\frac{R^{\,\prime}}{a}\right)^5K^{\,\prime}_2(4n-2\dot{\theta}^{\,\prime})
 \right]\,+\,O(i^2)\,+\,O(e^6)
 \;\,,
 \ea
 where we made use of notation (\ref{notation}) and (\ref{notationprime}).

 \section{Details of the calculation of $\,de/dt\,$}\label{AppendixB}\label{AppE}

  In notation (\ref{notation} - \ref{omegaprime}), our expression (\ref{30}) becomes
 \begin{eqnarray}
 \nonumber
 \frac{de}{dt}\,=\;-\;\frac{(1-e^2)^{1/2}}{e}\;n\;
 \sum_{l=2}^{\infty}\sum_{m=0}^{l}\,\frac{(l - m)!}{(l + m)!}\;\left(2- \delta_{0m}\right)\,\sum_{p=0}^{l}\sum_{q=-\infty}^{\infty}G^{\,2}_{lpq}(e)
 \left[\,(l-2p+q)\,(1-e^2)^{1/2} \right.\;
 ~\\
 \label{31b}\\
 \nonumber
 \left. \,-\;(l-2p)\,\right]\;
 \left[\,
 \frac{\,M^{\,\prime}}{M\,}\;\left(\frac{R}{a}\right)^{\textstyle{^{2l+1}}}\,F^{\,2}_{lmp}(i)\;K_{l}(\omega_{lmpq})
 \;+\;
 \frac{M\,}{\,M^{\,\prime}}\;\left(\frac{R^{\,\prime}}{a}\right)^{\textstyle{^{2l+1}}}\,F^{\,2}_{lmp}(i^{\,\prime})\;K^{\,\prime}_{l}(\omega^{\,\prime}_{lmpq})
 \,\right]
 ~~_{\textstyle{_{\textstyle ,}}}
 \end{eqnarray}
 its quadrupole part being
 \begin{eqnarray}
 \nonumber
 \left(\frac{de}{dt}\right)_{l=2}\,=\;-\;n\;
 \sum_{m=0}^{2}\,\frac{(2 - m)!}{(2 + m)!}\;\left(2- \delta_{0m}\right)\,\sum_{p=0}^{l}\sum_{q=-\infty}^{\infty}G^{\,2}_{2pq}(e) \qquad\;\qquad\;\qquad\;\qquad\;\qquad\;\qquad\\
 \nonumber\\
 \nonumber
 \frac{(2-2p+q)\,(1-e^2)\,-\;(2-2p)\,(1-e^2)^{1/2}}{e}\;
 \left[\,
 \frac{\,M^{\,\prime}}{M\,}\;\left(\frac{R}{a}\right)^{\textstyle{^{2l+1}}}\,F^{\,2}_{lmp}(i)\;K_{l}(\omega_{lmpq})
 \right.   \quad\,\quad
 ~\\
 \left.
 \;+\;
 \frac{M\,}{\,M^{\,\prime}}\;\left(\frac{R^{\,\prime}}{a}\right)^{\textstyle{^{2l+1}}}\,F^{\,2}_{lmp}(i^{\,\prime})\;K^{\,\prime}_{l}(\omega^{\,\prime}_{lmpq})
 \,\right]
 ~~_{\textstyle{_{\textstyle .}}}
 \label{32}
 \end{eqnarray}

 To write expression (\ref{32}) in the leading order over the inclination, we shall need the squares of the two relevant $\,F_{lmp}(i)\,$ functions:
 $\;F_{201}^2\,=\,\frac{\textstyle 1}{\textstyle 4}\,+\,O(i^2)\;$ and $\;F_{220}^2\,=\,9\,+\,O(i^2)\;$, all the other $\,F^2_{lmp}(i)\,$ being of  order $\,O(i^2)\,$ or higher. This way, we shall be interested in the following sets of integers: $\;(lmpq)\,=\,(201q)\;$ and $\;(lmpq)\,=\,(220q)\;$. The relevant eccentricity functions are given by equations (\ref{G}) above.

 For  $\;(lmpq)\,=\,(2010)\;$, both the expressions $\;(2-2p+q)\;$ and $\;(2-2p)\;$ vanish~--- and so does the $\,(de/dt)_{2010}\;$ input, up to higher-order terms in the inclinations:
 \begin{eqnarray}
 \left(\frac{de}{dt}\right)_{2010}\,=\;0\;+\;O(i^2)\;+\;O({i^{\;\prime\;}}^2)~~_{\textstyle{_{\textstyle .}}}
 \label{33}
 \end{eqnarray}
 Thence, of the sets $\;(lmpq)\,=\,(201q)\;$, only those with $\,q\,=\,-2,\,-1,\,1,\,2\,$ are important:
 \begin{eqnarray}
 \nonumber
 \left(\frac{de}{dt}\right)_{201,-1}=\,\left(\frac{de}{dt}\right)_{2011}=\;-\;\frac{9}{16}\,n\,e\,\left(1\,+\,\frac{5}{4}\,e^2\right)\,
 \left[
 \frac{M^{\,\prime}}{M\,}\,\left(\frac{R}{a}\right)^{\textstyle{^{5}}}K_{2}(n)
 \right.\qquad\qquad\;\qquad\;\qquad\\
 \label{34}\\
 \nonumber
 \left.
 \,+\;
 \frac{M\,}{M^{\,\prime}}\,\left(\frac{R^{\,\prime}}{a}\right)^{\textstyle{^{5}}}K^{\,\prime}_{2}(n)
 \right]
 \,+\,O(e^5)\,+\,O(i^2)\,+\,O({i^{\,\prime\;}}^2)~~_{\textstyle{_{\textstyle ,}}}\qquad\qquad
 \end{eqnarray}
 \begin{eqnarray}
 \nonumber
 \left(\frac{de}{dt}\right)_{201,-2}=\,\left(\frac{de}{dt}\right)_{2012}=\qquad\qquad\qquad\qquad\qquad\qquad\qquad\qquad\;\qquad\qquad\qquad\;\qquad\;\qquad\\
 \\
 \nonumber
 -\;\frac{81}{32}\;n\,e^3\,
 \left[
 \frac{M^{\,\prime}}{M\,}\;\left(\frac{R}{a}\right)^{\textstyle{^{5}}}K_{2}(2n)
 \;+\;
 \frac{M\,}{M^{\,\prime}}\;\left(\frac{R^{\,\prime}}{a}\right)^{\textstyle{^{5}}}K^{\,\prime}_{2}(2n)
 \right]
 \;+\;O(e^5)\;+\;O(i^2)\,+\,O({i^{\,\prime\;}}^2)~~_{\textstyle{_{\textstyle .}}}
 \end{eqnarray}
 Of the sets $\,(lmpq)\,=\,(220q)\,$, we shall be interested in the ones with $\,q\,=\,-1,\,0,\,1,\,2\;:$
 \begin{eqnarray}
 \nonumber
 \left(\frac{de}{dt}\right)_{220,-1}\;=\;\frac{3}{16}\;n\;e\;\left(1\,-\,\frac{e^2}{4}\right)\;
 \left[
 \frac{\,M^{\,\prime}}{M\,}\;\left(\frac{R}{a}\right)^{\textstyle{^{5}}}\;K_{2}(n-2\dot{\theta})    \right.\qquad\qquad\qquad\qquad\qquad\qquad
 ~\\
 \\
 \left.
 \;+\;
 \frac{M\,}{\,M^{\,\prime}}\;\left(\frac{R^{\,\prime}}{a}\right)^{\textstyle{^{5}}}\;K^{\,\prime}_{2}(n-2\dot{\theta}^{\,\prime})
 \right]
 \,+\,O(e^5)\,+\,O(i^2)\,+\,O({i^{\,\prime\;}}^2)~~_{\textstyle{_{\textstyle ,}}}\;\;\;\;\qquad
 \nonumber
 \end{eqnarray}
 \begin{eqnarray}
 \nonumber
 \left(\frac{de}{dt}\right)_{2200}\;=\;\frac{3}{4}\;n\;e\;\left(1\,-\,\frac{21}{4}\,e^2\right)\;
 \left[
 \frac{\,M^{\,\prime}}{M\,}\;\left(\frac{R}{a}\right)^{\textstyle{^{5}}}\;K_{2}(2n-2\dot{\theta})        \right.\quad\qquad\qquad\qquad\qquad\qquad
 ~\\
 \\
 \left.
 \;+\;
 \frac{M\,}{\,M^{\,\prime}}\;\left(\frac{R^{\,\prime}}{a}\right)^{\textstyle{^{5}}}\;K^{\,\prime}_{2}(2n-2\dot{\theta}^{\,\prime})
 \right]
 \,+\,O(e^5)\,+\,O(i^2)\,+\,O({i^{\,\prime\;}}^2)~~_{\textstyle{_{\textstyle ,}}}\;\qquad
 \nonumber
 \end{eqnarray}
 \begin{eqnarray}
 \nonumber
 \left(\frac{de}{dt}\right)_{2201}=\;-\;\frac{147}{16}\;n\,e\;\left(1\,-\,\frac{\textstyle 179}{\textstyle 28}\,e^2\right)\,
 \left[
 \frac{M^{\,\prime}}{M\,}\,\left(\frac{R}{a}\right)^{\textstyle{^{5}}}\,K_{2}(3n-2\dot{\theta})   \right. \quad\qquad\qquad\qquad\qquad
 ~\\
 \\
 \left.
 \;+\;
 \frac{M\,}{M^{\,\prime}}\,\left(\frac{R^{\,\prime}}{a}\right)^{\textstyle{^{5}}}\,K^{\,\prime}_{2}(3n-2\dot{\theta}^{\,\prime})
 \right]
 \,+\,O(e^5)\,+\,O(i^2)\,+\,O({i^{\,\prime\;}}^2)~_{\textstyle{_{\textstyle ,}}}\quad
 \nonumber
 \end{eqnarray}
 \begin{eqnarray}
 \nonumber
 \left(\frac{de}{dt}\right)_{2202}=\;-\;\frac{867}{8}\;n\,e^3\;
 \left[
 \frac{M^{\,\prime}}{M\,}\,\left(\frac{R}{a}\right)^{\textstyle{^{5}}}\,K_{2}(4n-2\dot{\theta})  \right. \qquad\qquad\qquad\qquad\qquad
 ~\\
 \label{37}\\
 \left.
 \;+\;
 \frac{M\,}{M^{\,\prime}}\,\left(\frac{R^{\,\prime}}{a}\right)^{\textstyle{^{5}}}\,K^{\,\prime}_{2}(4n-2\dot{\theta}^{\,\prime})
 \right]
 \,+\,O(e^5)\,+\,O(i^2)\,+\,O({i^{\,\prime\;}}^2)~_{\textstyle{_{\textstyle ,}}}\qquad\qquad
 \nonumber
 \end{eqnarray}
 where notation (\ref{notation} - \ref{omegaprime}) was employed.

 \section{Details of the calculation of $\,di/dt\,$}\label{AppendixI}

 According to equation~(\ref{eq.didt}), the evolution rate of the inclination involves derivatives of the perturbing function with respect to $\,\omega\,$, $\,\omega^{\,\prime}\,$, $\,\Omega\,$, $\,\Omega^{\,\prime}\,$ and $\,i^{\,\prime}\,$. Nevertheless, in the secular expression (\ref{eq.explicitdidt}), only differentiations over $\,\omega\,$ and $\,\Omega\,$ remain. These are given in Appendices~\ref{AppB} and \ref{AppC}.

 To write expression (\ref{eq.explicitdidt}) in the leading order over the inclination, we should keep in mind that in this expression the squared inclination functions $\,F^{\,2}_{lmp}(i)\,$ are accompanied by a factor of either $\,\alpha_{lmp}\,=\,\left[(l-2p)\,\cos i-m\right]/\sin i\;$ or $\;\beta_{lmp}\,=\,\left[m\,\cos i-(l-2p)\right]/\sin i\,$. The functions $\,F^{\,2}_{201}(i)\,$ and $\,F^{\,2}_{220}(i)\,$ are both of order $\,O(i^{\,0})\,$, but for $\,(lmp)=(201)\,$, the two factors $\,\alpha_{201}\,$ and $\,\beta_{201}\,$ vanish. In the case $\,(lmp)=(220)\,$, we have
\begin{equation}
\alpha_{220}\,=\;\beta_{220}\,=\;-\;\sin i\;+\;O(i^{\,3})\,\;.
\end{equation}
 Here we also have to consider the functions $\,F^{\,2}_{lmp}(i)\,$ of order $\,O(i^{\,2})\,$, namely $\,\displaystyle F^{\,2}_{210}\,= \,\frac{9}{4}\,\sin i^{\,2}\,+\,O(i^{\,4})\,$ and $\,\displaystyle F^{\,2}_{211}\,=\,\frac{9}{4}\,\sin i^{\,2}\,+\,O(i^{\,4})\,$, all the other $\,F^2_{lmp}(i)\,$ being of order $\,O(i^{\,4})\,$. The corresponding factors are $\,\alpha_{210}\,=\,1/\sin i\,+\,O(i)\,$, $\;\alpha_{211}\,=\,-\,1/\sin i\,+\,O(i)\,$, $\;\beta_{210}\,=\,-1/\sin i\,+\,O(i)\,$, $\;\beta_{211}\,=\,1/\sin i\,+\,O(i)\,$.

 Therefore, we shall be interested in the following sets of integers: $\,(lmpq)\,=\,(220q)\,$, $\,(lmpq)\,=\,(210q)\,$ and $\,(lmpq)\,=\,(211q)$, the corresponding eccentricity functions being given by equations (\ref{G}) above. Below we provide the resulting contributions. Deriving these, we used $\,1/\sqrt{1-e^2}\,=\,1\,+\,e^2/2\,+\,3e^4/8\,+\,O(e^6)\,$ and then, in each contribution, truncated this expansion as necessary to keep the overall answer precise up to $\,e^4\,$, inclusively.
 \ba
 \left(\frac{di}{dt}\right)_{220,-2}=\;0\,\;,\qquad\qquad\qquad\;\qquad\qquad\qquad\qquad\qquad\qquad\;\qquad\;\qquad\qquad\;\qquad\;\qquad\;\qquad
 \label{}
 \ea
 \ba
 \left(\frac{di}{dt}\right)_{220,-1}=\;-\;\frac{3}{16}\,n\,e^2\,\left(1\,-\,\frac{1}{4}\,e^2\right)
 \,\sin i\;\frac{M^{\,\prime}}{M\;}\,\left(\frac{R}{a}\right)^5
 \,\left[\frac{\beta\,n\,a^2}{C\;\dot{\theta}}\,-\,\left(1+\frac{1}{2}\,e^2\right) \right]\,K_2(n-2\dot{\theta})\,+\,O(i^3)\,+\,O(e^6)\,\;,\quad
 \label{}
 \ea
 \ba
 \left(\frac{di}{dt}\right)_{2200}=\,-\,\frac{3}{4}\,n\,\left(1-5e^2+\frac{63}{8}e^4\right)\,\sin i\;\frac{M^{\,\prime}}{M}\,\left(\frac{R}{a}\right)^5\,\left[\frac{\beta\,n\,a^2}{C\;\dot{\theta}}\,-\,\left(1+\frac{1}{2}e^2+\frac{3}{8}e^4\right) \right]\,
 K_2(2n-2\dot{\theta})\,+\,O(i^3)\,+\,O(e^6)\;,\quad
 \label{}
 \ea
 \ba
 \left(\frac{di}{dt}\right)_{2201}=\,-\;\frac{147}{16}\,n\,e^2\,\left(1\,-\,\frac{123}{28}\,e^2\right)
 \,\sin i\;\frac{M^{\,\prime}}{M\;}\,\left(\frac{R}{a}\right)^5
 \,\left[\frac{\beta\,n\,a^2}{C\;\dot{\theta}}\,-\,\left(1+\frac{1}{2}\,e^2\right) \right]\,K_2(3n-2\dot{\theta})\,+\,O(i^3)\,+\,O(e^6)\,\;,\quad
 \label{}
 \ea
 \ba
 \left(\frac{di}{dt}\right)_{2202}=\,-\;\frac{867}{16}\,n\,e^4\,\sin i\;\frac{M^{\,\prime}}{M\;}\,\left(\frac{R}{a}\right)^5
 \,\left[\frac{\beta\,n\,a^2}{C\;\dot{\theta}}\,-\,1 \right]\,K_2(4n-2\dot{\theta})\,+\,O(i^3)\,+\,O(e^6)\,\;,\;\;\qquad\qquad\qquad
 \label{}
 \ea
  \ba
 \left(\frac{di}{dt}\right)_{210,-2}=\;0\,\;,\qquad\qquad\qquad\;\qquad\qquad\qquad\qquad\qquad\qquad\;\qquad\;\qquad\qquad\;\qquad\;\qquad\;\qquad
 \label{}
 \ea
 \ba
 \left(\frac{di}{dt}\right)_{210,-1}=\,-\,\frac{3}{16}\,n\,e^2\,\left(1\,-\,\frac{1}{4}\,e^2\right)\,\sin i\;\frac{M^{\,\prime}}{M}\,\left(\frac{R}{a}\right)^5\,\left[\frac{\beta\,n\,a^2}{C\;\dot{\theta}}\,+\,\left(1+\frac{1}{2}e^2\right)\right]\,K_2(n-\dot{\theta})
 \,+\,O(i^3)\,+\,O(e^6)\,\;,\qquad
 \label{}
 \ea
 \ba
\left(\frac{di}{dt}\right)_{2100}=\,-\,\frac{3}{4}\;n\,\left(1\,-\,5\,e^2\,+\,\frac{63}{8}\,e^4 \right)\,\sin i\;\frac{M^{\,\prime}}{M}\,\left(\frac{R}{a}\right)^5
 \,\left[\frac{\beta\,n\,a^2}{C\;\dot{\theta}}\,+\,\left(1+\frac{1}{2}e^2+\frac{3}{8}e^4\right)\right]\,K_2(2n-\dot{\theta})
 \,+\,O(i^3)\,+\,O(e^6)\,\;,\quad
 \label{}
 \ea
 \ba
 \left(\frac{di}{dt}\right)_{2101}=\;-\;\frac{147}{16}\;n\;e^2\,\left(1\,-\,\frac{123}{28}\,e^2\right)\;\sin i\;
 \frac{M^{\,\prime}}{M}\,\left(\frac{R}{a}\right)^5\,
 \left[\frac{\beta\,n\,a^2}{C\;\dot{\theta}}\,+\,\left(1+\frac{1}{2}e^2\right)\right]\,K_2(3n-\dot{\theta})
 \,+\,O(i^3)\,+\,O(e^6)\,\;,\quad
 \label{}
 \ea
 \ba
\left(\frac{di}{dt}\right)_{2102}=\;-\;\frac{867}{16}\;n\;e^4\,\sin i\;\frac{M^{\,\prime}}{M}\,\left(\frac{R}{a}\right)^5
 \,\left[\frac{\beta\,n\,a^2}{C\;\dot{\theta}}\,+\,1\right]\,K_2(4n-\dot{\theta})
 \,+\,O(i^{\,3})\,+\,O(e^{\,6})\,\;,\;\;\qquad\;\qquad\;
 \label{}
 \ea
 \ba
\left(\frac{di}{dt}\right)_{211,-2}=\;\frac{243}{64}\;n\;e^4\;\sin i\;\frac{M^{\,\prime}}{M}\,\left(\frac{R}{a}\right)^5
 \;\left[\frac{\beta\,n\,a^2}{C\;\dot{\theta}}\,+\,1\right]\;\,K_2(-2n-\dot{\theta})\,+\,O(i^3)\,+\,O(e^6)\,\;,\qquad\;\qquad
 \label{}
 \ea
 \ba
 \left(\frac{di}{dt}\right)_{211,-1}=\;\frac{27}{16}\;n\;e^2\,\left(1\,+\,\frac{9}{4}\,e^2\right)\;\sin i\;\frac{M^{\,\prime}}{M}\,\left(\frac{R}{a}\right)^5\,\left[\frac{\beta\,n\,a^2}{C\;\dot{\theta}}\,+\,\left(1+\frac{1}{2}e^2\right)\right]\;K_2(-n-\dot{\theta})\,+\,O(i^3)\,+\,O(e^6)\,\;,\quad
 \label{}
 \ea
 \ba
 \left(\frac{di}{dt}\right)_{2110}=\;\frac{3}{4}\;n\;\left(1+3e^2+6e^4\right)\;\sin i\;\frac{M^{\,\prime}}{M}\,\left(\frac{R}{a}\right)^5\,\left[\frac{\beta\,n\,a^2}{C\;\dot{\theta}}\,+\,\left(1+\frac{1}{2}e^2+\frac{3}{8}e^4\right)\right]\;K_2(-\dot{\theta})\,+\,O(i^3)\,+\,O(e^6)\,\;,
 \label{}
 \ea
 \ba
 \left(\frac{di}{dt}\right)_{2111}\,=\;\frac{27}{16}\;n\;e^2\,\left(1\,+\,\frac{9}{4}\,e^2\right)\;\sin i\;\frac{M^{\,\prime}}{M}\,\left(\frac{R}{a}\right)^5\,\left[\frac{\beta\,n\,a^2}{C\;\dot{\theta}}\,+\,\left(1+\frac{1}{2}e^2\right)\right]\;K_2(n-\dot{\theta})\,+\,O(i^3)\,+\,O(e^6)\,\;,\qquad\;
 \label{}
 \ea
 \ba
 \left(\frac{di}{dt}\right)_{2112}\,=\;\frac{243}{64}\;n\;e^4\,\sin i\;
 \frac{M^{\,\prime}}{M}\,\left(\frac{R}{a}\right)^5\,\left[\frac{\beta\,n\,a^2}{C\;\dot{\theta}}\;
 +\,1\right]\;K_2(2n-\dot{\theta})\,+\,O(i^{\,3})\,+\,O(e^{\,6})\,\;.\qquad\;\;
 \ea
 Without loss of precision, in all these formulae $\;\sin i\;$ may be changed to $\;i\;$.

 \section{Long-period oscillating terms in the inclination rate
 \label{sec.nonaveragedidt}}

 To derive the inclination rate $\,di/dt\,$ non-averaged over the long period oscillating terms, it is necessary to include the contribution due to the $\,J_2\,$ in the potential energy: \footnote{~This contribution comprises the $\,\{lmhj\}\,=\,\{20hj\}\,$ terms from the expansion for the potential energy (Frouard \& Efroimsky 2017a, eqn 115).}
 \ba
 \nonumber
 U(\erbold,\,\erbold^{\,*})\;=\;J_2\;\frac{\mathcal{G}\,M}{a}\,\left(\frac{R}{a}\right)^2\sum_{h=0}^2 F_{20h}(i) \sum_{j=-\infty}^\infty
 G_{2hj}(e)\;\cos v_{20hj}~\qquad~\qquad\qquad\qquad\qquad\qquad\qquad\\
 \nonumber\\
 -\;\sum_{l=2}^{\infty}\,\left(\frac{R}{a}\right)^{\textstyle{^{l+1}}}\frac{\mathcal{G}\,M^{\,\prime}}{a^*}\,\left(\frac{R}{a^*}\right)^{\textstyle{^l}}\sum_{m=0}^{l}\,\frac{(l - m)!}{(l + m)!}\;\left(2- \delta_{0m}\right)\sum_{p=0}^{l}F_{lmp}(\inc^*)~\qquad~\qquad~\qquad\qquad\quad
                                   \label{21J2}
                                   \label{A21J2}\\
                                    \nonumber
 \sum_{q=-\infty}^{\infty}G_{lpq}(e^*)
 \sum_{h=0}^{\it l}F_{lmh}(\inc)\sum_{j=-\infty}^{\infty}
 G_{lhj}(e)\;k_{l}(\omega_{lmpq})\,\cos\left[
 \left(v_{lmpq}^*-m\theta^*\right)-
 \left(v_{lmhj}-m\theta\right)-
 \epsilon_l(\omega_{lmpq}) \right]
 ~~_{\textstyle{_{\textstyle .}}}~~~
  \ea
 To calculate $\,di/dt\,$, we insert the above in the formula (\ref{eq.R}) for $\,\mathcal{R}\,$,
 then plug the result in the orbital equation (\ref{eq.didt}), and finally carry out an averaging over the mean anomaly $\mathcal{M}$.
 This entails:

 \ba
 \frac{di}{dt}\;=\;n\;\frac{\beta n a^2}{C\dot\theta\sin i}
   \,J_2\,\left(\frac{R}{a}\right)^2 \sum_{h=0}^2
   (2-2h)\,F_{20h}(i)\,
    G_{2hj}(e)\;\sin (2-2h)\omega
\nonumber\\
\nonumber\\
\nonumber\\
 +\;\frac{n\;\cos i}{(1-e^2)^{1/2}\;\sin i}J_2 \left(\frac{R}{a}\right)^2
\sum_{h=0}^2
   (2-2h)F_{20h}(i)
    G_{2hj}(e)\;\sin (2-2h)\omega
\nonumber\\
\nonumber\\
\nonumber\\
+\;n\frac{\sin(\omega-\omega^{\,\prime})}{(1-e^2)^{1/2}}J^{\,\prime}_2 \left(\frac{R^{\,\prime}}{a}\right)^2
\sum_{h=0}^2 \frac{dF_{20h}(i^{\,\prime})}{di^{\,\prime}} G_{2hj}(e)\;\cos (2-2h)\omega^{\,\prime}
\nonumber\\
\nonumber\\
\nonumber\\
 +\;n\frac{\cos(\omega-\omega^{\,\prime})\;\cos i^{\,\prime}}{(1-e^2)^{1/2}\;\sin i^{\,\prime}} J^{\,\prime}_2 \left(\frac{R^{\,\prime}}{a}\right)^2
\sum_{h=0}^2 (2-2h)\;F_{20h}(i^{\,\prime})\;G_{2hj}(e)\;\sin(2-2h)\omega^{\,\prime}
\nonumber\\
\nonumber\\
\nonumber\\
 +\;n\;\frac{\beta n a^2}{C\dot\theta} \frac{M^{\,\prime}}{M}\sum_{l=2}^\infty \left(\frac{R}{a}\right)^{2l+1}
\sum_{m=0}^l \frac{(l-m)!}{(l+m)!} (2-\delta_{0m}) \sum_{p=0}^l F_{lmp}(i) \sum_{q=-\infty}^\infty
G_{lpq}(e)
\nonumber\\
\times \sum_{h=0}^l \frac{(l-2h)-m\cos i}{\sin i}F_{lmh}(i) G_{2hr}(e)
k_l(\omega_{lmpq}) \sin[2(h-p)\omega -
\epsilon_l(\omega_{lmpq})]
\nonumber\\
\nonumber\\
\nonumber\\
 -\;n\;\frac{1}{(1-e^2)^{1/2}} \frac{M^{\,\prime}}{M} \sum_{l=2}^\infty \left(\frac{R}{a}\right)^{2l+1}
\sum_{m=0}^l \frac{(l-m)!}{(l+m)!}(2-\delta_{0m}) \sum_{p=0}^l F_{lmp}(i) \sum_{q=-\infty}^\infty
G_{lpq}(e) \nonumber\\
\times \sum_{h=0}^l \frac{m-(l-2h)\cos i}{\sin i} F_{lmh}(i) G_{2hr}(e) k_l(\omega_{lmpq})
\sin[2(h-p)\omega-\epsilon_l(\omega_{lmpq})]
\nonumber\\
\nonumber\\
\nonumber\\
 -\;n\;\frac{\sin(\omega-\omega^{\,\prime})}{(1-e^2)^{1/2}} \frac{M}{M^{\,\prime}}\sum_{l=2}^\infty \left(\frac{R^{\,\prime}}{a}\right)^{2l+1}
\sum_{m=0}^l \frac{(l-m)!}{(l+m)!}(2-\delta_{0m}) \sum_{p=0}^l F_{lmp}(i^{\,\prime}) \sum_{q=-\infty}^\infty
G_{lpq}(e) \nonumber\\
\times \sum_{h=0}^l \frac{dF_{lmh}(i^{\,\prime})}{di^{\,\prime}} G_{2hr}(e) k^{\,\prime}_l(\omega_{lmpq})
\cos[2(h-p)\omega^{\,\prime}-\epsilon^{\,\prime}_l(\omega^{\,\prime}_{lmpq})]
\nonumber\\
\nonumber\\
\nonumber\\
 -\;n\;\frac{\cos(\omega-\omega^{\,\prime})}{(1-e^2)^{1/2}} \frac{M}{M^{\,\prime}}\sum_{l=2}^\infty \left(\frac{R^{\,\prime}}{a}\right)^{2l+1}
\sum_{m=0}^l\frac{(l-m)!}{(l+m)!}(2-\delta_{0m})\sum_{p=0}^l F_{lmp}(i^{\,\prime}) \sum_{q=-\infty}^\infty
G_{lpq}(e) \nonumber\\
\times \sum_{h=0}^l \frac{m-(l-2h)\cos i^{\,\prime}}{\sin i^{\,\prime}}\;F_{lmh}(i^{\,\prime})\;G_{2hr}(e)\;k^{\,\prime}_l(\omega^{\,\prime}_{lmpq})
 \;\sin[2(h-p)\omega^{\,\prime}-\epsilon^{\,\prime}_l(\omega_{lmpq})]\;\;,
 \ea
 with $j=2h-2$ and $r=2(h-p)+q$.

 \end{document}